\documentclass[twoside,twocolumn,english,prc,amsmath,myheadings]{article}
\usepackage{mathpazo}
\usepackage[T1]{fontenc}
\usepackage[latin9]{inputenc}
\usepackage[a4paper]{geometry}
\usepackage{color}
\usepackage{verbatim}
\usepackage{float}
\usepackage{amssymb}
\usepackage{graphicx}
\usepackage{subscript}

\makeatletter
\newcommand{\lyxaddress}[1]{
	\par {\raggedright #1
	\vspace{1.4em}
	\noindent\par}
}

\def\@oddhead{\rightmark \hfill HI collisions from 62.4 GeV down to 4 GeV in the EPOS4 framework \hfill \thepage}
\def\@evenhead{\thepage \hfill K. Werner et al.\hfill}
\def\fnum@table{\tablename~{\bf\thetable}}
\def\fnum@figure{\figurename~{\bf\thefigure}}
\def\tablename{\footnotesize{\bf Table}}
\def\figurename{\footnotesize{\bf Figure}}

\usepackage{dcolumn}
\usepackage[font=small,labelfont=bf]{caption}

\def\citet{\cite}

\AtBeginDocument{
  
}

\usepackage{babel}

\makeatother

\usepackage{babel}
\begin{document}
\twocolumn[   \begin{@twocolumnfalse}  
\title{Heavy ion collisions from $\sqrt{s_{NN}}$ of 62.4~GeV down to 4~GeV in the EPOS4
framework}
\author{{\normalsize{}K.$\,$Werner$^{(1)}$, J. Jahan $^{(2)}$, I. Karpenko
$^{(3)}$, T. Pierog $^{(4)}$, M. Stefaniak $^{(5,6)}$}\textsuperscript{}{\normalsize{},
D. Vintache $^{(1)}$}}
\maketitle

\lyxaddress{\begin{center}
$^{(1)}$SUBATECH, Nantes University -- IN2P3/CNRS -- IMT Atlantique,
44300 Nantes, France\\
$^{(2)}$Department of Physics, University of Houston, Houston, TX
77204, USA\\
$^{(3)}$Faculty of Nuclear Sciences and Physical Engineering, Czech
Technical University in Prague, \\
 B\v{r}ehov\'{a} 7, Prague, Czech Republic\\
$^{(4)}$Institute for Astroparticle Physics, Karlsruhe Institute
of Technology, Karlsruhe, Germany\\
$^{(5)}$Department of Physics, The Ohio State University, Columbus,
Ohio 43210, USA\\
$^{(6)}$GSI Helmholtz Centre for Heavy Ion Research, 64291 Darmstadt,
Germany
\par\end{center}}
\begin{abstract}
The EPOS4 project is an attempt to construct a realistic model for
describing relativistic collisions of different systems, from proton-proton
($pp$) to nucleus-nucleus ($AA$), at energies from several TeV per
nucleon down to several GeV. We argue that a parallel scattering formalism
(as in EPOS4) is relevant for primary scatterings in AA collisions
above 4 GeV, whereas sequential scattering (cascade) is appropriate
below. We present briefly the basic elements of EPOS4, and then investigate
heavy ion collisions from 62.4 GeV down to 4 GeV, to understand
how physics changes with energy, studying in particular the disappearance
of the fluid component at low energies.


~~
\end{abstract}
\end{@twocolumnfalse}]

\section{Introduction\label{=======Introduction=======}}

Scatterings of two nuclei are fundamentally different at high energies
(several TeV per nucleon) compared to low energies (several GeV). At
high energies, one can clearly separate ``primary scatterings'',
happening instantaneously (at $t=0$) and ``secondary scatterings''
of the primary particles (for $t>0$), which may lead to thermalized
matter (quark gluon plasma, QGP) and hadronic rescattering after hadronization.
At low energies, one cannot separate these two stages anymore, and
at very low energies (< 4 GeV), heavy ion collisions can be treated
by purely sequential hadronic scatterings, also referred to as hadronic
cascade \cite{Bas98,Ble99,PHSD:bratkovskaya2011,PHSD:cassing2008,PHSD:cassing2009,PHSD:moreau2019,PHSD:song2015,SMASH:2016}.
So the physics picture changes drastically when going from several
TeV down to a few GeV, and we try to understand these changes. We will
start at high energies and we will see how (and when) the ``high
energy features'' disappear when lowering the collision energies.

Most important for the discussion of primary scatterings at very high
energies (several TeV per nucleon) is the observation that nucleon-nucleon
scatterings must happen in parallel, and not sequentially, based on
very elementary considerations concerning time scales. To take this
\textquotedbl parallel scattering scenario\textquotedbl{} into account,
EPOS4 brings together ancient knowledge about S-matrix theory (to
deal with parallel scatterings) and modern concepts of perturbative
QCD and saturation, going much beyond the usual factorization approach;
see Refs. \cite{werner:2023-epos4-overview,werner:2023-epos4-heavy,werner:2023-epos4-smatrix,werner:2023-epos4-micro}. 

Let us look more in detail at the relevant time scales (see \cite{werner:2023-epos4-smatrix}
for details). The particle (hadron) formation time $\tau_{\mathrm{form}}$
has to be compared to the collision time $\tau_{\mathrm{collision}}$
(the duration of an $AA$ collision) and to the interaction time \textbf{$\tau_{\mathrm{interaction}}$}
(time between two nucleon-nucleon interactions). We define the high
energy threshold $E_{\mathrm{HE}}$ 
(in the sense of energy $\sqrt{s_{NN}}$ per nucleon-nucleon pair)
by the identity 
\begin{equation}
\tau_{\mathrm{form}}=\tau_{\mathrm{collision}},
\end{equation}
and the low energy threshold $E_{\mathrm{LE}}$ by the identity 
\begin{equation}
\tau_{\mathrm{form}}=\tau_{\mathrm{interaction}}.s
\end{equation}
Considering central rapidity hadrons ($\gamma_{\mathrm{hadron}}=1$),
a formation time $\tau_{\mathrm{form}}=1\,\mathrm{fm}/c$, and a big
nucleus with $R=6.5\,\mathrm{fm}$, we get \cite{werner:2023-epos4-smatrix}
\begin{equation}
E_{\mathrm{LE}}\approx4\,\mathrm{GeV},\quad E_{\mathrm{HE}}\approx24\,\mathrm{GeV}.
\end{equation}
Beyond $E_{\mathrm{HE}}$ particle production starts only after the
two nuclei have passed through each other, which means all the nucleon-nucleon
collisions should happen in parallel, instantaneously, there is no
time sequence. Below $E_{\mathrm{LE}}$ a hadronic cascade is appropriate.
Between the two thresholds, one needs some ``partially parallel scattering
scenario'', which is not yet implemented in EPOS4. We will employ
the full \textquotedbl parallel scattering scenario\textquotedbl{}
down to lowest energies, and we will investigate where precisely and
how it breaks down. Since PbPb at 5.02 ATeV and AuAu at 200 GeV have
already been discussed in \cite{werner:2023-epos4-smatrix}, we will
focus in this paper on energies below 200 GeV. 

In the overview \cite{werner:2023-epos4-overview} and in detail in
\cite{werner:2023-epos4-heavy,werner:2023-epos4-smatrix} it is shown
how such a ``parallel scattering scheme'' for primary scatterings 
can be constructed, based
on S-matrix theory, which we will sketch very briefly in the following.

An early realization is the Gribov-Regge (GR) approach \cite{Gribov:1967vfb,Gribov:1968jf,GribovLipatov:1972,Abramovsky:1973fm}
for $pp$ and $AA$ scatterings. This S-matrix approach has a modular
structure, it is based on so-called ``cut Pomerons'', representing
elementary parton-parton scatterings, with the associated mathematical
object $G=\mathrm{cut}T$, with $T$ being the Fourier transform (with
respect to the momentum transfer) of the corresponding T-matrix, divided
by $2s$, with $s$ referring to the Mandelstam variable. This so-called
``impact parameter representation'' with $G=G(b)$ with an impact
parameter $b$ makes formulas simpler. Although the GR approach is
an excellent tool to deal with parallel scatterings, a serious drawback
is the fact that the energy-momentum sharing between the multiple
scatterings is not taken care of. And obviously GR has no answer to
the question of how to connect the cut Pomeron expression $G$ and
the corresponding QCD expression $G_{\mathrm{QCD}}$ for parton-parton
scattering.

In \cite{Drescher:2000ha}, a possible solution has been proposed,
by taking into account energy-momentum sharing (let us call this approach
``GR\textsuperscript{+}'') and based on the hypothesis ``$G$ is
equal to $G_{\mathrm{QCD}}$'', where the latter is essentially a
cut parton ladder based on DGLAP parton evolutions \cite{GribovLipatov:1972,AltarelliParisi:1977,Dokshitzer:1977}.
A detailed discussion about the calculation of $G_{\mathrm{QCD}}$
can be found in \cite{werner:2023-epos4-heavy}. Unfortunately, it
turned out that implementing energy-momentum sharing has a very negative
side effect: it ruins seriously elementary geometric properties such as
binary scaling in AA scattering. In EPOS4, the first step towards
a solution of the problem is a detailed understanding of what causes
the problem and that it is fundamental, and not just a wrong parameter
choice. Using 
\begin{equation}
G=G_{\mathrm{QCD}}
\end{equation}
leads unavoidably to contradictions. In a second step, a solution
could be presented. Let us look at the arguments of $G$ and $G_{\mathrm{QCD}}$:
Both depend on $b$ (not written explicitly) and on the lightcone
momenta of the external legs $x^{+}$ and $x^{-}$. But most importantly,
$G_{\mathrm{QCD}}$ also depends on some low virtuality cutoff fo
the parton evolution, and this cutoff is now considered to not anymore
be simply a constant, but a dynamical variable, named saturation scale
$Q_{\mathrm{sat}}^{2}$. So we have $G_{\mathrm{QCD}}=G_{\mathrm{QCD}}(Q_{\mathrm{sat}}^{2},x^{+}x^{-})$.
The fundamental relation between $G$ and $G_{\mathrm{QCD}}$ is now
\begin{equation}
G(x^{+},x^{-})\!=\frac{n}{R_{\mathrm{deform}} (N_{\mathrm{conn}},x^{+},x^{-}) }G_{\mathrm{QCD}}(Q_{\mathrm{sat}}^{2},x^{+},x^{-}),\label{fundamental}
\end{equation}
with $Q_{\mathrm{sat}}^{2}=Q_{\mathrm{sat}}^{2}(N_{\mathrm{conn}},x^{+},x^{-}) $ and with (being crucial)
\begin{equation}
G\:\mathrm{independent\,of}\,N_{\mathrm{conn}},\label{fundamental-1}
\end{equation}
with the so-called connection number
$N_{\mathrm{conn}}$ counting the number of Pomerons
being connected to the same projectile and target nucleon 
as the given Pomeron.
The quantity $R_{\mathrm{deform}}^{(N_{\mathrm{conn}})}$ is the deformation
of the distribution of $x^{+}x^{-}$ (the Pomeron's energy squared)
in the case of $N_{\mathrm{conn}}>1$ 
compared to the case $N_{\mathrm{conn}}=1$.
This deformation (a consequence of several Pomerons competing
for energy sharing) destroys factorization and binary scaling unless
one uses eqs. (\ref{fundamental},\ref{fundamental-1}).
Finally, $n$ is a normalization constant.

The primary scattering will produce ``parton-ladders\textquotedblright .
The link between primary and secondary scatterings is the ``core-corona
procedure\textquotedblright : The parton ladders are treated as classical
relativistic (kinky) strings. So in general, we have a large number
of (partly overlapping) strings. Based on the momenta and the density
of string segments (referred to as prehadrons in the following), 
one separates at some early proper time $\tau_{0}$
the core (going to be treated as fluid) from the corona (escaping
hadrons, including jet hadrons).

We will first discuss in Sec. \ref{=======the-role-of-core=======}
about particle production and the role of core, corona, and remnants
for different energies, down to 4 GeV.

Then in Secs. \ref{=======results-pt=======}, \ref{=======results-yields=======},
and \ref{=======results-v2=======},
we will show very detailed results from 62.4 GeV down to 7.7 GeV,
considering the transverse momentum dependencies of yields and flow harmonics $v_2$ 
\begin{itemize}
\item
for all energies,
\item 
for all possible centrality choices, and
\item
for all hadron species where the corresponding data exist, namely 
pions, kaons, (anti)protons, but also hyperons ($\Lambda$, $\Xi$, $\Omega$) and $\Phi$ mesons.
\end{itemize}
This is the most complete collection of model/data comparisons concerning 
gold-gold collisions in this energy domain, 
in particular concerning differential yields ($p_t$ dependencies).

Over the last decade, a few other models involving a fluid dynamic picture have 
been developed for the energy range $\sqrt{s_{\rm NN}}=7.7-62.4$~GeV, 
where the UrQMD cascade \cite{Karpenko:2015xea}, a decelerating string picture
\cite{Shen:2017bsr, Shen:2022oyg} and a 3D extension to Monte Carlo Glauber 
\cite{Shen:2020jwv} have been employed for the initial state. 
These models succeeded in reproducing the collision energy dependence of the shape 
and the magnitude of the rapidity density of hadrons, 
and (except \cite{Shen:2017bsr, Shen:2022oyg}) 
of $p_t$-integrated yields and elliptic flow coefficients. 
In Ref. \cite{Karpenko:2015xea},  transverse momentum spectra 
of several hadron species at selected centralities and energies were shown.

\section{The role of core, corona, and remnants for different energies \label{=======the-role-of-core=======}}

Multiple scattering diagrams (representing the primary scatterings)
are the origin of particle production, which means first the production
of so-called prehadrons, having the quantum numbers of hadrons, but
not necessarily being final. Prehadrons originate from Pomerons (via
kinky strings) or from remnants. A detailed description can be found
in section 4 of Ref. 
\cite{werner:2023-epos4-heavy}, where we also discuss the different
types of Pomerons and their relation with pQCD, including technical
details for the pQCD computations.

The pQCD part (parton ladders) is a crucial element of a Pomeron at
LHC energies. But with decreasing energy, it becomes more and more
likely that these Pomerons are replaced by purely soft ones, see Sec.
3 of \cite{werner:2023-epos4-micro}. At $200$ GeV, the relative
weight of ``normal'' compared to soft Pomerons is roughly 1:1, at
even lower energies, the soft dominates. Also the Pomerons get less
energetic, producing fewer particles.

Based on these abovementioned prehadrons, we employ a so-called core-corona
procedure (see \cite{Werner:2007bf}, for a more recent discussion
in the EPOS4 framework see \cite{werner:2023-epos4-micro}), to distinguish
core from corona particles, at some given (early) proper-time $\tau_{0}$.
The core prehadrons constitute ``bulk matter'' and will be treated
via hydrodynamics. The corona prehadrons become simply hadrons and
propagate with reduced energy (due to some energy loss). Corona particles
are either very energetic (then they move out even from the center),
or they are close to the surface, or we have a combination of both.
\begin{figure}[h]
\centering{}\includegraphics[bb=45bp 30bp 570bp 730bp,clip,scale=0.49]
{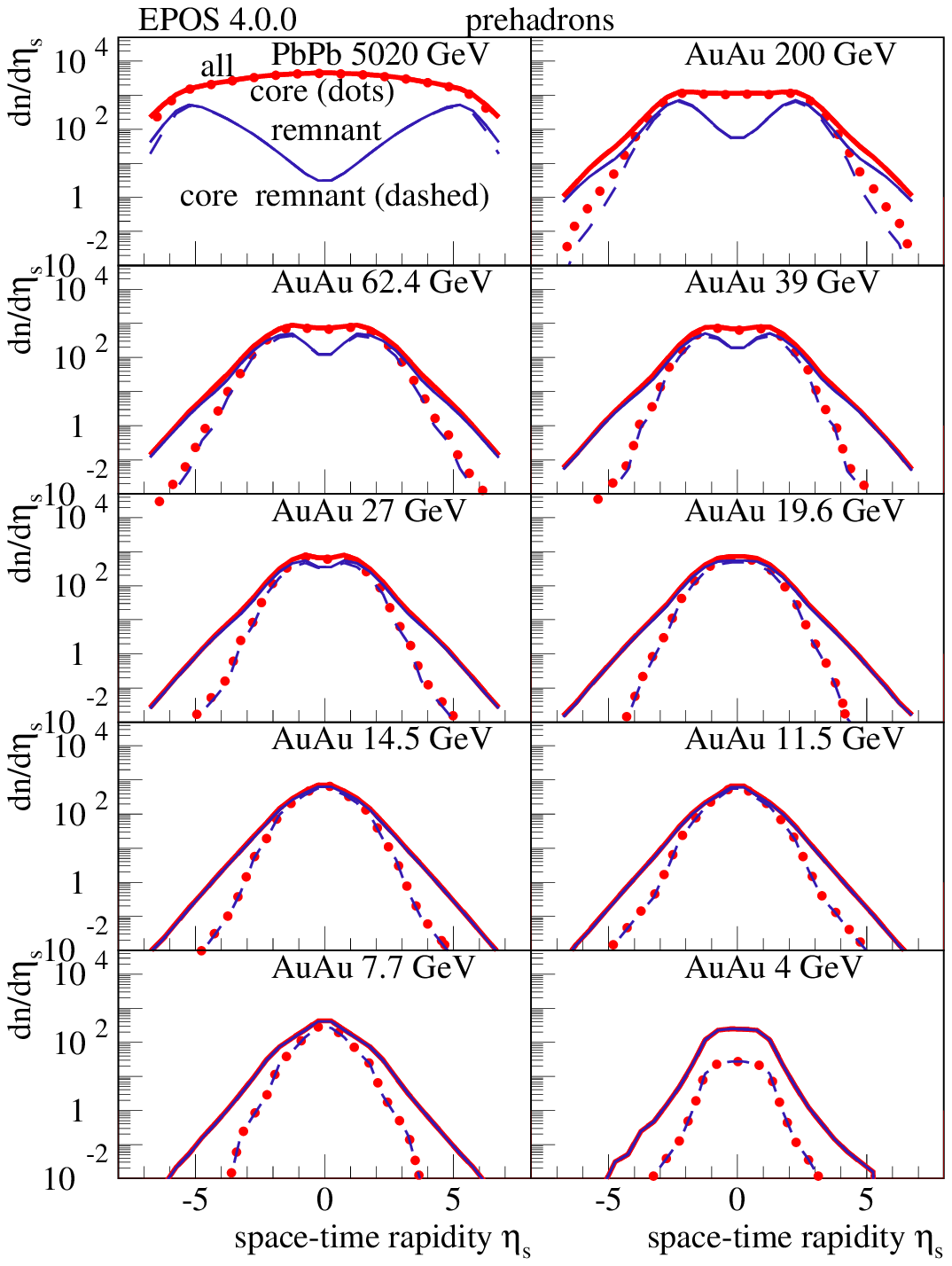}
\caption{The prehadron yield as a function of space-time rapidity ($\eta_{s}=\frac{1}{2}\ln((t+z)/(t-z))$,
with $t$ being the time and $z$ the longitudinal coordinate), for
central $AA$ collisions at different energies. The curves refer to
all prehadrons (red full), all core prehadrons (red dotted), prehadrons
from remnant decay (blue thin full), and core prehadrons from remnant decay
(blue thin dashed).} \label{prehadrons-aa}
\end{figure}

We will try to understand the relative importance of the core part
and of the fraction coming from remnant decay. In Fig. \ref{prehadrons-aa},
we show results for central (0-5\%) PbPb collisions at 5.02 TeV, and
central (0-5\%) AuAu collisions from 200 down to 4 GeV (per nucleon).
We plot four different curves: all prehadrons (red full), all core
prehadrons (red dotted), prehadrons from remnant decay (blue thin full),
and core prehadrons from remnant decay (blue thin dashed). Looking at the
results for PbPb at 5.02 TeV, we observe that almost all prehadrons
are core prehadrons, so the core dominates. Prehadrons from remnants
are preferentially produced at large rapidities, but also here almost
all are core prehadrons. Going down in energy, starting already at
200 GeV, we see that the core still dominates around $\eta=0$, but
at lage values of $|\eta|$, the core contribution drops dramatically.
We also observe that the remnant contributions become more and more
important, and very dominant below 20 GeV. Nevertheless, close
to $\eta=0$, the core remains dominant - down to 7.7 GeV. There is
actually little change from 39 down to 7.7 GeV, but things change
from 7.7 to 4 GeV: the core part decreases strongly.

\begin{figure}[h]
\centering{}\includegraphics[bb=45bp 30bp 570bp 730bp,clip,scale=0.49]
{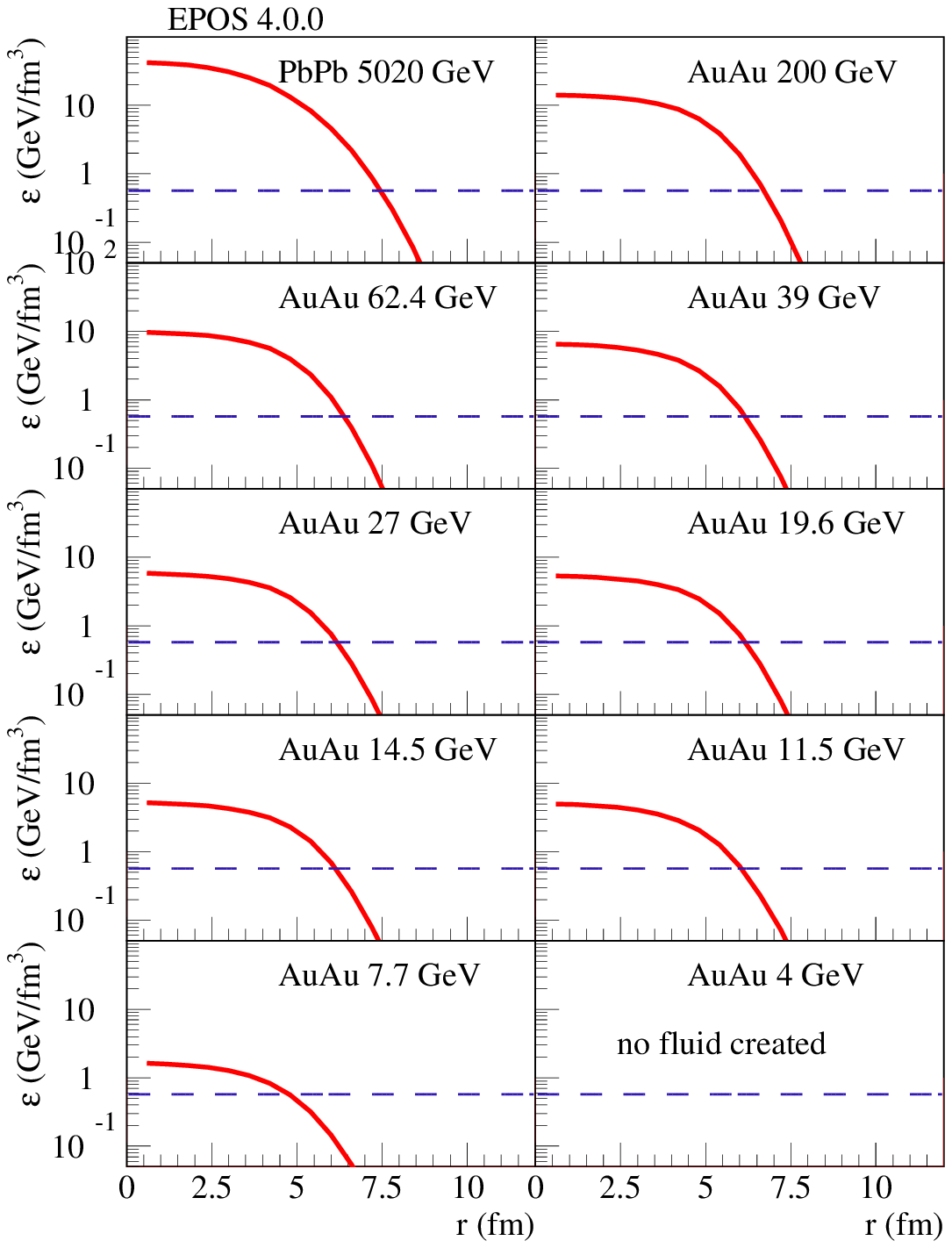}
\caption{Energy density at the initial proper-time $\tau_{0}$ as a function
of the transverse coordinate $r$ for central $AA$ collisions at
different energies. The blue dashed line is the freeze-out energy
density. \label{energy-density}}
\end{figure}

In EPOS, $t=0$ is defined as the time corresponding to maximal overlap of the colliding nuclei. Fluidization takes place at a fixed proper time $\tau=\tau_0$, which we later call initial proper time. We set $\tau_0=1.5$~fm/c for all collision energies except for 7.7 GeV, where $\tau_0=2.0$~fm/c is used. Such setting ensures that the fluid stage starts after all the primary nucleon-nucleon scatterings have taken place.
Having identified core pre-hadrons, we compute the corresponding energy-momentum
tensor $T^{\mu\nu}$ and the flavor flow vector at some position $x$
at initial proper time $\tau=\tau_{0}$ as 
\begin{equation}
T^{\mu\nu}(x)=\sum_{i}\frac{p_{i}^{\mu}p_{i}^{\nu}}{p_{i}^{0}}g(x-x_{i})
\end{equation}
and 
\begin{equation}
N_{q}^{\mu}(x)=\sum_{i}\frac{p_{i}^{\mu}}{p_{i}^{0}}\,q_{i}\,g(x-x_{i}),
\end{equation}
with $q_{i}\in{u,d,s}$ being the net flavor content and $p_{i}$
the four-momentum of prehadron $i$. The function $g$ is some Gaussian
smoothing kernel (see \cite{werner:2023-epos4-micro}). The Lorentz
transformation into the comoving frame provides the energy density
$\varepsilon$ and the flow velocity components $v^{i}$, which will
be used as the initial condition for a hydrodynamical evolution \cite{Werner:2013tya}.
This is based on the hypothesis that equilibration happens rapidly
and affects essentially the space components of the energy-momentum
tensor. In Fig. \ref{energy-density}, we plot 
the energy density at the initial
proper-time $\tau_{0}$ as a function of the transverse coordinate
$r$ in AA collisions at different energies.
We also indicate as a blue dashed line the freeze-out
energy density $\varepsilon_{\mathrm{FO}}$. At the highest energies,
the energy density is way above the critical value $\varepsilon_{\mathrm{FO}}$,
but with decreasing beam energy, the energy density approaches this
value. At 7.7 GeV, the energy density is slightly above $\varepsilon_{\mathrm{FO}}$,
and finally at 4 GeV, there is no fluid created.

As a next step, a viscous~hydrodynamic~expansion follows. Starting
from the initial proper time $\tau_{0}$, the core part of the system
evolves according to the equations of relativistic viscous hydrodynamics
\cite{Werner:2013tya,Karpenko_2014}, where we use presently $\eta/s=0.08$.
The ``core-matter'' hadronizes on some hyper-surface defined by
a constant energy density $\epsilon_{\mathrm{FO}}$ (presently $0.57\mathrm{\,GeV/fm^{3}}$).
In earlier versions \cite{Werner:2011-hydro-pp-900GeV}, we used a
so-called Cooper-Frye procedure. This is problematic in particular
for small systems: not only energy and flavor conservation become
important, but we also encounter problems due to the fact that we
get small ``droplets'' with huge baryon chemical potential, with
strange results for heavy baryons. In EPOS4, we will systematically
use microcanonical hadronization, with major technical improvements
compared to earlier versions (see \cite{werner:2023-epos4-micro}).

In the following, we want to study core and corona contributions to
hadron production. We will distinguish: 
\begin{description}
\item [{(A)}] The ``\textbf{core+corona}'' contribution: primary interactions
(S-matrix approach for parallel scatterings), plus core-corona separation,
hydrodynamic evolution and microcanonical hadronization of the core, but without
hadronic rescattering. 
\item [{(B)}] The ``\textbf{core}'' contribution: as (A), but considering
only core particles. 
\item [{(C)}] The ``\textbf{corona}'' contribution: as (A), but considering
only corona particles. 
\item [{(D)}] The ``\textbf{full}'' EPOS4 scheme: as (A), but in addition
hadronic rescattering. 
\end{description}
\noindent In cases (A), (B), and (C) , we need to exclude the hadronic
afterburner, because the latter affects both core and corona particles,
so in the full approach, the core and corona contributions are not
visible anymore. In the following, we will focus on energies below
200 GeV, since the corresponding plots for PbPb at 5.02 ATeV and AuAu
at 200 GeV have already been discussed in \cite{werner:2023-epos4-smatrix}. 

\begin{figure}[h]
\includegraphics[bb=30bp 30bp 570bp 620bp,clip,scale=0.47]
{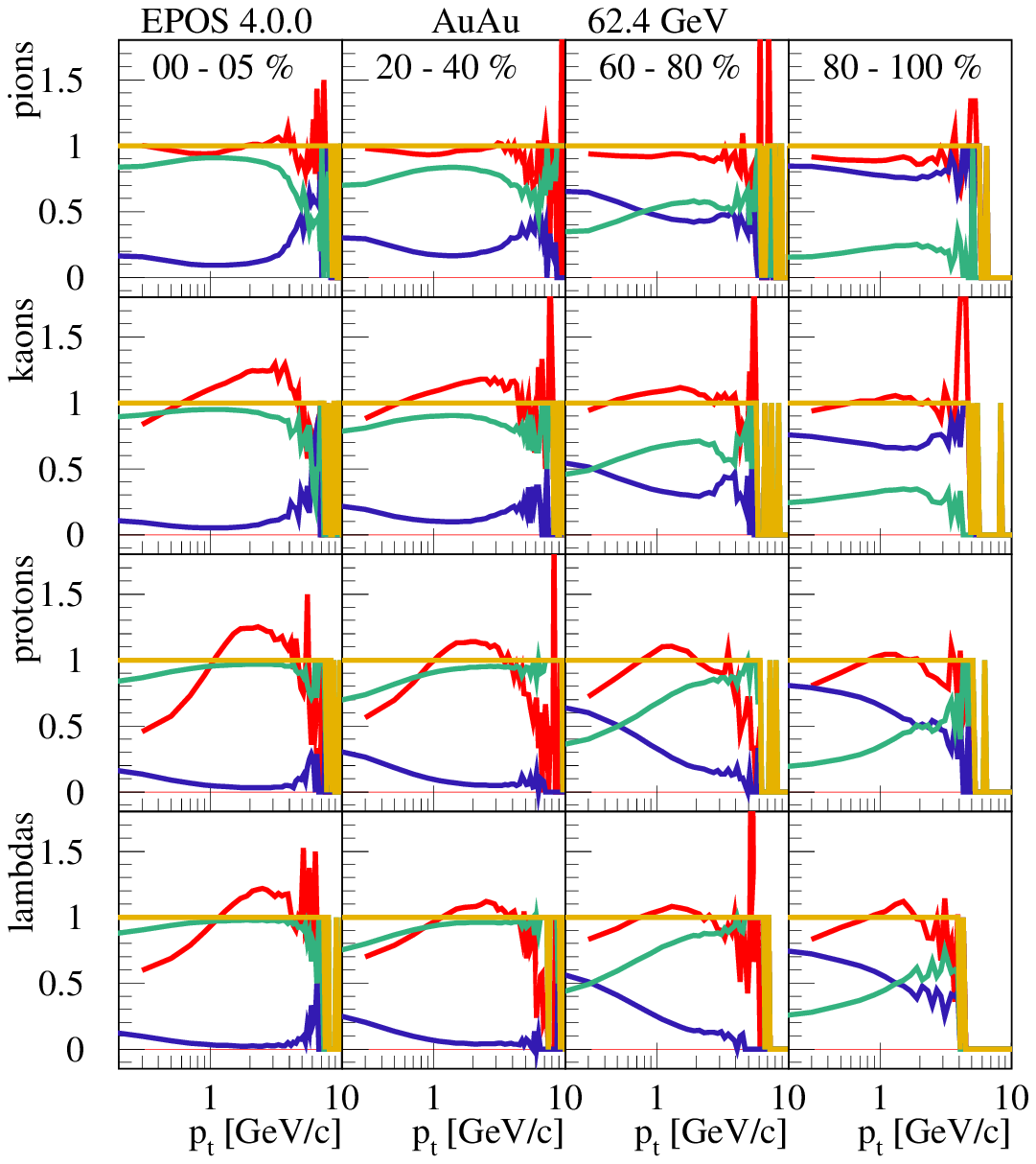} 
\caption{The X / core+coron ratio, with X being the corona contribution
(blue), the core (green), and the full contribution (red),
for 4 centrality classes and four different particle species, for
AuAu at 62.4 GeV.\label{core-corona-2}}
\end{figure}

\begin{figure}[h]
\includegraphics[bb=30bp 30bp 570bp 620bp,clip,scale=0.47]
{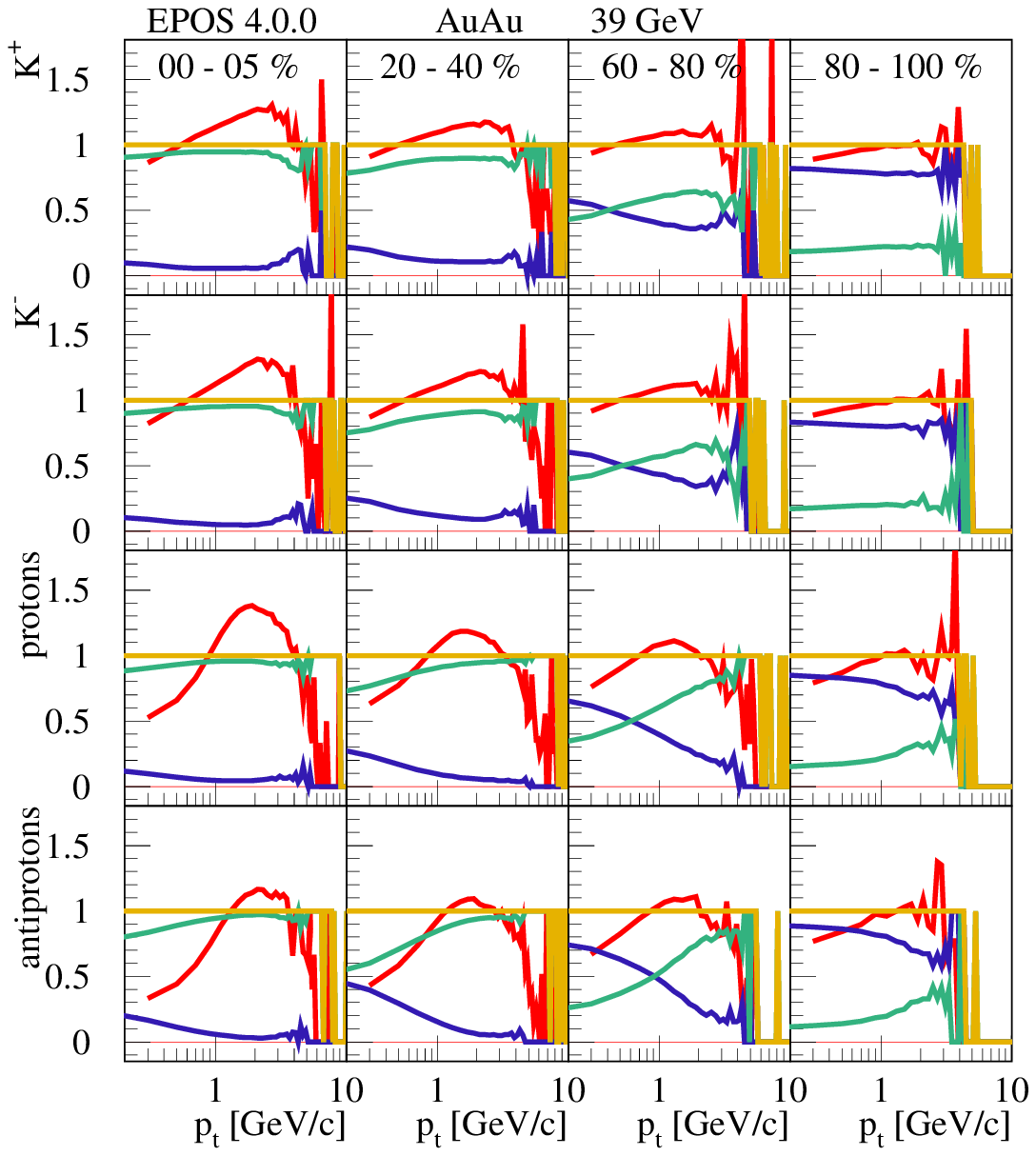}
\caption{Same as Fig. \ref{core-corona-2}, but for 39 GeV, and we consider
$K^{+},$$K^{-}$, $p$, and $\bar{p}$ hadrons.\label{core-corona-4a}}
\end{figure}

\begin{figure}[h]
\includegraphics[bb=30bp 30bp 570bp 620bp,clip,scale=0.47]
{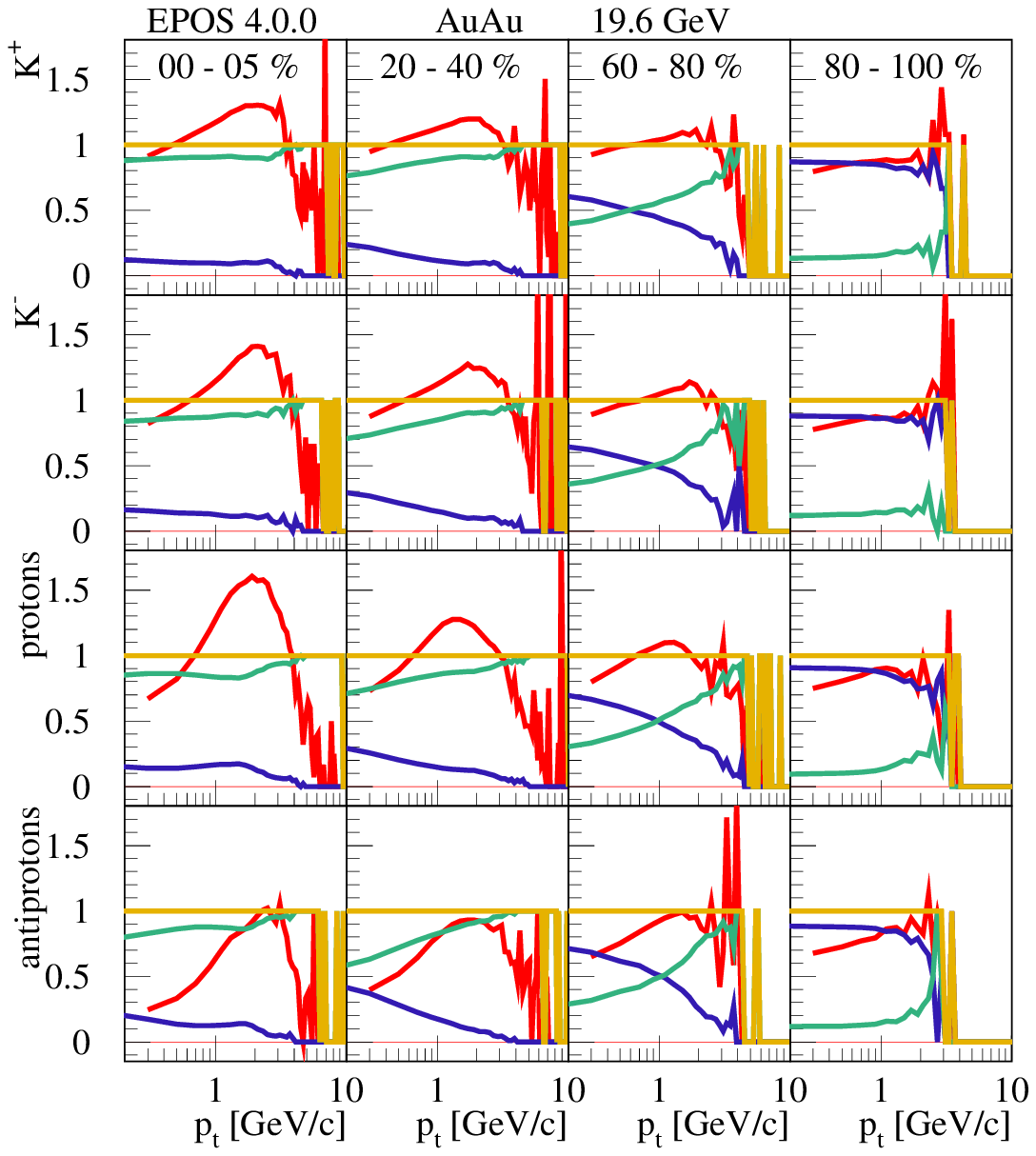}
\caption{Same as Fig. \ref{core-corona-2}, but for 19.6 GeV, and we consider
$K^{+},$$K^{-}$, $p$, and $\bar{p}$ hadrons.\label{core-corona-3}}
\end{figure}

In Figs. \ref{core-corona-2}, \ref{core-corona-4a}, and \ref{core-corona-3}, 
we show ratios X / core+corona versus $p_{t}$, with X being the corona
contribution (blue), the core (green), and the full contribution
(red), for AuAu collisions at different energies. 
In Fig. \ref{core-corona-2}, we show results at 62.4 GeV, for (from top to bottom)
pions ($\pi^{\pm}$), kaons ($K^{\pm}$), protons ($p$ and $\bar{p}$),
and lambdas ($\Lambda$ and $\bar{\Lambda}$). The four columns represent
four different centrality classes, namely 0-5\%, 20-40\%, 60-80\%,
80-100\%. Looking at the green (core) and blue (corona)
curves, we observe that the core contribution increases from peripheral
to central collisions. Concerning the $p_{t}$ dependence, we observe
a maximum of the green core curves around 1-2$\,\mathrm{GeV/c}$,
at very low $p_{t}$ the core contribution goes down, so even
at very small $p_{t}$ values the corona contributes. At higher energies
(see \cite{werner:2023-epos4-smatrix}), one observes a crossing of
the green core and the blue corona curves (core = corona) between around
2 GeV/c (mesons, peripheral) and 5 GeV/c (baryons, central), and for
larger values the corona contribution dominates clearly. Here, the
crossing is absent. There are two reasons. The spectra are softer
compared to higher energies, and around 3-6 GeV/c, the core and corona
curves are almost parallel. And since high $p_{t}$ becomes rare,
we cannot reach very high $p_{t}$ in the Monte Carlo simulations. 

The red curve , full over core+corona, represents the
effect of the hadronic cascade in the full case. The pions are
not much affected, but for kaons and even more for protons and lambdas,
rescattering makes the spectra harder. We should keep in mind that
rescattering involves particles from fluid hadronization, but also
corona particles from hard processes. Concerning the baryons, rescattering
reduces (considerably) low $p_{t}$ yields, due to baryon-antibaryon
annihilation.
In Figs. \ref{core-corona-4a} and \ref{core-corona-3}, we show  results for
AuAu collisions at 39 and 19.6 GeV. 
The high $p_{t}$ particles are getting
rare, we hardly get (for simulations with reasonable CPU times) beyond
5 GeV/c. Another ``low energy effect'': the difference between particles
and antiparticles becomes more and more important, this is why we
consider separately $K^{+}$, $K^{-}$, $p$, and $\bar{p}$. The
biggest effect can be seen when comparing the effect of hadronic rescattering
(red curves) for protons and antiprotons: with decreasing energy,
the proton curve goes slightly up, and the antiproton curve goes significantly
down (they are rare, and most of them are annihilated). Interesting
observation: also at lower energies, the core ratios (green
curves) get down at low $p_{t}$, but stay close to unity at intermediate
$p_{t}$.

Corresponding plots for AuAu collisions at 27 GeV, 11.5
GeV, and 7.7 GeV can be found in the appendix.

When discussing these results at low energies (7.7-19.6 GeV), we
should keep in mind our discussion in Sec. \ref{=======Introduction=======},
where we estimated that our parallel scattering scenario should
be valid beyond 24 GeV per nucleon, and below we have to take into
account the fact that particle production starts before the two nuclei
have passed through each other (not yet done in EPOS4).

But anyway, in the following sections, we will explore to what extent
the model ``works'' (and can explain data), and where it fails,
and what we can learn from that.\vspace{2cm}

\section{Results concerning p\protect\textsubscript{t} spectra\protect\textsubscript{}
\label{=======results-pt=======}}

In this section, we show simulation results compared to data. We will
not add too many comments to each curve, the main purpose is to check
if the concepts discussed in the previous sections give a coherent
picture (and reproduce the data) or not.\\

The number of plots is huge, and it would be tempting to show only a selection of results and provide a simple message. And it would be tempting to optimize parameters to have ``nice'' plots. But the EPOS4 philosophy is different: It is an attempt to 
\begin{itemize}
\item
have a single unique approach, with version number (here 4.0.0), and  
\item
make a maximum of tests, as complete as possible, for all kinds of observables and at various energies. \\
\end{itemize}

Modeling heavy ion collisions is complex, consisting of several stages, with considerable uncertainties at each step. And there is nothing like the simple ``smoking gun'' to prove or disprove certain concepts. Flow harmonics for example are popular observables showing the effect of flow asymmetries, but the flow affects  also the $p_t$ spectra, more or less strongly depending on the mass of the particle. So rather than looking at few selected curves  and expecting precision results, the idea is here more to get a global view and to see to what extent a given theoretical scenario can provide an overall good description of a very large set of data.\\

We will see later, that $p_t$ spectra and elliptical flow $v_2$, although both are affected by flow, they are at a different degree sensitive to particular details of the model, so we get complementary information, and therefore both observables should be studied.\\ 
   
This ``global view'' strategy requires a huge amount of tests, and therefore the``tuning'' of parameters and options  is far from being optimized, But we expect reasonable agreement in the energy range (beyond 24 GeV) where the parallel scattering approach is applicable.\\

In the following we discuss $p_t$ spectra.
We will start with the highest energy, and then move down till 7.7
GeV per nucleon. To go this way is convenient in the sense that from
the theory point of view, the high energy case contains in principle
everything, we do not need to add ``features'' at low energies,
simply certain phenomena ``die out''.\\

We will focus on energies from 39 GeV down to 7.7
GeV, the corresponding plots for PbPb at 5.02 ATeV and AuAu at 200
GeV have already been discussed in \cite{werner:2023-epos4-smatrix}. 

\pagebreak{}

\subsection{Results for 39 GeV}

\begin{figure}[H]
\centering{}\includegraphics[bb=30bp 35bp 450bp 580bp,clip,scale=0.6]
{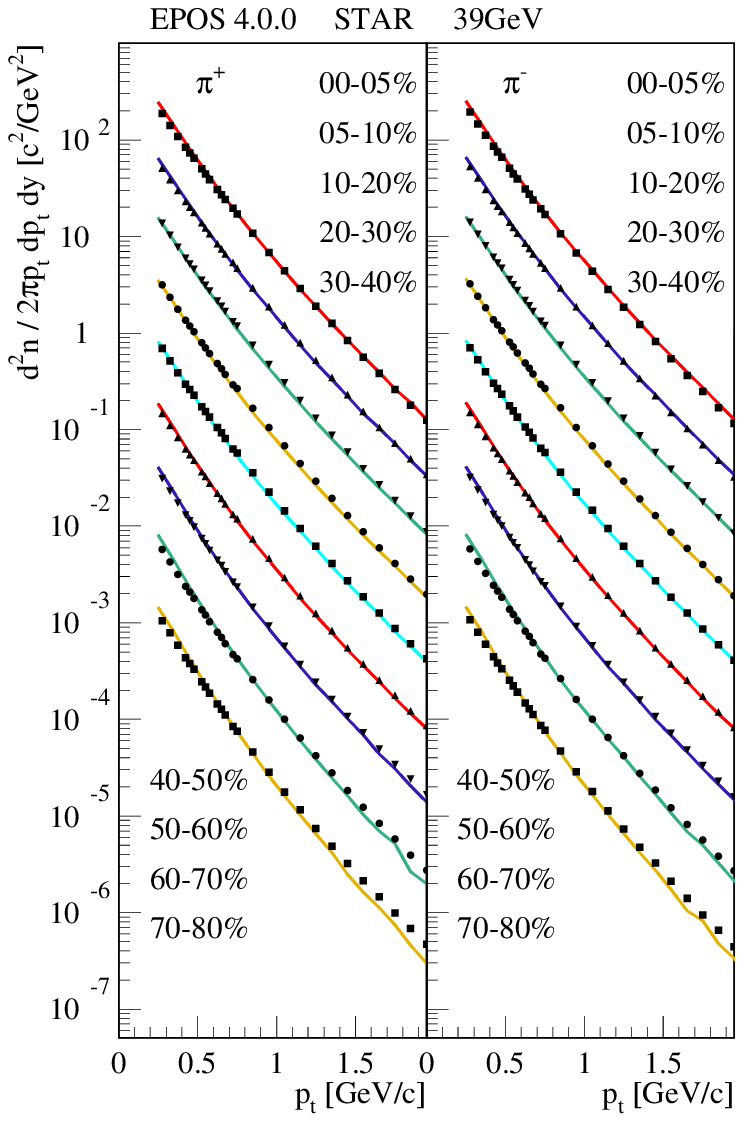}\\

\caption{Transverse momentum distributions of $\pi^{+}$, $\pi^{-}$ in AuAu
collisions at 39 GeV for different centrality classes. EPOS4 simulation
(lines) are compared to data from STAR \cite{STAR:2017}. From top
to bottom, we multiply the curves by $3^{-i}$, $i=0,1,2,3,...$.
\label{39-transverse-momentum-1}}
\end{figure}

\begin{figure}[H]
\begin{centering}
\includegraphics[bb=30bp 35bp 450bp 580bp,clip,scale=0.6]
{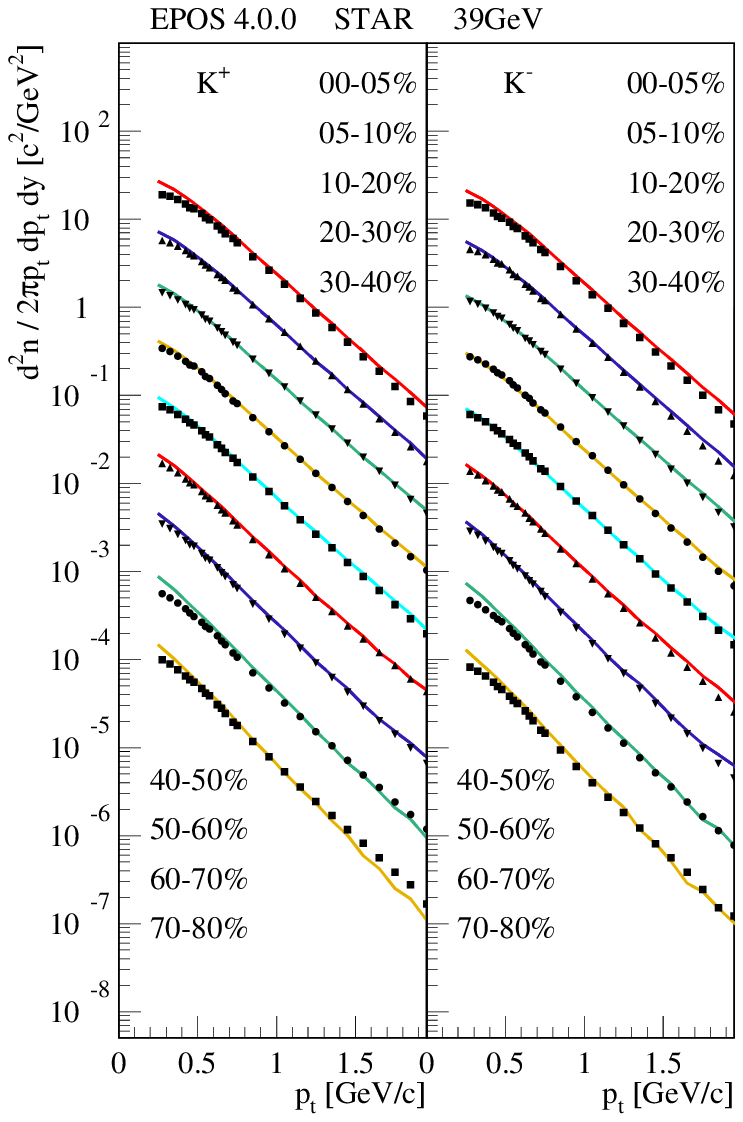} 
\par\end{centering}
\centering{}\includegraphics[bb=30bp 30bp 450bp 580bp,clip,scale=0.6]
{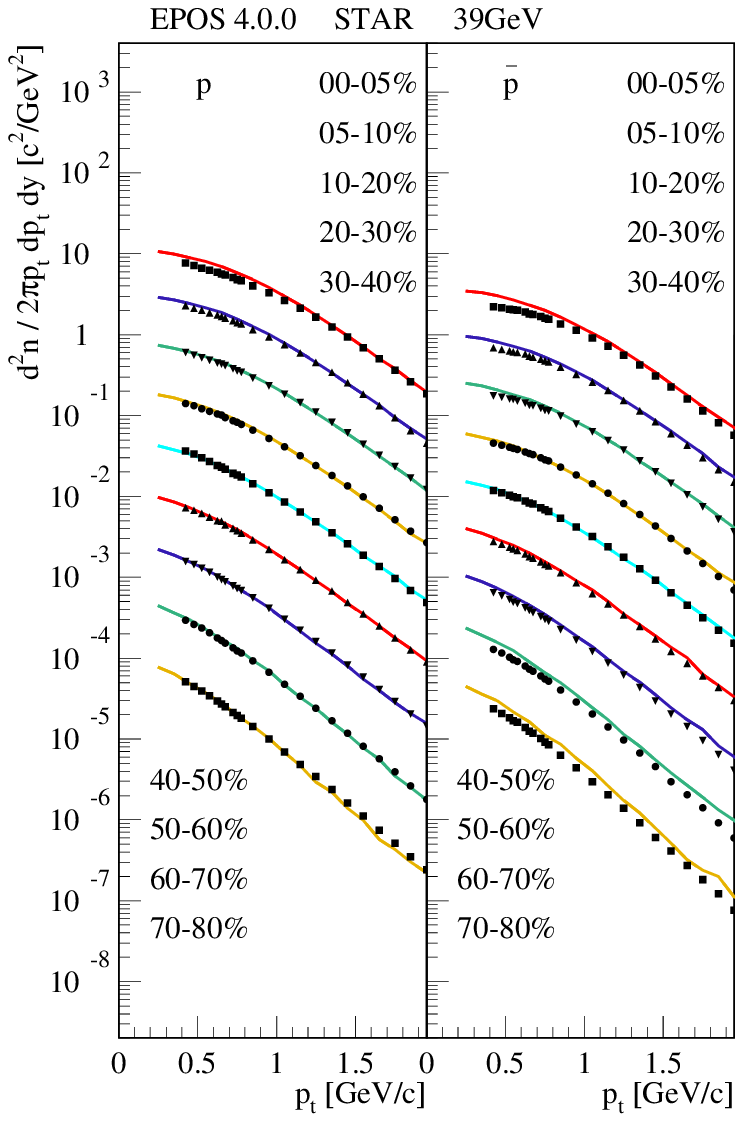}
\caption{Same as Fig. \ref{39-transverse-momentum-1}, but for $K^{+}$, $K^{-}$,
$p$, $\bar{p}$. \label{39-transverse-momentum-2}}
\end{figure}

In Figs. \ref{39-transverse-momentum-1} and \ref{39-transverse-momentum-2},
we show transverse momentum distributions of $\pi^{+}$, $\pi^{-}$,
$K^{+}$, $K^{-}$, $p$, $\bar{p}$ in AuAu collisions at 39 GeV
for different centrality classes. EPOS4 simulation (lines) are compared
to data from STAR \cite{STAR:2017}. From top to bottom, we multiply
the curves by $3^{-i}$, $i=0,1,2,3,...$.\\

\begin{figure}[H]
\begin{centering}
\includegraphics[bb=30bp 35bp 450bp 580bp,clip,scale=0.6]
{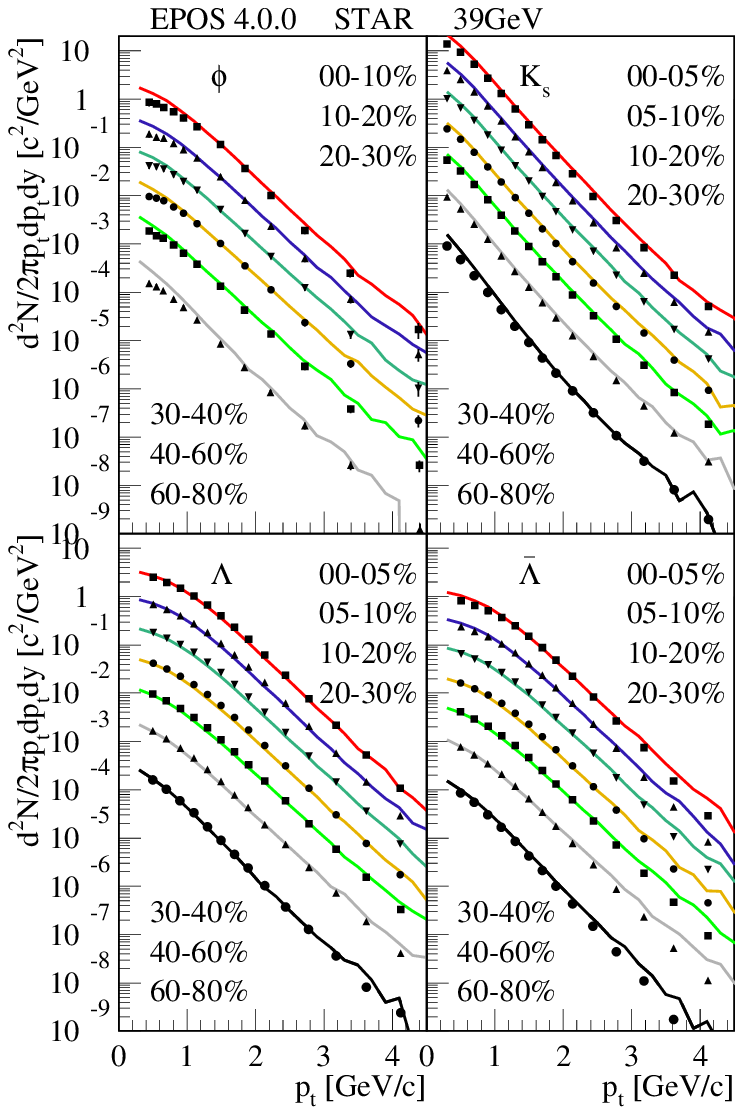} 
\par\end{centering}
\centering{}\includegraphics[bb=30bp 30bp 450bp 580bp,clip,scale=0.6]
{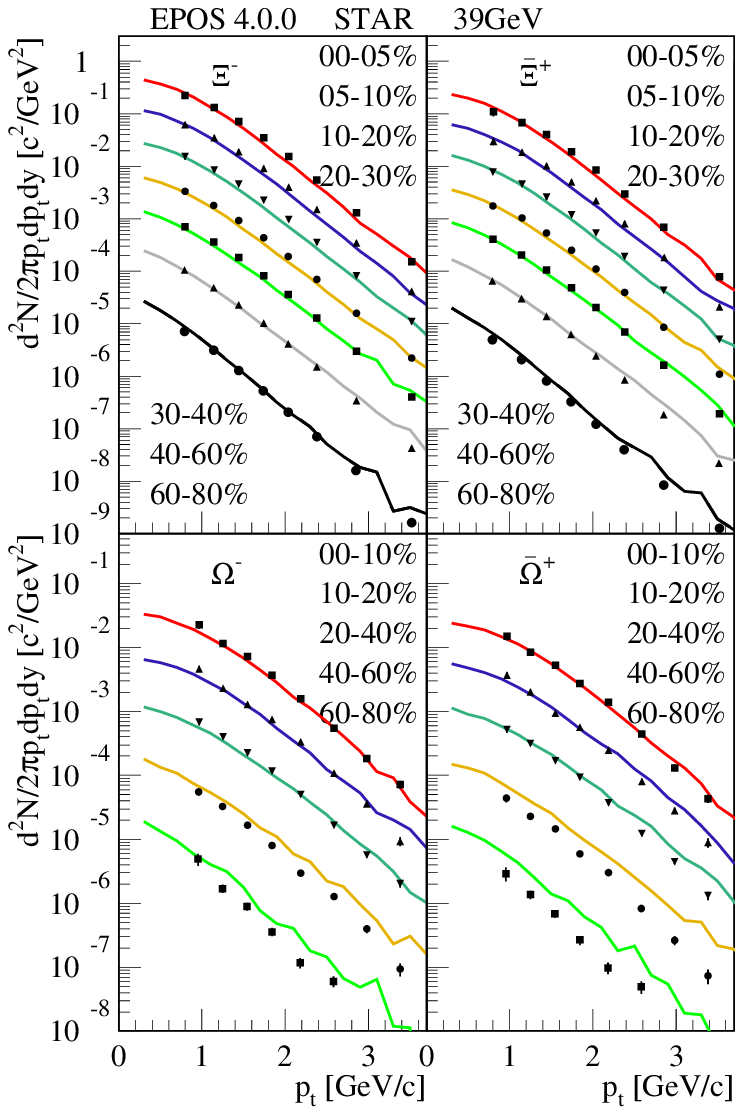}
\caption{Transverse momentum distributions of $\phi$, $K_{0}$, $\Lambda$,
$\bar{\Lambda}$ $\Xi^{-}$, $\bar{\Xi}^{+}$, $\Omega^{-}$, $\bar{\Omega}^{+}$
in AuAu collisions at 39 GeV at central rapidity for different centralities.
EPOS4 simulation (lines) are compared to data from STAR \cite{STAR:2020}.
\label{39-transverse-momentum-3}}
\end{figure}

In Fig. \ref{39-transverse-momentum-3}, we show transverse momentum
distributions of $\phi$, $K_{0}$, $\Lambda$, $\bar{\Lambda}$ $\Xi^{-}$,
$\bar{\Xi}^{+}$, $\Omega^{-}$, $\bar{\Omega}^{+}$ in AuAu collisions
at 39 GeV at central rapidity for different centralities. EPOS4 simulation
(lines) are compared to data from STAR \cite{STAR:2020}.\\

In general, the simulation results are close to the data, concerning
identified particles as pions, kaons, and protons, and as well as
hyperons, although for kaons and antiprotons at low $p_{t}$ the
simulations are slightly above the data. Concerning the $\phi$ meson,
the simulations are somewhat above the data at low $p_{t}$

\pagebreak{}

\subsection{Results for 27 GeV}

\begin{figure}[H]
\centering{}\includegraphics[bb=30bp 35bp 450bp 580bp,clip,scale=0.6]
{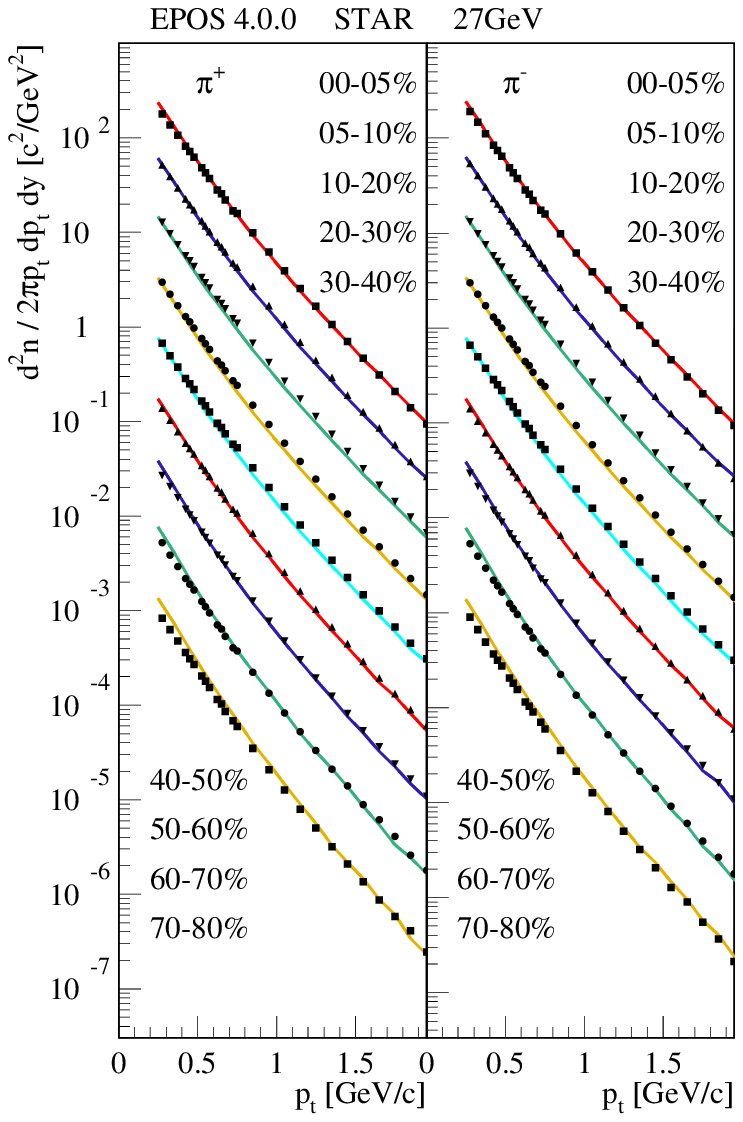}\\

\caption{Transverse momentum distributions of $\pi^{+}$, $\pi^{-}$ in AuAu
collisions at 27 GeV for different centrality classes. EPOS4 simulation
(lines) are compared to data from STAR \cite{STAR:2017}. From top
to bottom, we multiply the curves by $3^{-i}$, $i=0,1,2,3,...$.
\label{27-transverse-momentum-1}}
\end{figure}

\begin{figure}[H]
\begin{centering}
\includegraphics[bb=30bp 35bp 450bp 580bp,clip,scale=0.6]
{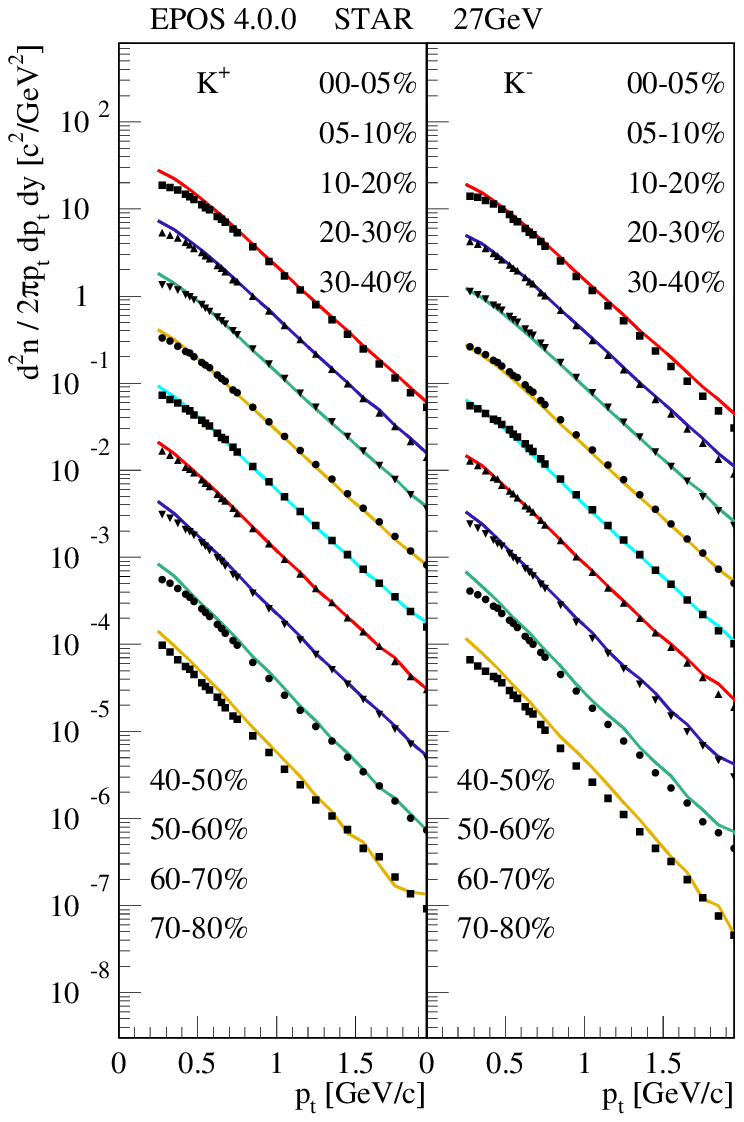} 
\par\end{centering}
\centering{}\includegraphics[bb=30bp 30bp 450bp 580bp,clip,scale=0.6]
{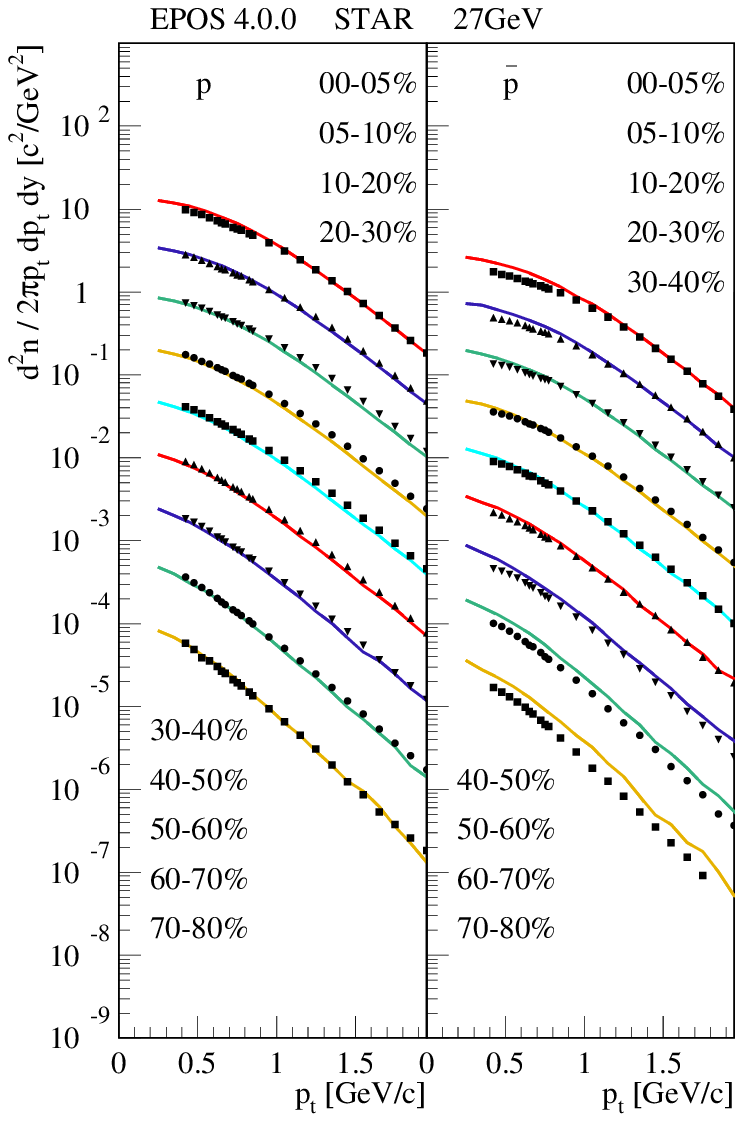}
\caption{Same as Fig. \ref{27-transverse-momentum-1}, but for $K^{+}$, $K^{-}$,
$p$, $\bar{p}$. \label{27-transverse-momentum-2}}
\end{figure}

In Fig. \ref{27-transverse-momentum-1} and \ref{27-transverse-momentum-2},
we show transverse momentum distributions of $\pi^{+}$, $\pi^{-}$,
$K^{+}$, $K^{-}$, $p$, $\bar{p}$ in AuAu collisions at 27 GeV
for different centrality classes. EPOS4 simulation (lines) are compared
to data from STAR \cite{STAR:2017}. From top to bottom, we multiply
the curves by $3^{-i}$, $i=0,1,2,3,...$.\\

\begin{figure}[H]
\begin{centering}
\includegraphics[bb=30bp 35bp 450bp 580bp,clip,scale=0.6]
{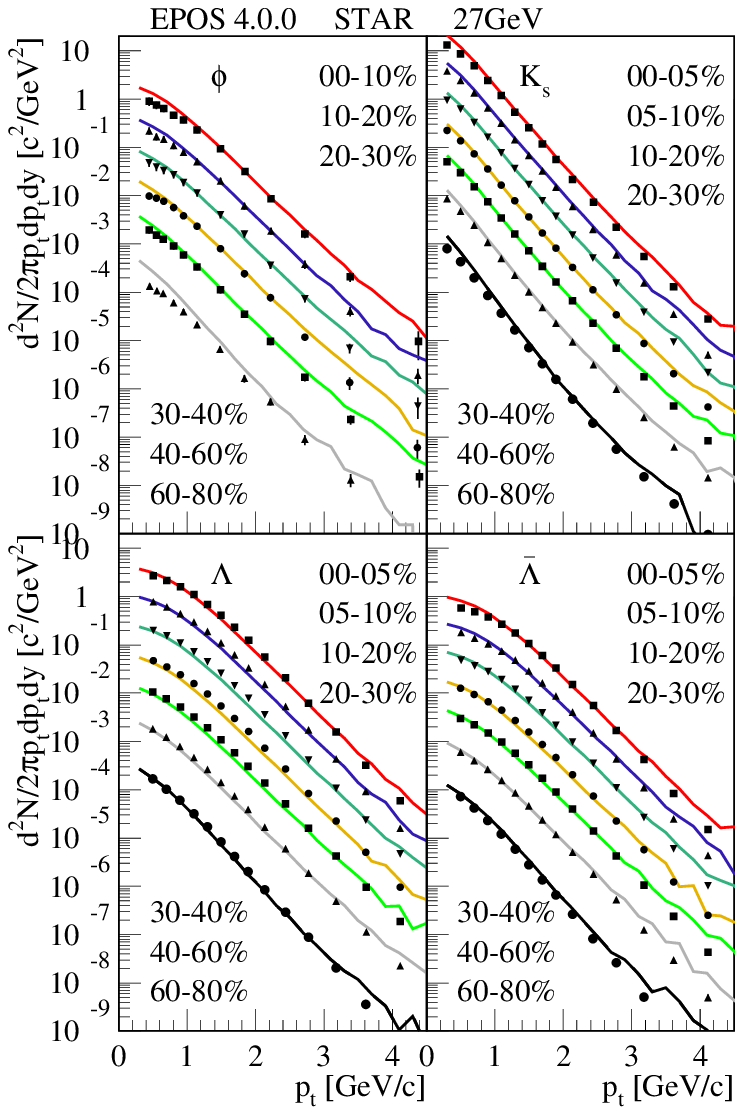} 
\par\end{centering}
\centering{}\includegraphics[bb=30bp 30bp 450bp 580bp,clip,scale=0.6]
{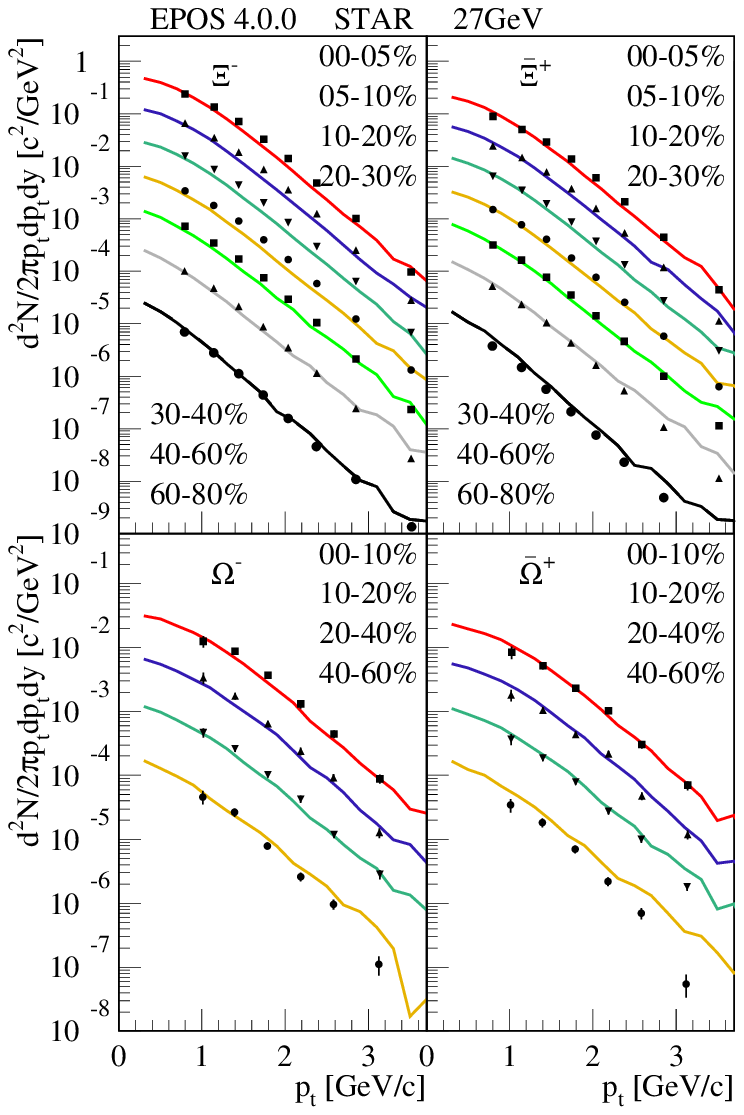}
\caption{Transverse momentum distributions of $\phi$, $K_{0}$, $\Lambda$,
$\bar{\Lambda}$ $\Xi^{-}$, $\bar{\Xi}^{+}$, $\Omega^{-}$, $\bar{\Omega}^{+}$
in AuAu collisions at 27 GeV at central rapidity for different centralities.
EPOS4 simulation (lines) are compared to data from STAR \cite{STAR:2020}.
\label{27-transverse-momentum-3}}
\end{figure}

In Fig. \ref{27-transverse-momentum-3}, we show transverse momentum
distributions of $\phi$, $K_{0}$, $\Lambda$, $\bar{\Lambda}$ $\Xi^{-}$,
$\bar{\Xi}^{+}$, $\Omega^{-}$, $\bar{\Omega}^{+}$ in AuAu collisions
at 27 GeV at central rapidity for different centralities. EPOS4 simulation
(lines) are compared to data from STAR \cite{STAR:2020}.\\

In general, the simulation results are relatively close to the data,
concerning identified particles as pions, kaons, and protons, and
as well hyperons. But compared to 39 GeV, for kaons, antiprotons,
and the $\phi$ meson, the deviation (simulation compared to data)
at low $p_{t}$ gets bigger.

\pagebreak{}

\subsection{Results for 19.6 GeV}

\begin{figure}[H]
\centering{}\includegraphics[bb=30bp 35bp 450bp 580bp,clip,scale=0.6]
{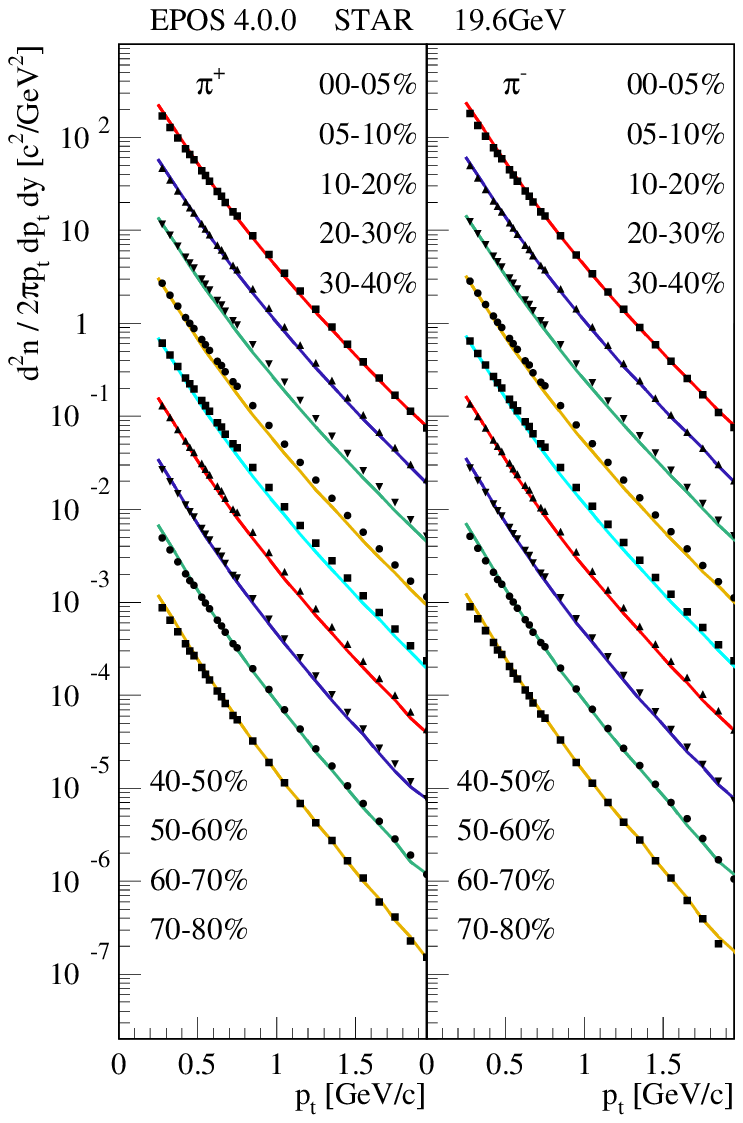}\\

\caption{Transverse momentum distributions of $\pi^{+}$, $\pi^{-}$ in AuAu
collisions at 19.6 GeV for different centrality classes. EPOS4 simulation
(lines) are compared to data from STAR \cite{STAR:2017}. From top
to bottom, we multiply the curves by $3^{-i}$, $i=0,1,2,3,...$.
\label{19-transverse-momentum-1}}
\end{figure}

\begin{figure}[H]
\begin{centering}
\includegraphics[bb=30bp 35bp 450bp 580bp,clip,scale=0.6]
{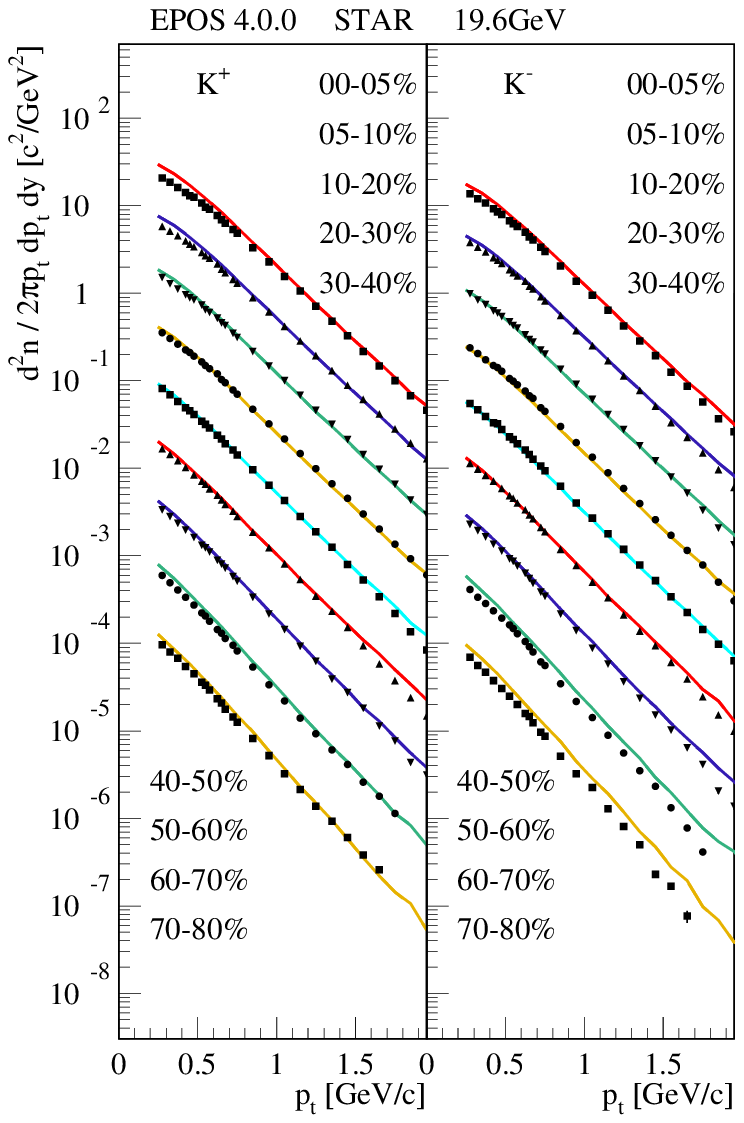} 
\par\end{centering}
\centering{}\includegraphics[bb=30bp 30bp 450bp 580bp,clip,scale=0.6]
{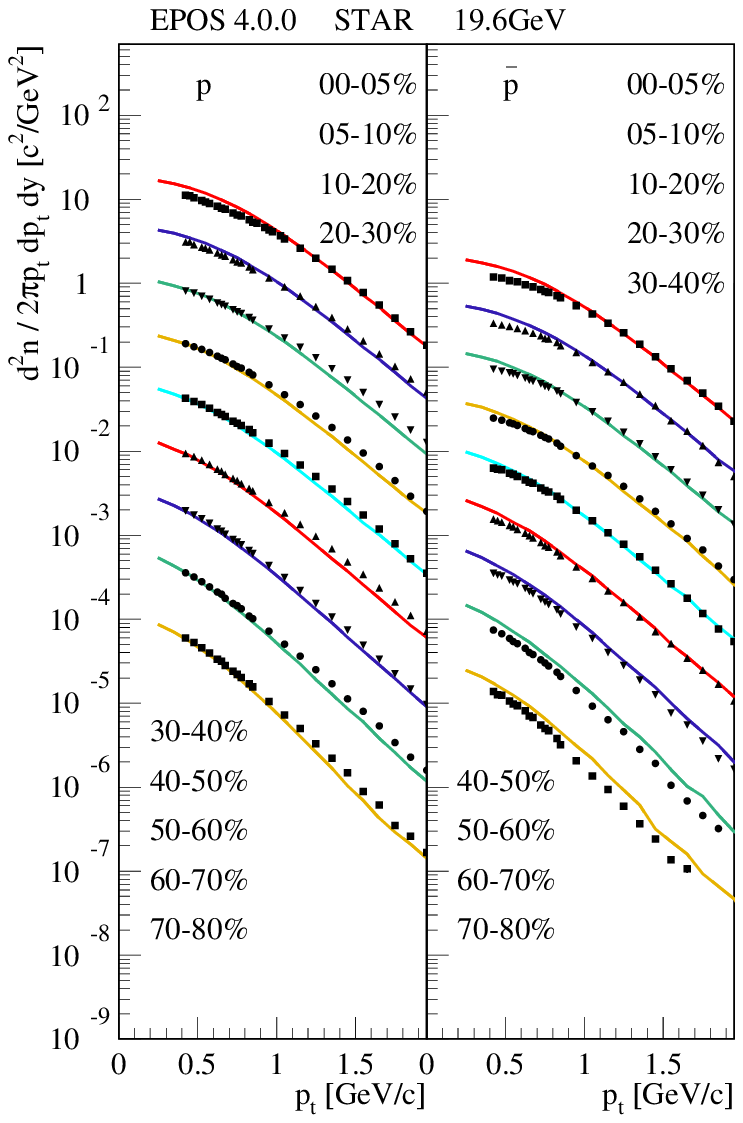}
\caption{Same as Fig. \ref{19-transverse-momentum-1}, but for $K^{+}$, $K^{-}$,
$p$, $\bar{p}$. \label{19-transverse-momentum-2}}
\end{figure}

In Figs. \ref{19-transverse-momentum-1} and \ref{19-transverse-momentum-2},
we show transverse momentum distributions of $\pi^{+}$, $\pi^{-}$,
$K^{+}$, $K^{-}$, $p$, $\bar{p}$ in AuAu collisions at 19.6 GeV
for different centrality classes. EPOS4 simulation (lines) are compared
to data from STAR \cite{STAR:2017}. From top to bottom, we multiply
the curves by $3^{-i}$, $i=0,1,2,3,...$. \\

\begin{figure}[H]
\begin{centering}
\includegraphics[bb=30bp 35bp 450bp 580bp,clip,scale=0.6]
{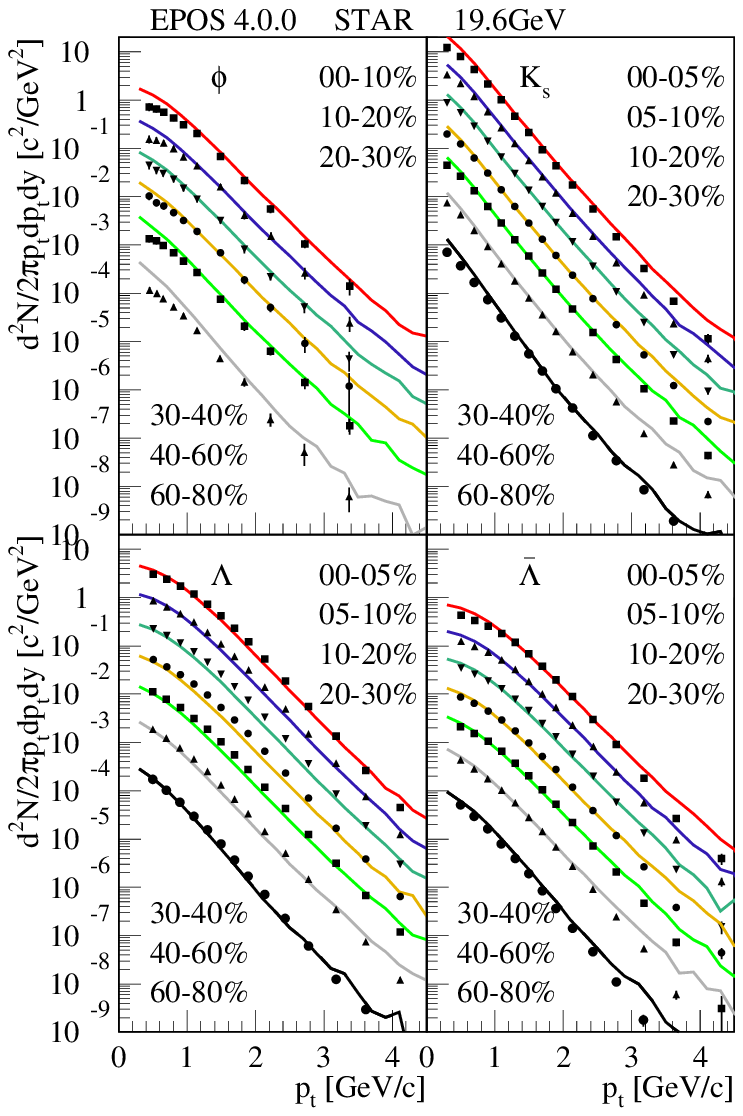} 
\par\end{centering}
\centering{}\includegraphics[bb=30bp 30bp 450bp 580bp,clip,scale=0.6]
{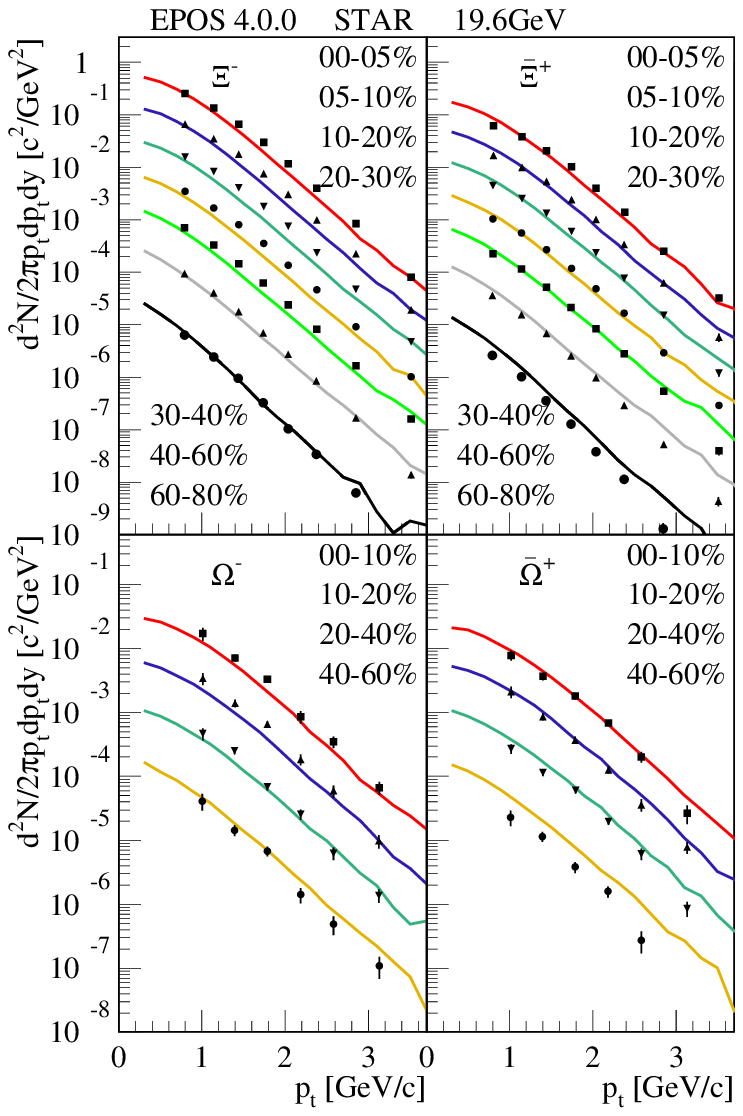}
\caption{Transverse momentum distributions of $\phi$, $K_{0}$, $\Lambda$,
$\bar{\Lambda}$ $\Xi^{-}$, $\bar{\Xi}^{+}$, $\Omega^{-}$, $\bar{\Omega}^{+}$
in AuAu collisions at 19.6 GeV at central rapidity for different centralities.
EPOS4 simulation (lines) are compared to data from STAR \cite{STAR:2020}.
\label{19-transverse-momentum-3}}
\end{figure}

In Fig. \ref{19-transverse-momentum-3}, we show transverse momentum
distributions of $\phi$, $K_{0}$, $\Lambda$, $\bar{\Lambda}$ $\Xi^{-}$,
$\bar{\Xi}^{+}$, $\Omega^{-}$, $\bar{\Omega}^{+}$ in AuAu collisions
at 19.6 GeV at central rapidity for different centralities. EPOS4
simulation (lines) are compared to data from STAR \cite{STAR:2020}.\\

In general, the simulation results are relatively close to the data,
concerning identified particles as pions, kaons, and protons, and
as well hyperons. Similar to what we have already seen at 27 GeV,
for kaons, antiprotons, and the $\phi$ meson, the simulation is somewhat
above the data.\\

\pagebreak{}

\subsection{Results for 11.5 GeV}

\begin{figure}[H]
\centering{}\includegraphics[bb=30bp 35bp 450bp 580bp,clip,scale=0.6]
{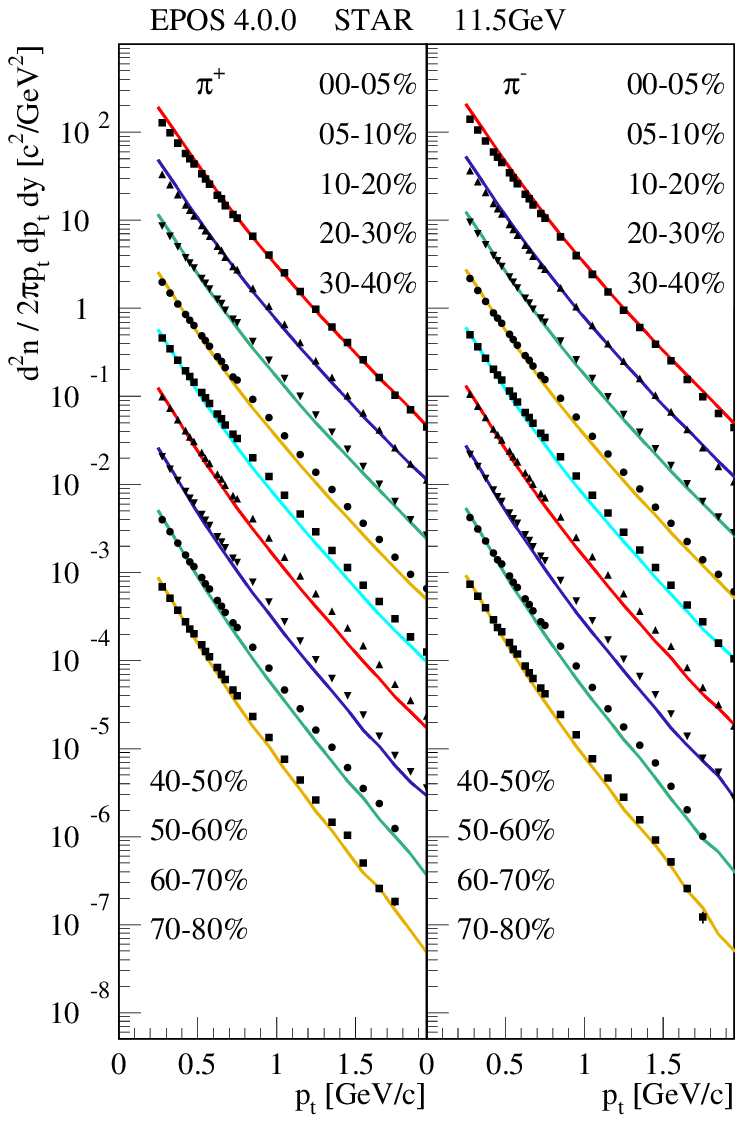}\\
 \caption{Transverse momentum distributions of $\pi^{+}$, $\pi^{-}$ in AuAu
collisions at 11.5 GeV for different centrality classes. EPOS4 simulation
(lines) are compared to data from STAR \cite{STAR:2017}. From top
to bottom, we multiply the curves by $3^{-i}$, $i=0,1,2,3,...$.
\label{11-transverse-momentum-1}}
\end{figure}

\begin{figure}[H]
\begin{centering}
\includegraphics[bb=30bp 35bp 450bp 580bp,clip,scale=0.6]
{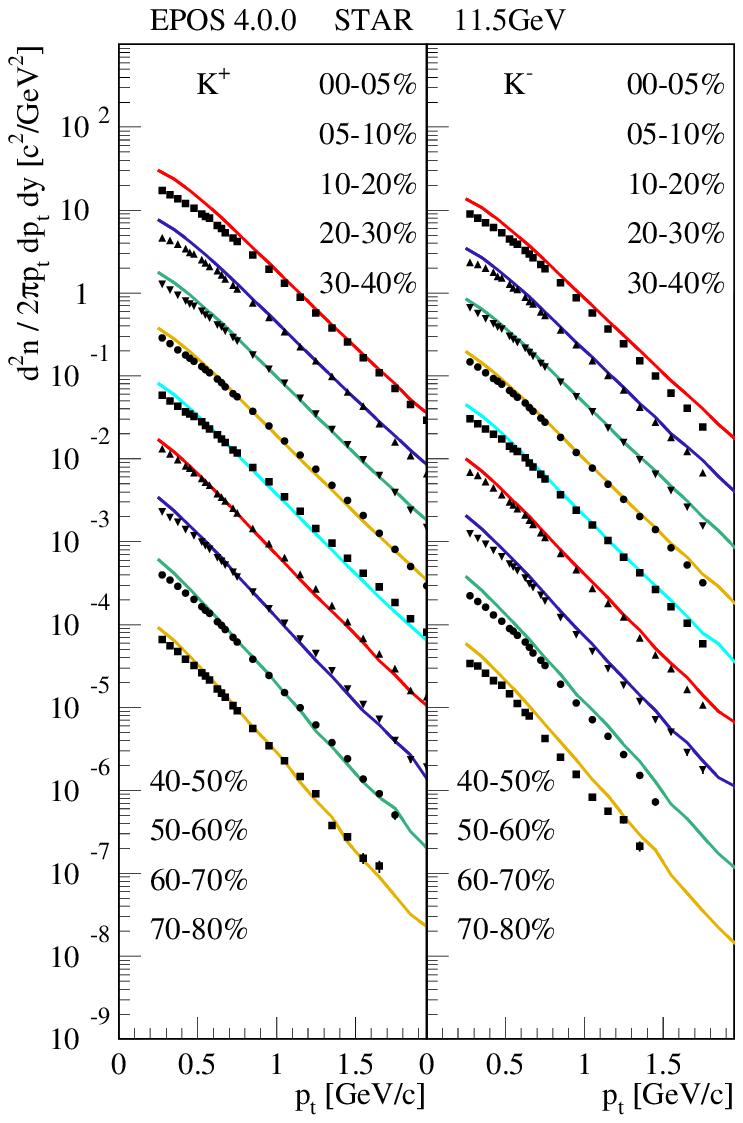} 
\par\end{centering}
\centering{}\includegraphics[bb=30bp 30bp 450bp 580bp,clip,scale=0.6]
{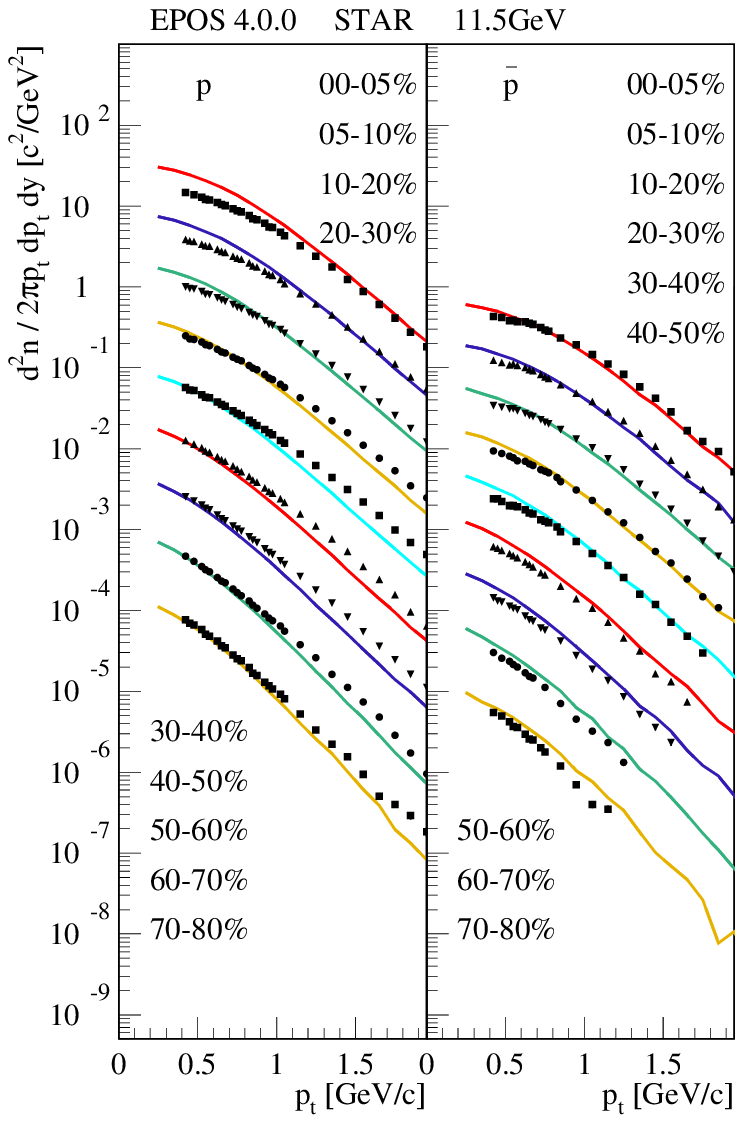}
\caption{Same as Fig. \ref{11-transverse-momentum-1}, but for $K^{+}$, $K^{-}$,
$p$, $\bar{p}$. \label{11-transverse-momentum-2}}
\end{figure}

In Figs. \ref{11-transverse-momentum-1} and \ref{11-transverse-momentum-2},
we show transverse momentum distributions of $\pi^{+}$, $\pi^{-}$,
$K^{+}$, $K^{-}$, $p$, $\bar{p}$ in AuAu collisions at 11.5 GeV
for different centrality classes. EPOS4 simulation (lines) are compared
to data from STAR \cite{STAR:2017}. From top to bottom, we multiply
the curves by $3^{-i}$, $i=0,1,2,3,...$.\\

\begin{figure}[H]
\begin{centering}
\includegraphics[bb=30bp 35bp 450bp 580bp,clip,scale=0.6]
{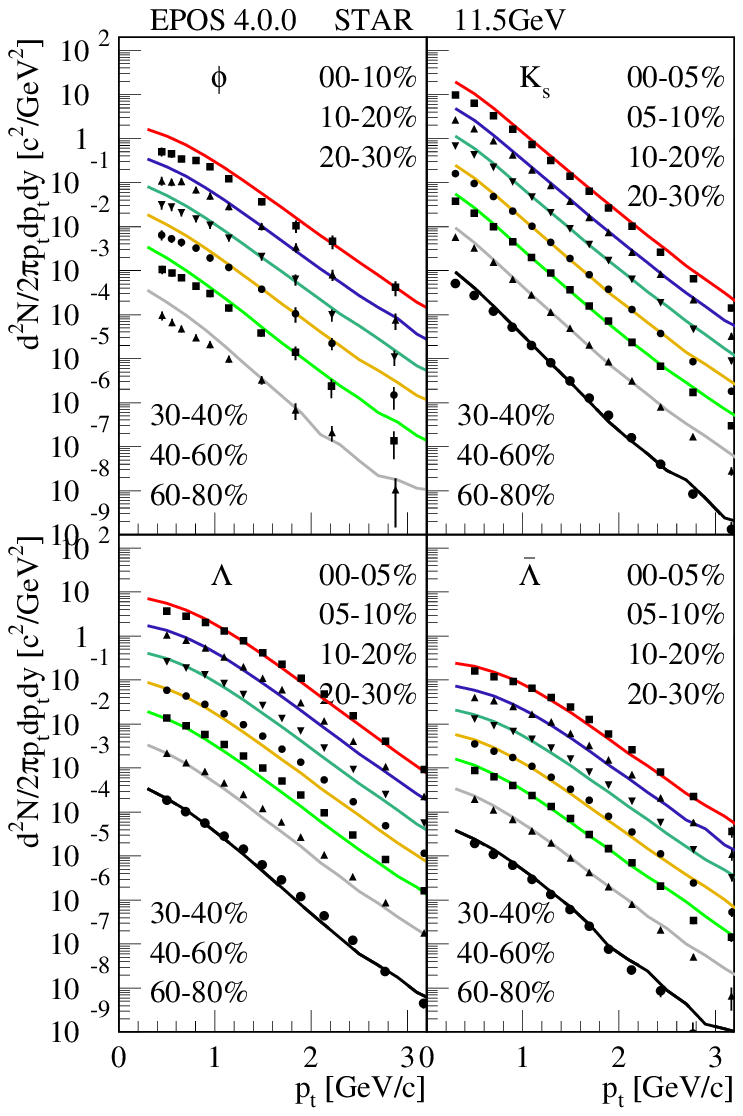} 
\par\end{centering}
\centering{}\includegraphics[bb=30bp 30bp 450bp 580bp,clip,scale=0.6]
{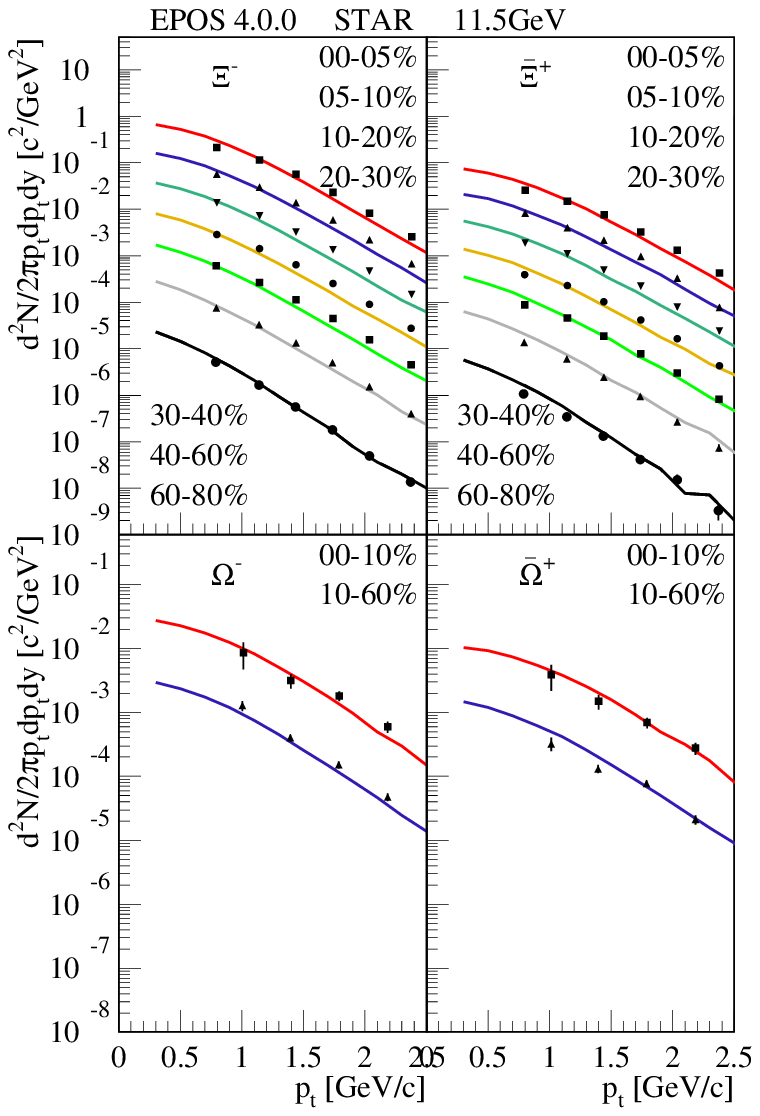}
\caption{Transverse momentum distributions of $\phi$, $K_{0}$, $\Lambda$,
$\bar{\Lambda}$ $\Xi^{-}$, $\bar{\Xi}^{+}$, $\Omega^{-}$, $\bar{\Omega}^{+}$
in AuAu collisions at 11.5 GeV at central rapidity for different centralities.
EPOS4 simulation (lines) are compared to data from STAR \cite{STAR:2020}.
\label{11-transverse-momentum-3}}
\end{figure}

In Fig. \ref{11-transverse-momentum-3}, we show transverse momentum
distributions of $\phi$, $K_{0}$, $\Lambda$, $\bar{\Lambda}$ $\Xi^{-}$,
$\bar{\Xi}^{+}$, $\Omega^{-}$, $\bar{\Omega}^{+}$ in AuAu collisions
at 11.5 GeV at central rapidity for different centralities. EPOS4
simulation (lines) are compared to data from STAR \cite{STAR:2020}.\\

Here we see for the first time (compared to higher energies) significant
deviations between simulation and data. The most striking is a large
proton excess at low $p_{t}$. And (as already seen earlier) a $\phi$
excess. Surprisingly, the hyperons are doing well.

\pagebreak{}

\subsection{Results for 7.7 GeV\textcolor{red}{\Huge{}{} \label{=======results-7GeV=======}}}

\begin{figure}[H]
\centering{}\includegraphics[bb=30bp 35bp 450bp 580bp,clip,scale=0.6]
{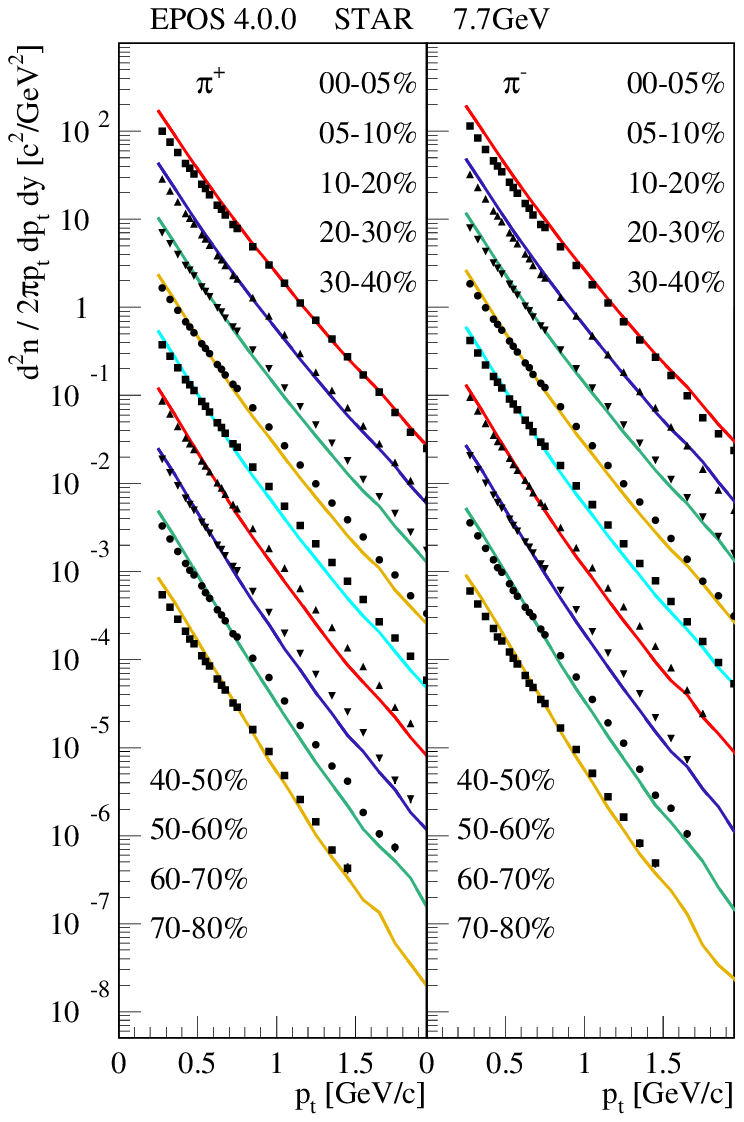}\\
 \caption{Transverse momentum distributions of $\pi^{+}$, $\pi^{-}$ in AuAu
collisions at 7.7 GeV for different centrality classes. EPOS4 simulation
(lines) are compared to data from STAR \cite{STAR:2017}. From top
to bottom, we multiply the curves by $3^{-i}$, $i=0,1,2,3,...$.
\label{7-transverse-momentum-1}}
\end{figure}

\begin{figure}[H]
\begin{centering}
\includegraphics[bb=30bp 35bp 450bp 580bp,clip,scale=0.6]
{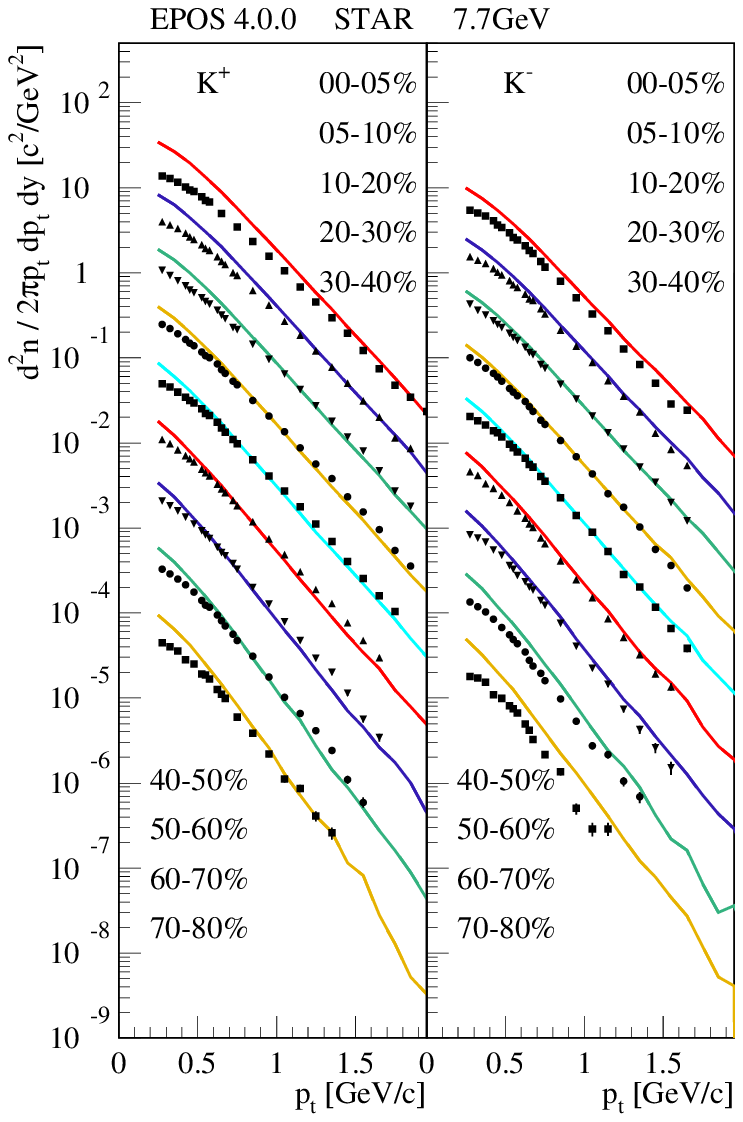} 
\par\end{centering}
\centering{}\includegraphics[bb=30bp 30bp 450bp 580bp,clip,scale=0.6]
{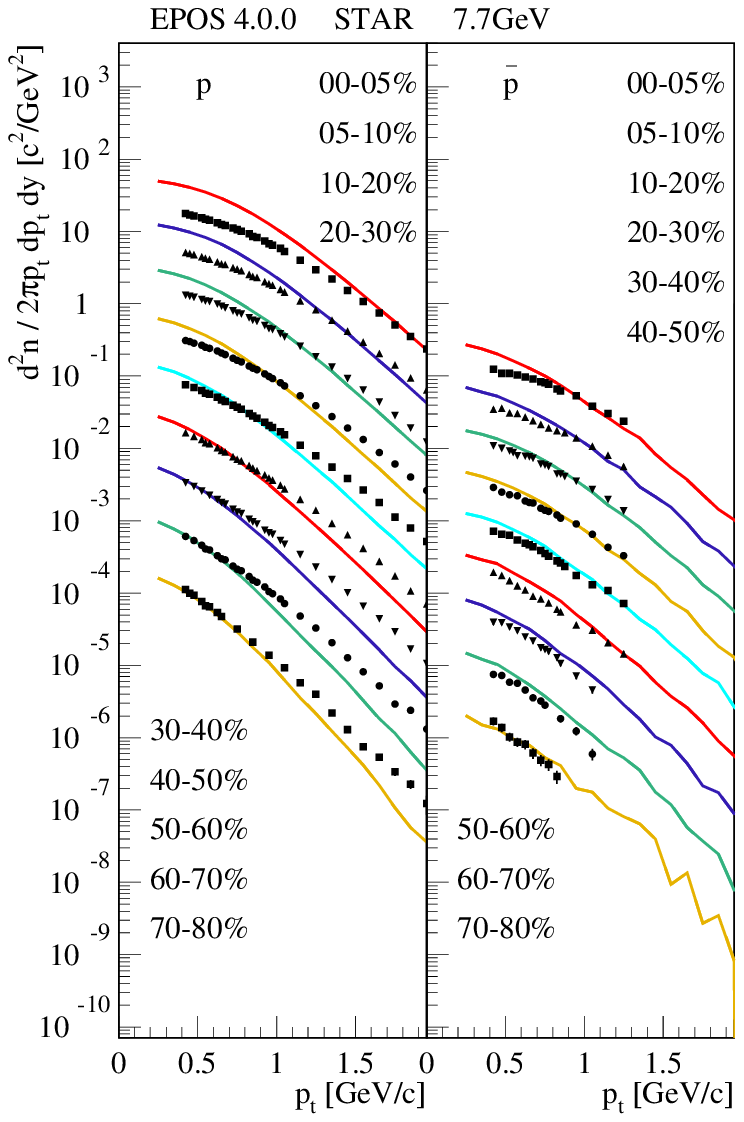}
\caption{Same as Fig. \ref{7-transverse-momentum-1}, but for $K^{+}$, $K^{-}$,
$p$, $\bar{p}$. \label{7-transverse-momentum-2}}
\end{figure}

In Figs. \ref{7-transverse-momentum-1} and \ref{7-transverse-momentum-2},
we show transverse momentum distributions of $\pi^{+}$, $\pi^{-}$,
$K^{+}$, $K^{-}$, $p$, $\bar{p}$ in AuAu collisions at 7.7 GeV
for different centrality classes. EPOS4 simulation (lines) are compared
to data from STAR \cite{STAR:2017}. From top to bottom, we multiply
the curves by $3^{-i}$, $i=0,1,2,3,...$. \\

\begin{figure}[H]
\begin{centering}
\includegraphics[bb=30bp 35bp 450bp 580bp,clip,scale=0.6]
{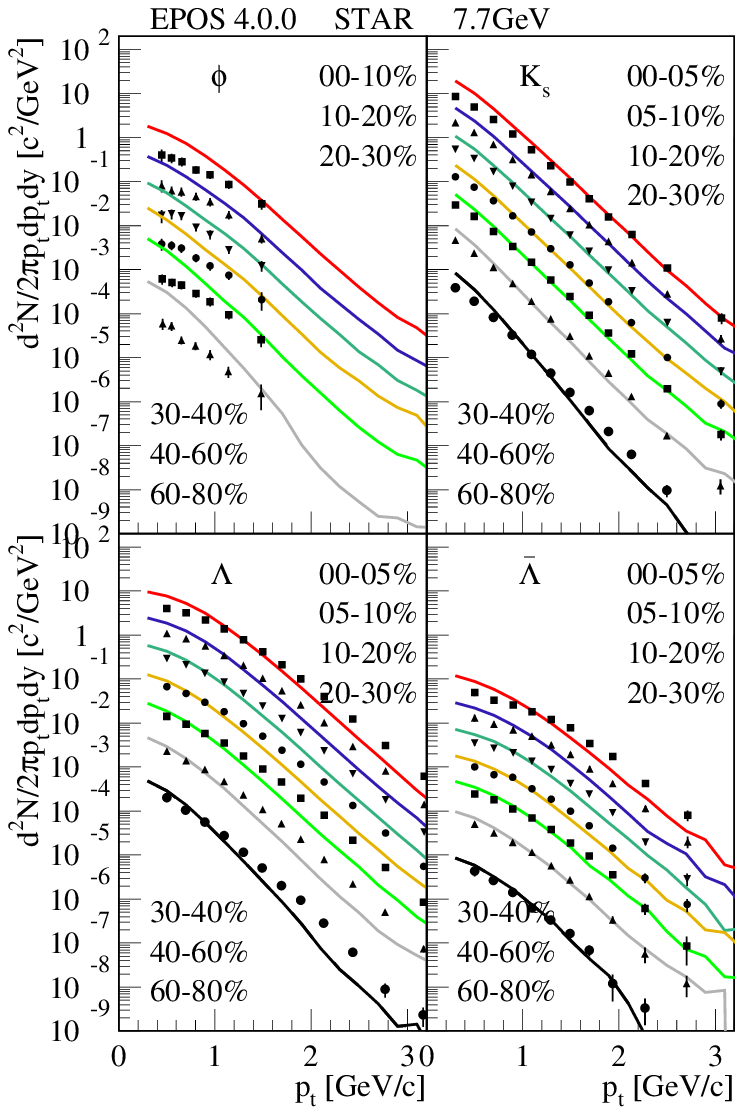} 
\par\end{centering}
\centering{}\includegraphics[bb=30bp 30bp 450bp 580bp,clip,scale=0.6]
{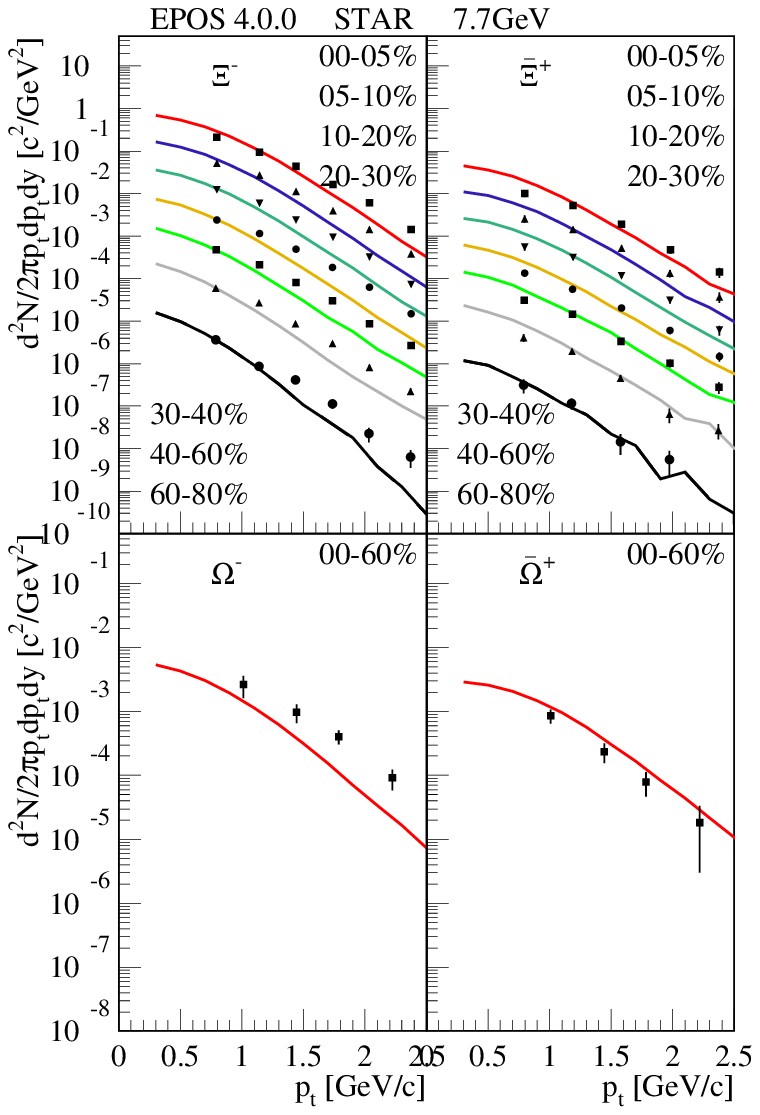}
\caption{Transverse momentum distributions of $\phi$, $K_{0}$, $\Lambda$,
$\bar{\Lambda}$ $\Xi^{-}$, $\bar{\Xi}^{+}$, $\Omega^{-}$, $\bar{\Omega}^{+}$
in AuAu collisions at 19.6 GeV at central rapidity for different centralities.
EPOS4 simulation (lines) are compared to data from STAR \cite{STAR:2020}.
\label{7-transverse-momentum-3}}
\end{figure}

In Fig. \ref{7-transverse-momentum-3}, we show transverse momentum
distributions of $\phi$, $K_{0}$, $\Lambda$, $\bar{\Lambda}$ $\Xi^{-}$,
$\bar{\Xi}^{+}$, $\Omega^{-}$, $\bar{\Omega}^{+}$ in AuAu collisions
at 19.6 GeV at central rapidity for different centralities. EPOS4
simulation (lines) are compared to data from STAR \cite{STAR:2020}.
\\

Here, at 7.7 GeV, essentially all spectra from the simulation are
too soft, the yields at low $p_{t}$ too high, with the biggest excess
observed for protons and $K^{+}$ mesons. So here the model does not
work.

\section{Results concerning integrated yields\label{=======results-yields=======}}

In this section, we show the integrated yields $dN/dy$ for every particle species as a function of $\sqrt{s_{NN}}$, at mid-rapidity in the most central AuAu collisions. We summarise thus the discussion from the previous section, by displaying how particle production changes with collision energy for every hadronic species, compared to STAR data from \cite{STAR:2017, STAR:2020}. 
\\

\begin{figure}[h]
\includegraphics[bb=30bp 25bp 650bp 580bp,clip,scale=0.2]
{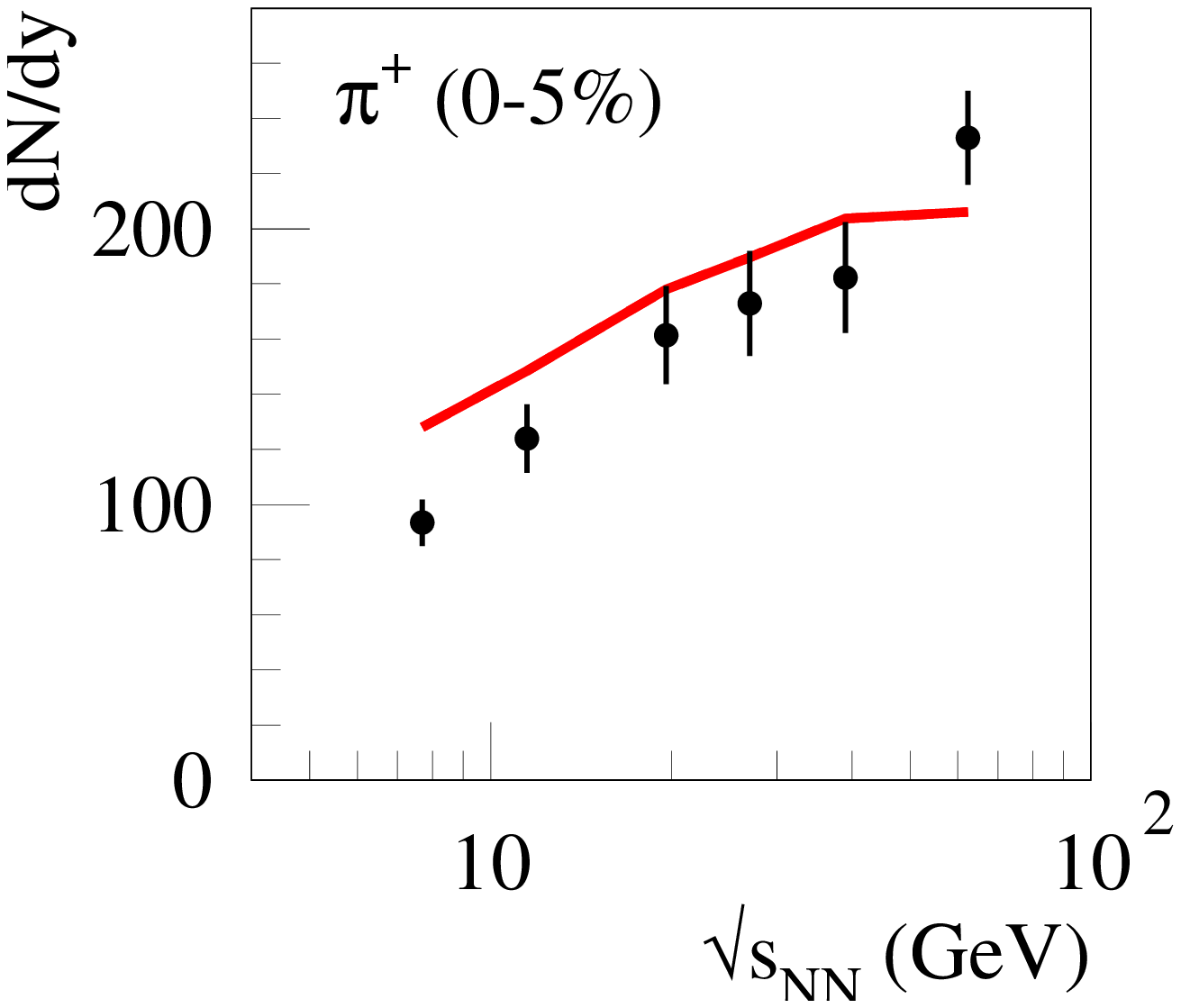} 
\includegraphics[bb=30bp 25bp 650bp 580bp,clip,scale=0.2]
{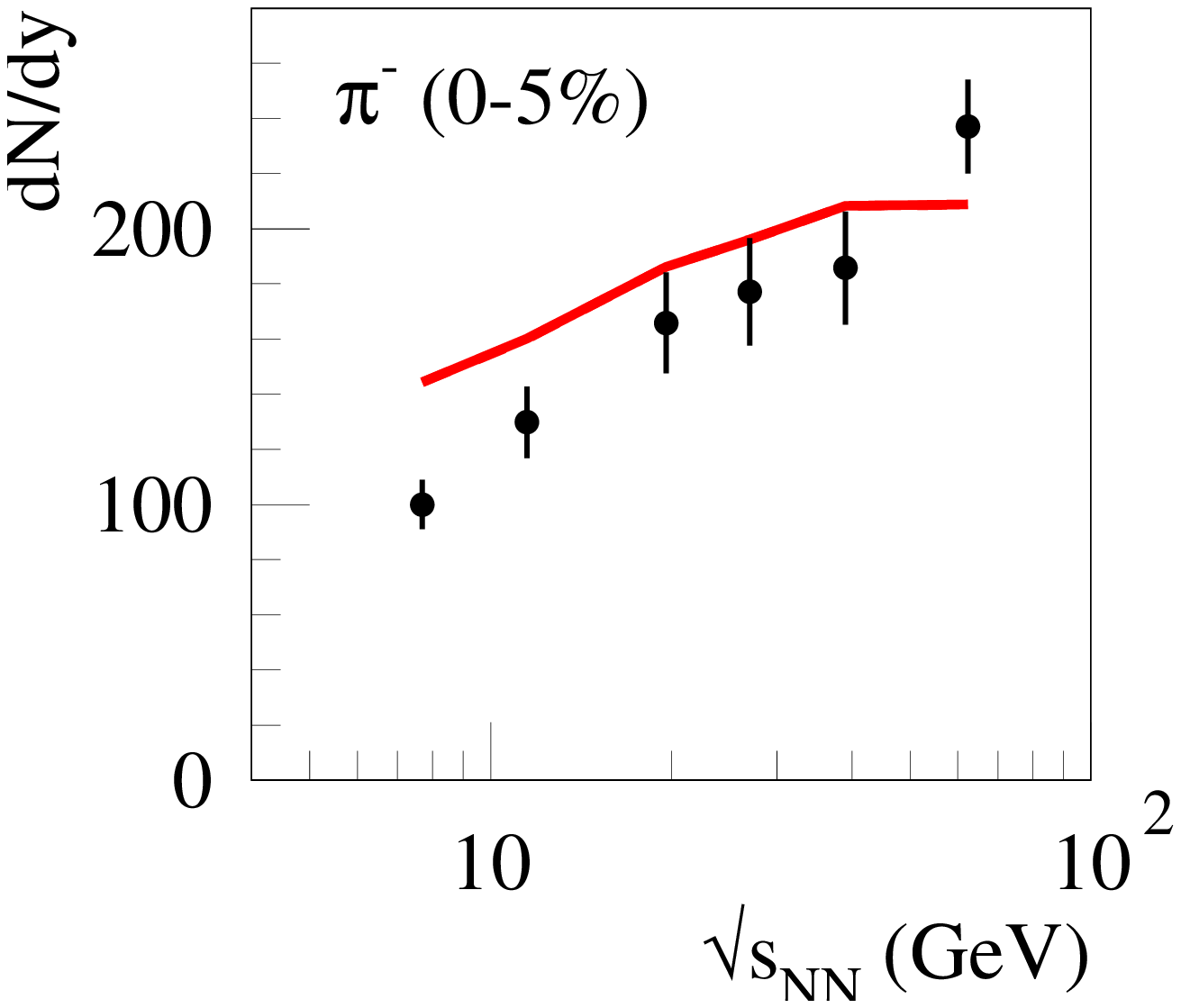}
\includegraphics[bb=30bp 25bp 650bp 580bp,clip,scale=0.2]
{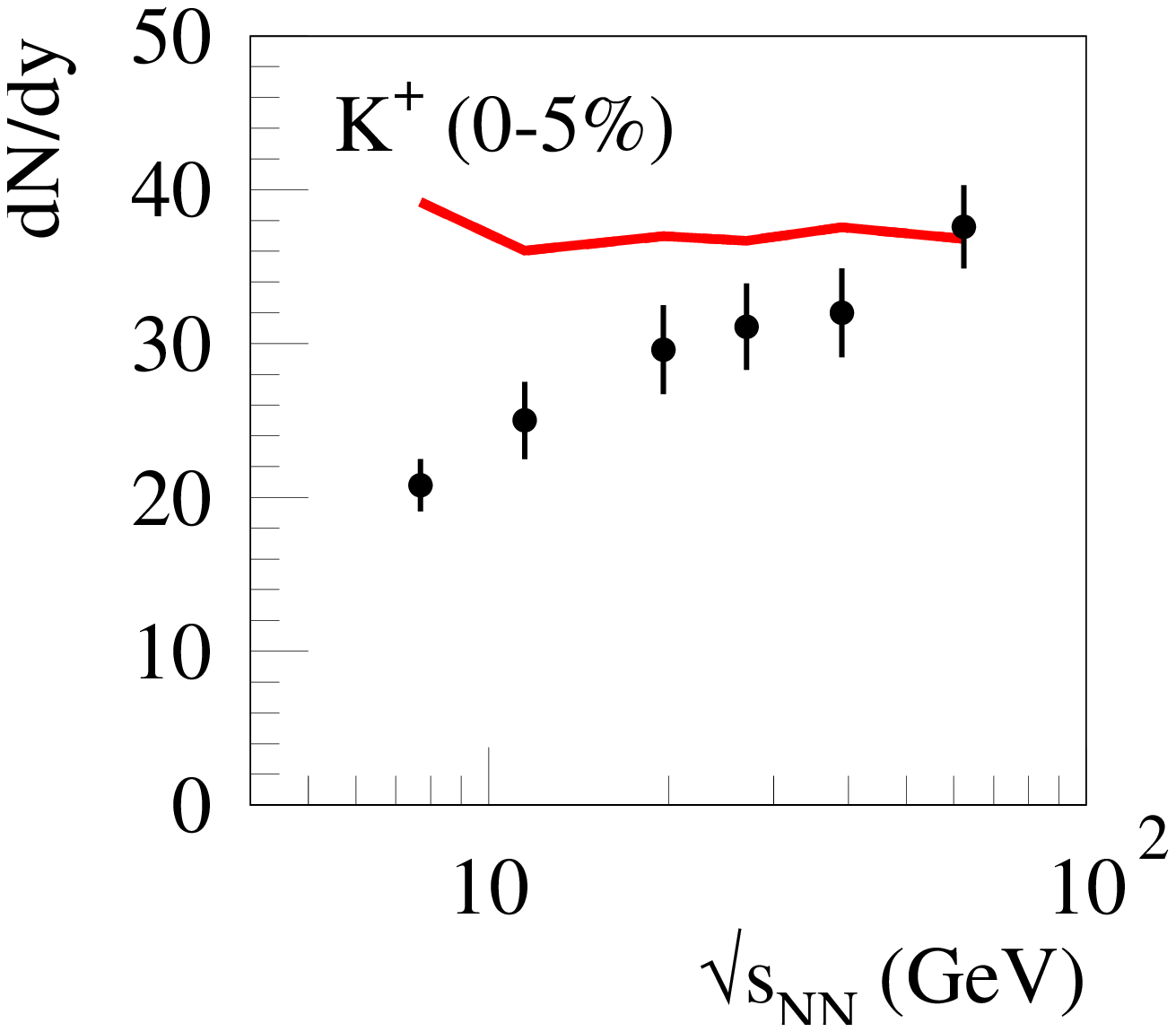} 
\includegraphics[bb=30bp 25bp 650bp 580bp,clip,scale=0.2]
{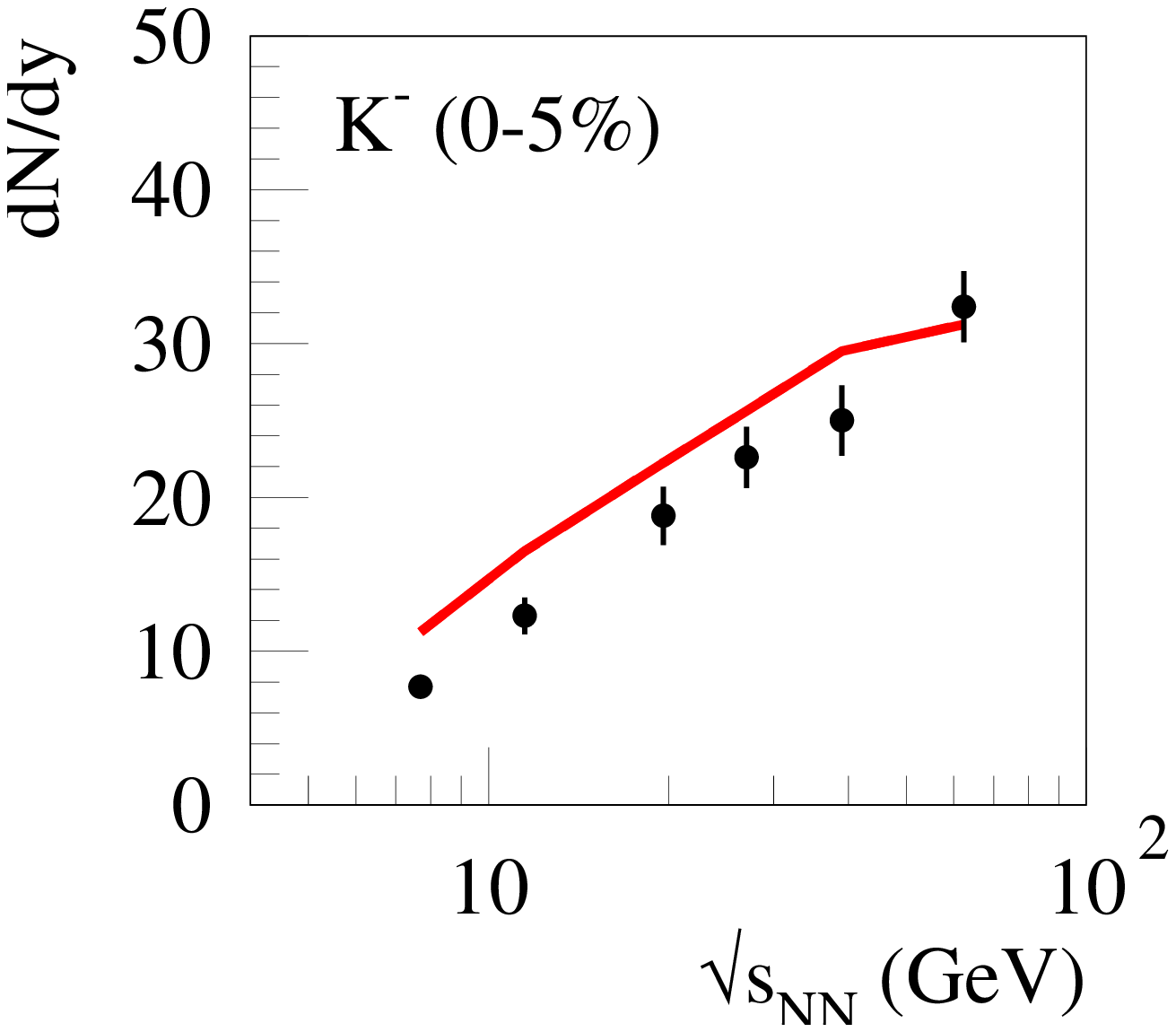}
\includegraphics[bb=30bp 25bp 650bp 580bp,clip,scale=0.2]
{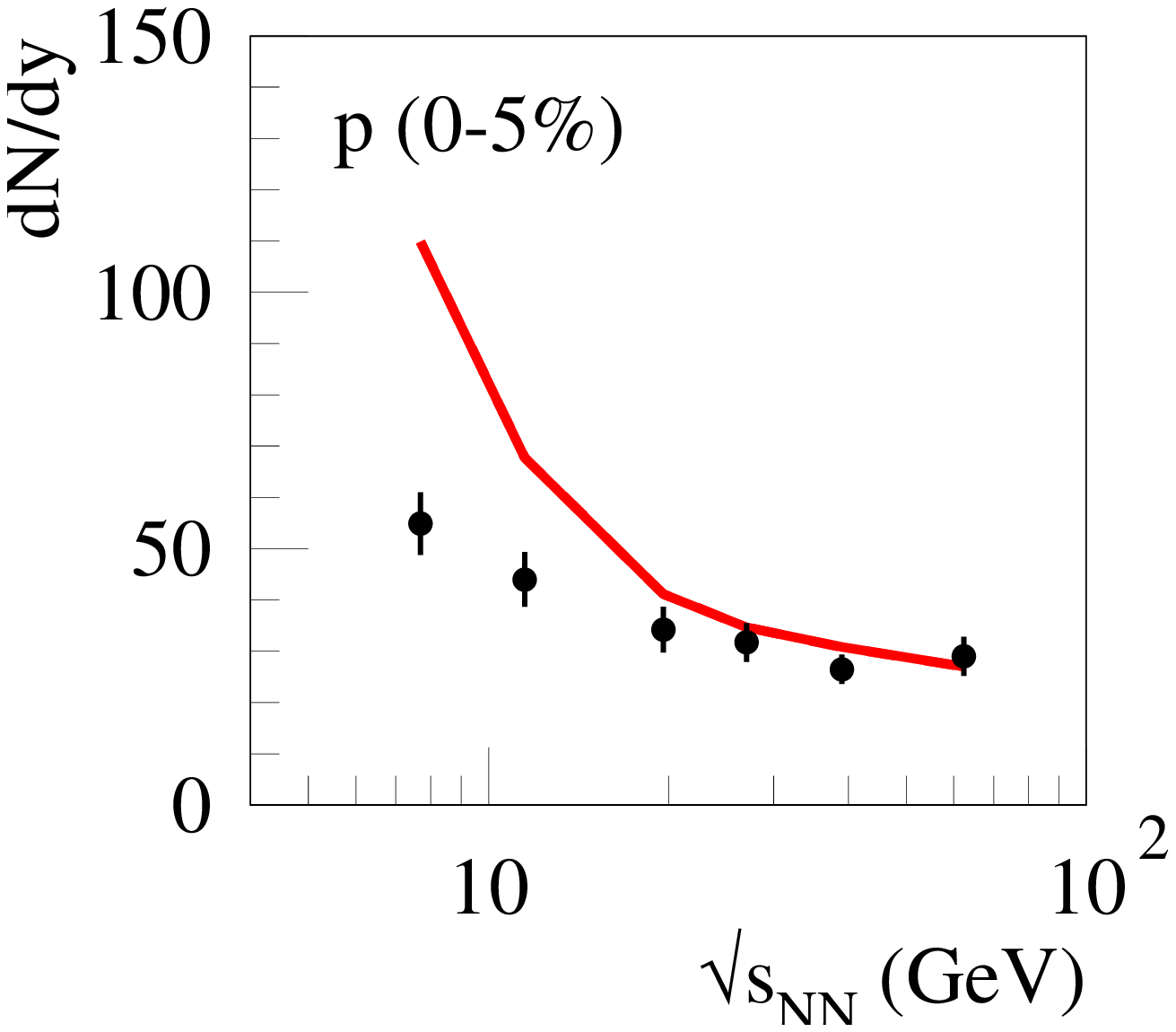} 
\includegraphics[bb=30bp 25bp 650bp 580bp,clip,scale=0.2]
{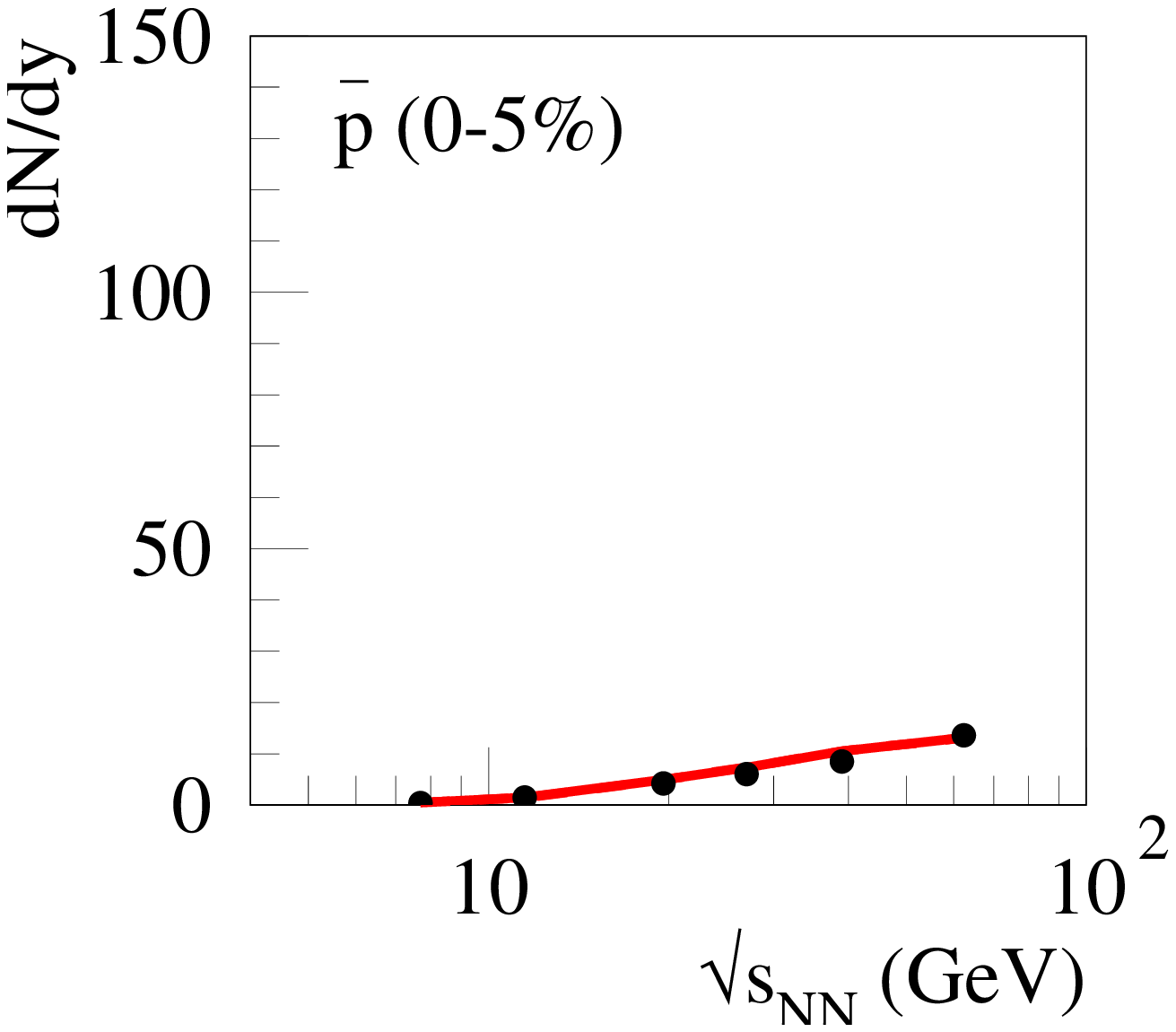}
\caption{Integrated yields of $\pi^{+}$, $\pi^{-}$, $K^+$, $K^-$, $p$, and $\overline{p}$  at mid-rapidity ($|y|<0.1$) in the most central AuAu collision class (0-5\%), as a function of collision energy $\sqrt{s_{NN}}$. EPOS4 simulation (lines) are compared to data from STAR \cite{STAR:2017}.
\label{integrated-yield-pi}}
\end{figure}

\begin{figure}[h]
\includegraphics[bb=30bp 25bp 650bp 580bp,clip,scale=0.2]
{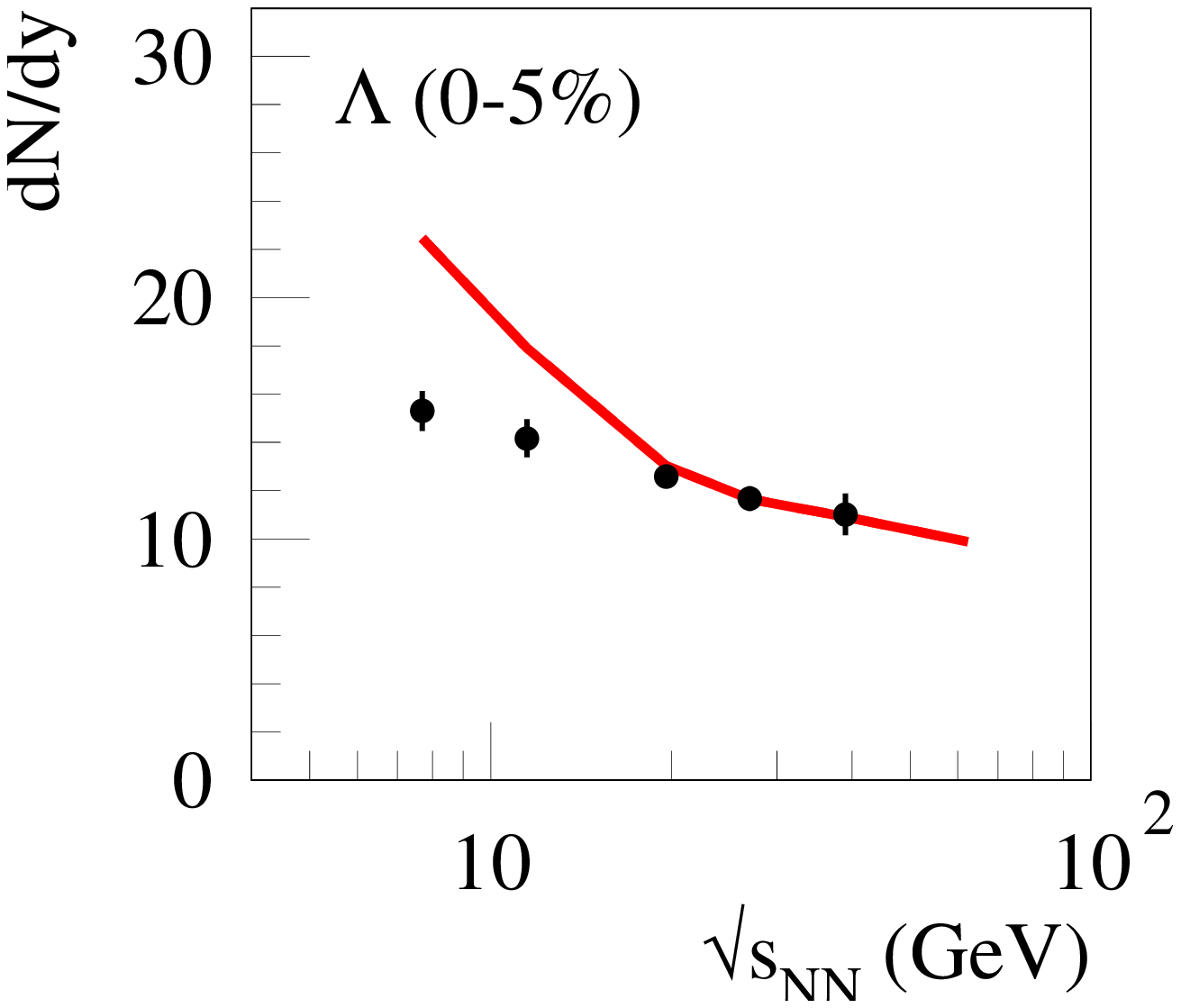} 
\includegraphics[bb=30bp 25bp 650bp 580bp,clip,scale=0.2]
{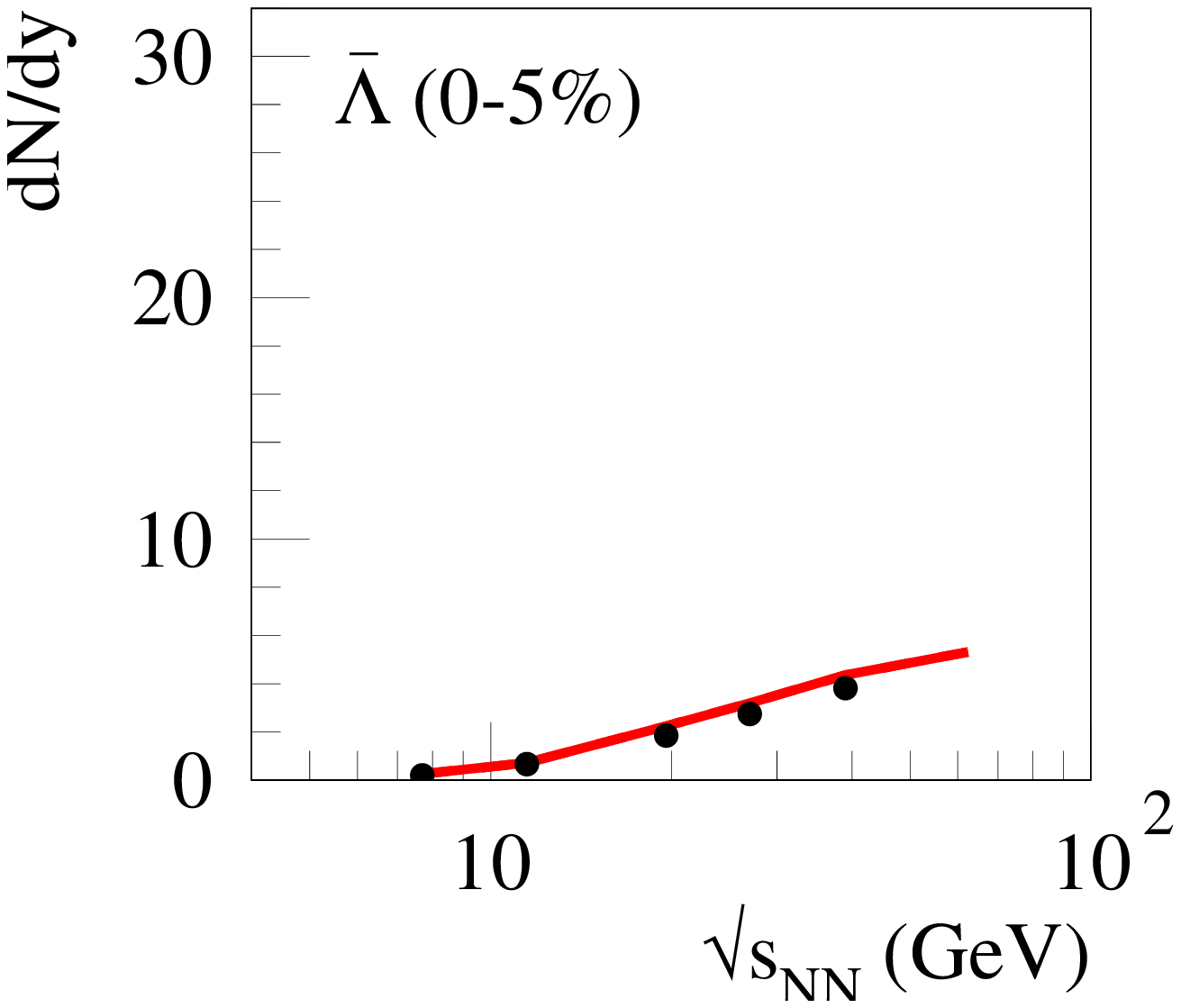}
\includegraphics[bb=30bp 25bp 650bp 580bp,clip,scale=0.2]
{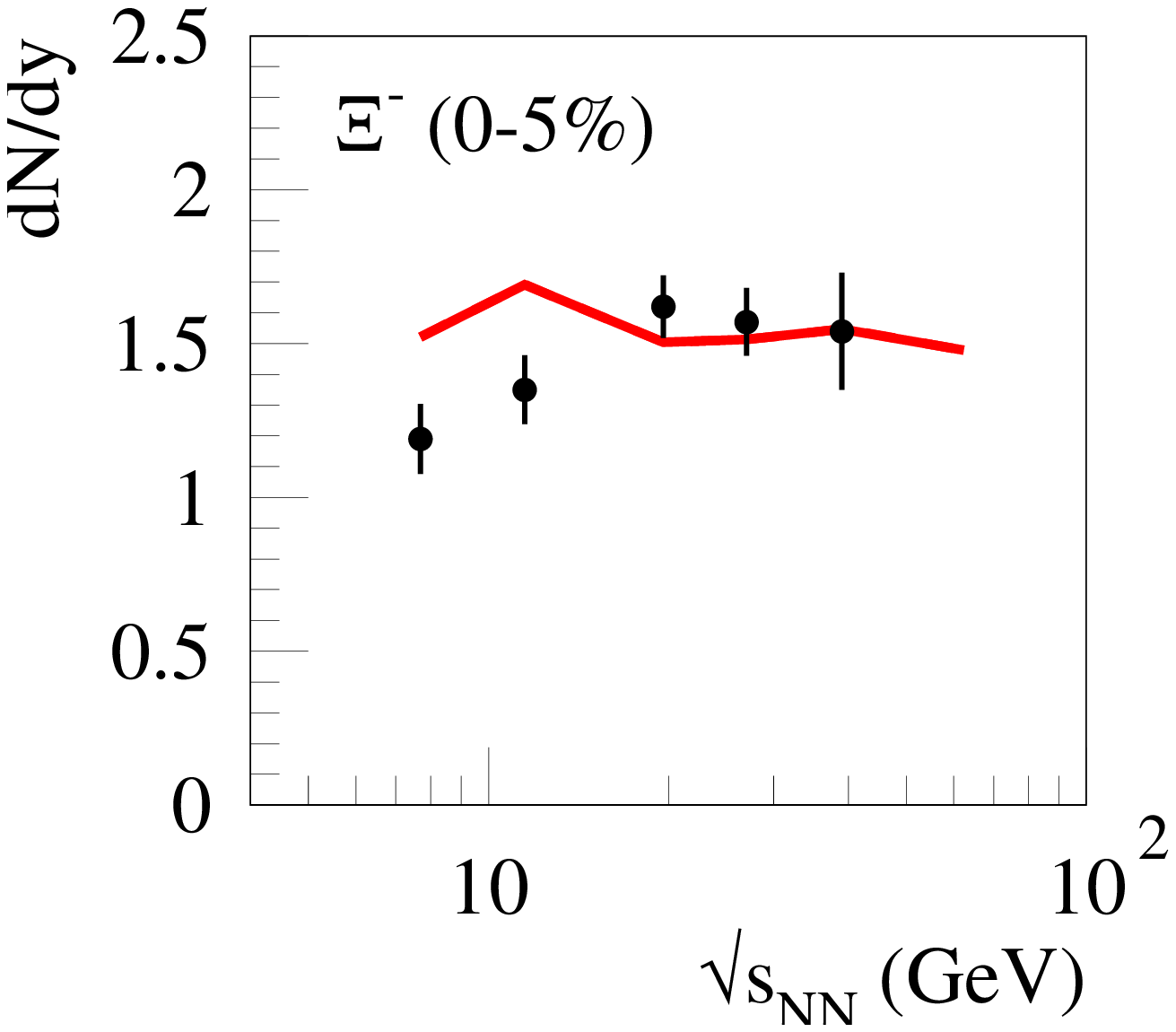} 
\includegraphics[bb=30bp 25bp 650bp 580bp,clip,scale=0.2]
{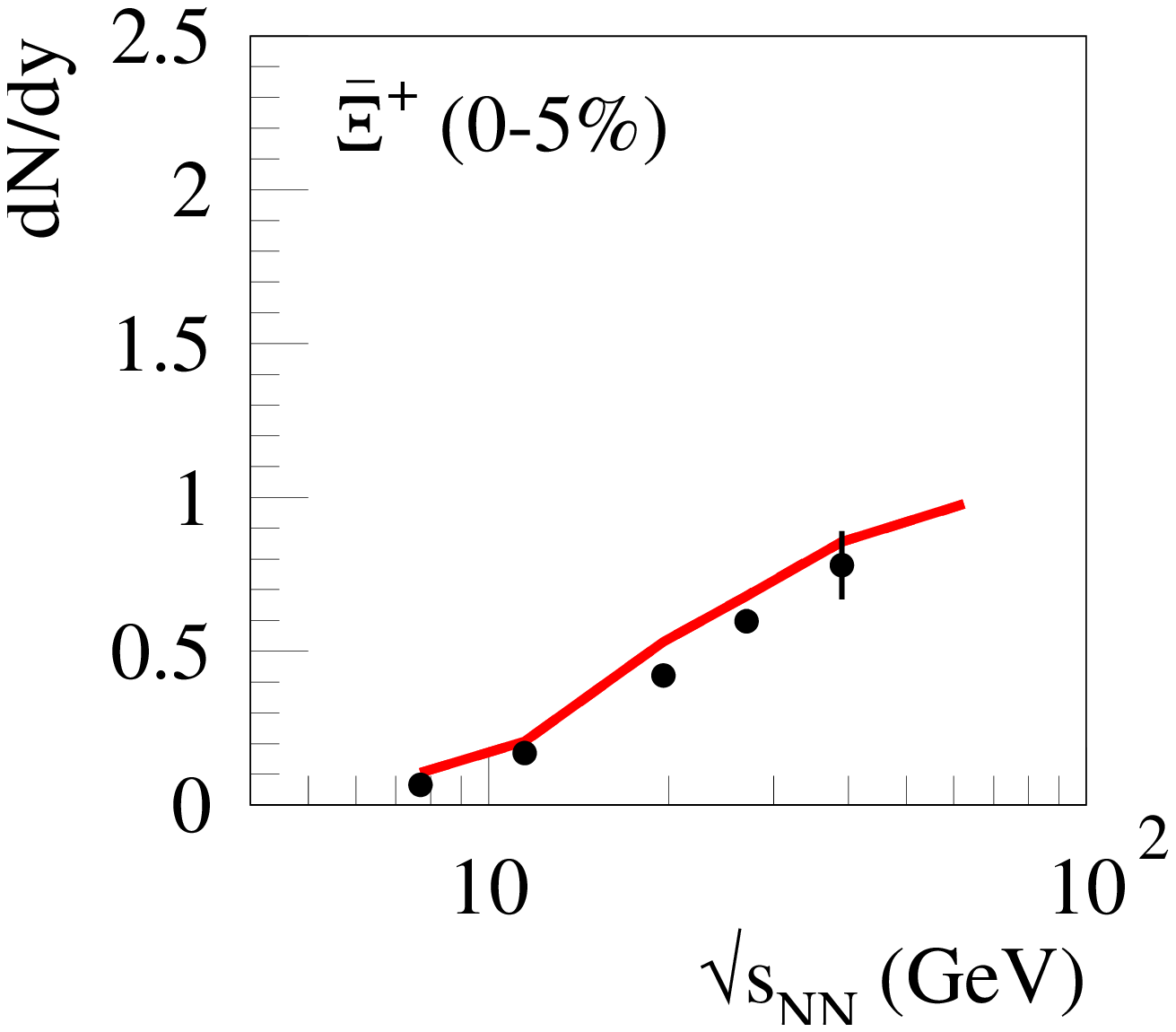}
\includegraphics[bb=30bp 25bp 650bp 580bp,clip,scale=0.2]
{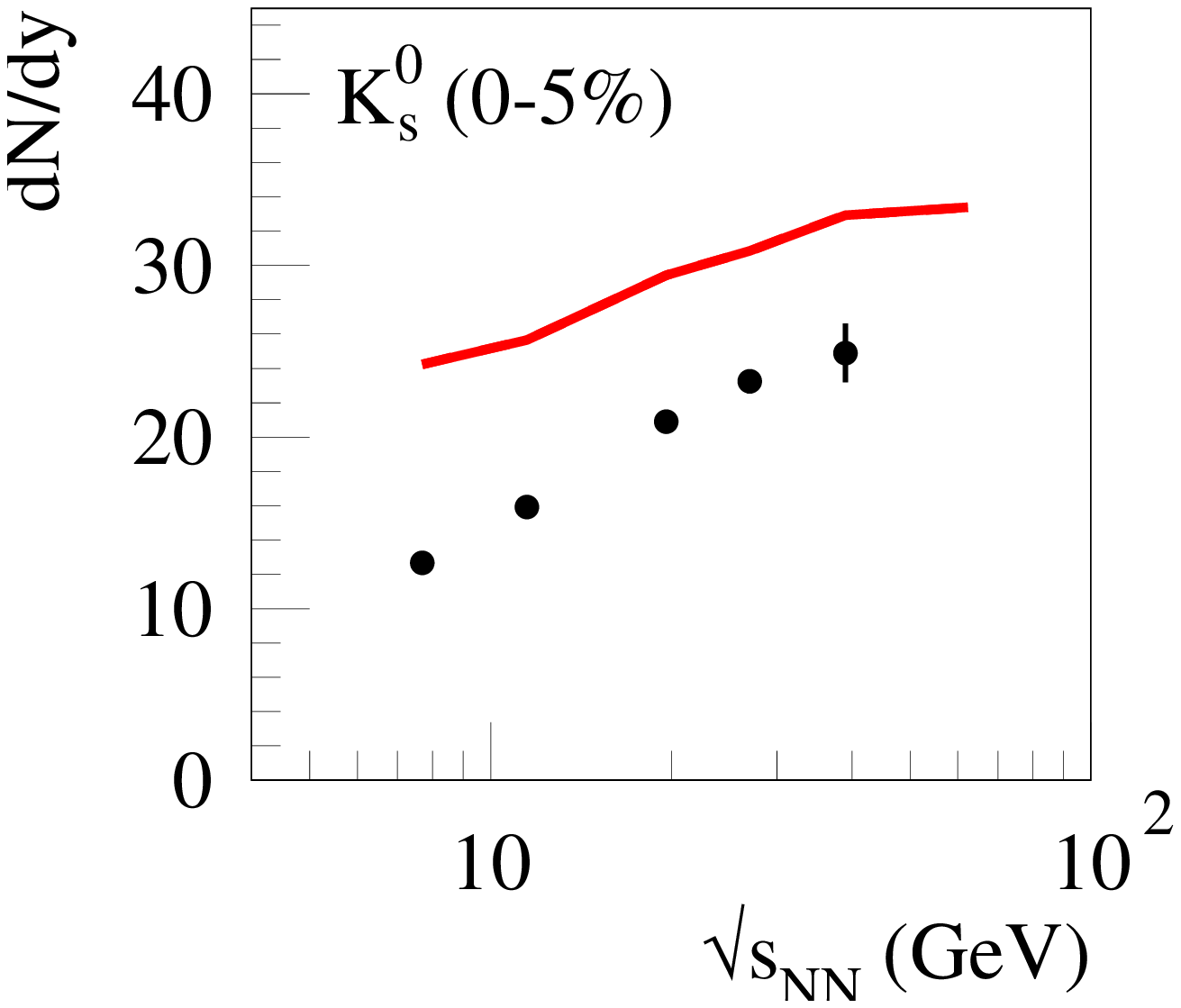} 
\includegraphics[bb=30bp 25bp 650bp 580bp,clip,scale=0.2]
{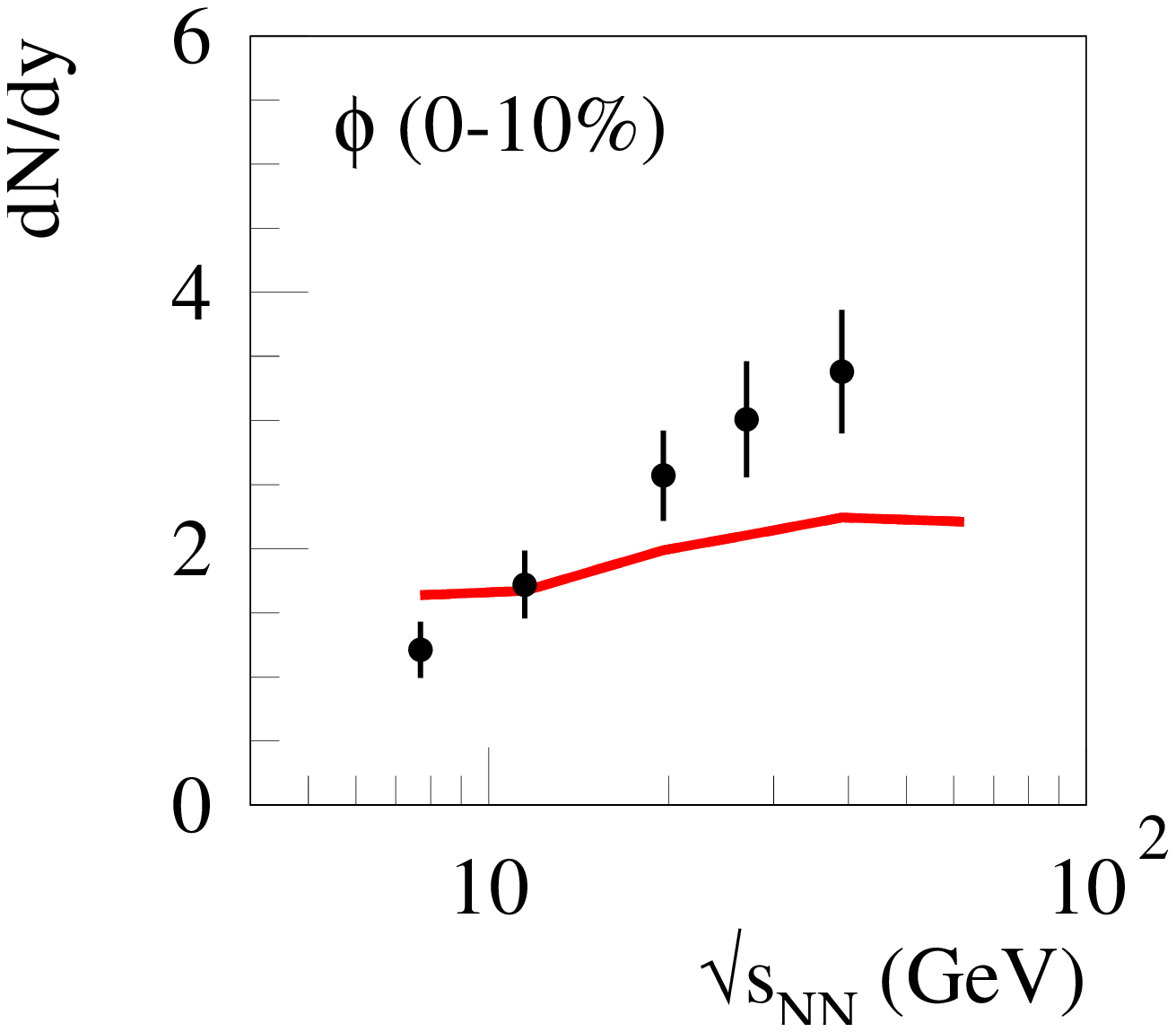}
\caption{Integrated yields of $\Lambda$, $\overline{\Lambda}$, $\Xi^-$, $\overline{\Xi}^+$, $\phi$, $K^0_s$ at mid-rapidity ($|y|<0.1$) in the most central AuAu collision class (0-5\%), as a function of collision energy $\sqrt{s_{NN}}$. EPOS4 simulation (lines) are compared to data from STAR \cite{STAR:2017}.
\label{integrated-yield-Lam}}
\end{figure}

We compare in Fig. \ref{integrated-yield-pi}
the yields from EPOS4 (lines) of $\pi^+$, $\pi^-$ $K^+$, $K^-$, $p$, and $\overline{p}$, measured within $|y|<0.1$, with STAR data from \cite{STAR:2017}. 
We do the same in Fig. \ref{integrated-yield-Lam}
for $\Lambda$, $\overline{\Lambda}$, $\Xi^-$, $\overline{\Xi}^+$, $\phi$, and $K^0_s$, measured within $|y|<0.5$, compared with STAR data from \cite{STAR:2020}. 
For all hadronic species, the events considered are from the 0-5\% centrality class, except for the $\phi$ mesons for which we consider 0-10\% centrality events due to the available data. 
We only omit here $\Omega^-$ and $\overline{\Omega}^+$, as the data published by STAR in \cite{STAR:2020} is not using the same centrality classes for all energies, making it impossible to display meaningful results about the evolution of integrated yields with collision energy.
\\

When looking at the whole ensemble of results, we see that particle production in EPOS4 simulations above 19.6 GeV is, overall, in good qualitative and quantitative agreement with STAR data. As expected, it reflects the conclusions from the last section.  
At lower energies, we observe a clear overproduction of protons, as well as $\Lambda$ and $\Xi^-$ baryons, unlike their respective anti-baryon counterparts. 
The yield of $K^+$ mesons is also largely overestimated at low energies, and displays notably almost no dependence in energy.
\\

Finally, the yield of $K^0_s$ shown in Fig. \ref{integrated-yield-Lam} is overestimated for the whole range of collision energies displayed here, compatible with the $p_t$ spectra at low $p_t$. However, its energy dependence follows qualitatively well the data.\\ 

On the contrary, the $\phi$ meson is the only species which is underproduced in collision with energies 19.6 GeV and above. But this is in contradiction with the $p_t$ spectra shown earlier, where we see an overproduction.

\section{Results concerning v\protect\textsubscript{2}\textcolor{red}{\Huge{}\label{=======results-v2=======}}}

\begin{figure}[h]
\begin{centering}
\includegraphics[bb=40bp 35bp 470bp 495bp,clip,scale=0.6]
{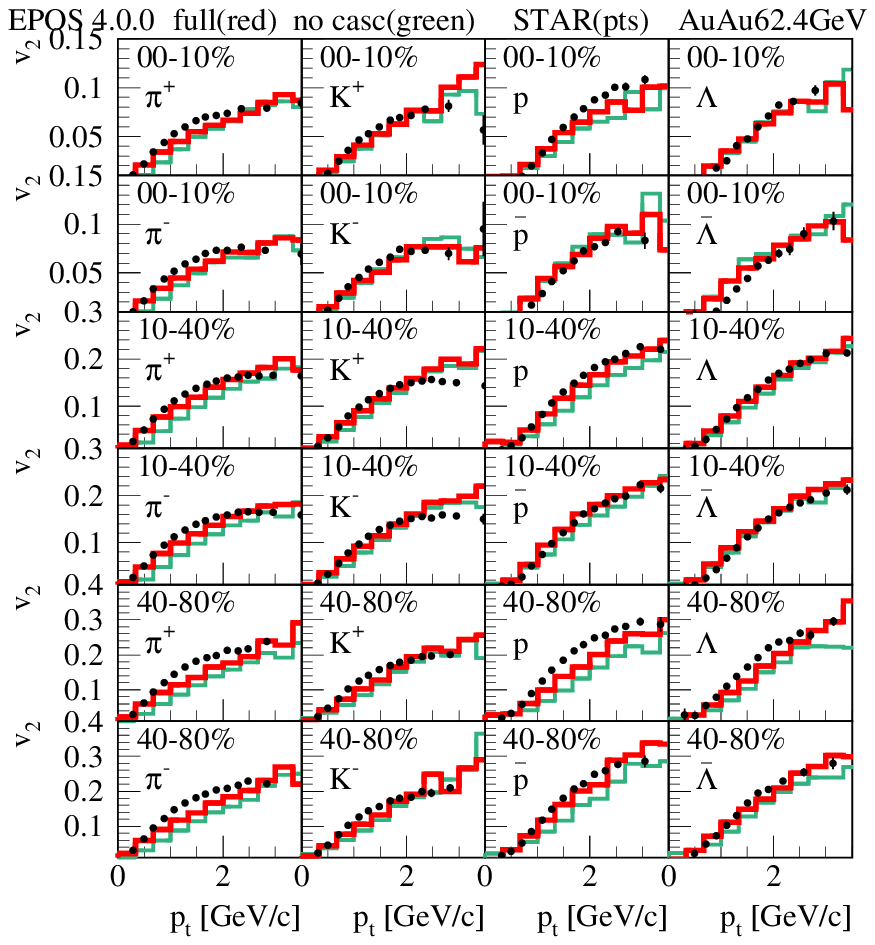} 
\par\end{centering}
\begin{centering}
\includegraphics[bb=30bp 35bp 470bp 495bp,clip,scale=0.6]
{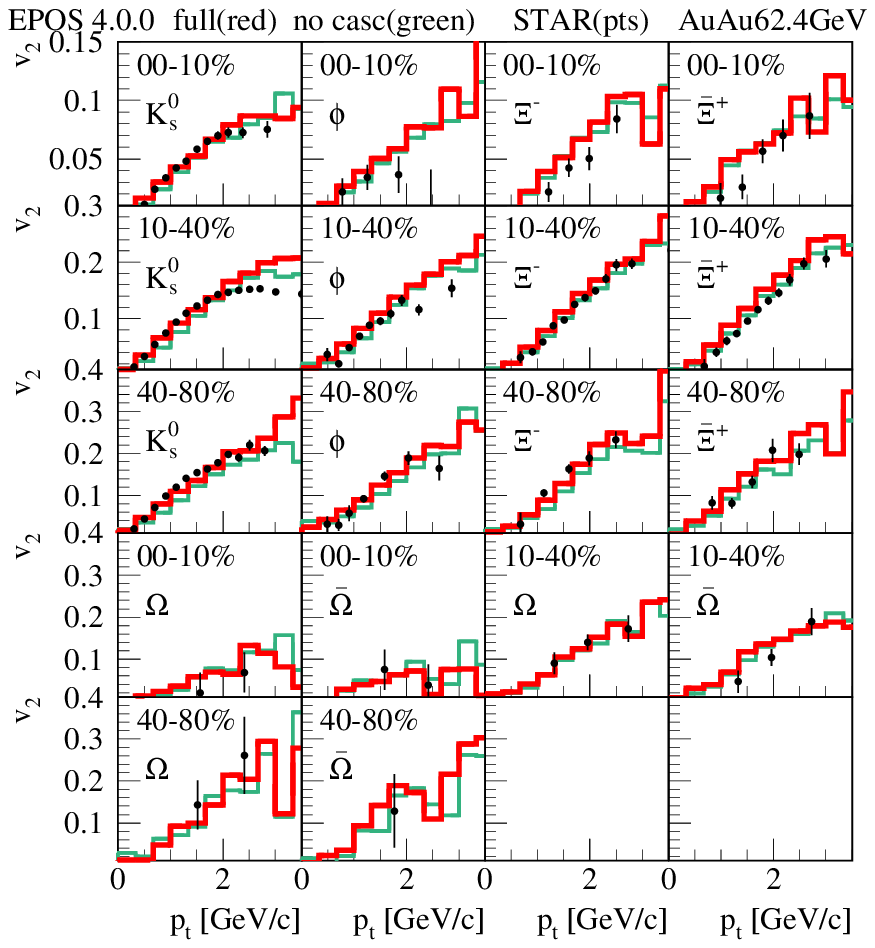} 
\par\end{centering}
\centering{}\caption{Transverse momentum dependence of $v_{2}$ of identified particles
in AuAu collisions at 62.4 GeV at central rapidity for different centralities.
EPOS4 simulation, full simulations (thick red lines) and without hadronic
cascade (thin green lines), are compared to data from STAR \cite{STAR:2015-V2-identified-7to62}
(dots). \label{v2-62}}
\end{figure}

\begin{figure}[h]
\begin{centering}
\includegraphics[bb=40bp 35bp 470bp 495bp,clip,scale=0.6]
{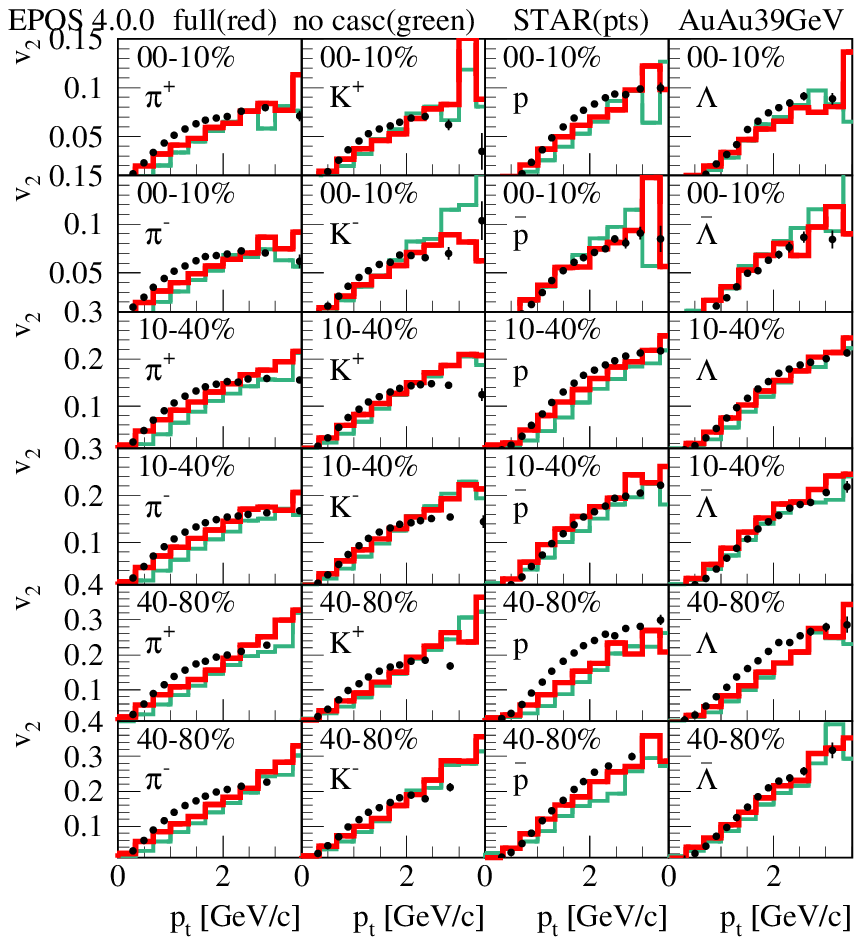}
\par\end{centering}
\begin{centering}
\includegraphics[bb=30bp 35bp 470bp 495bp,clip,scale=0.6]
{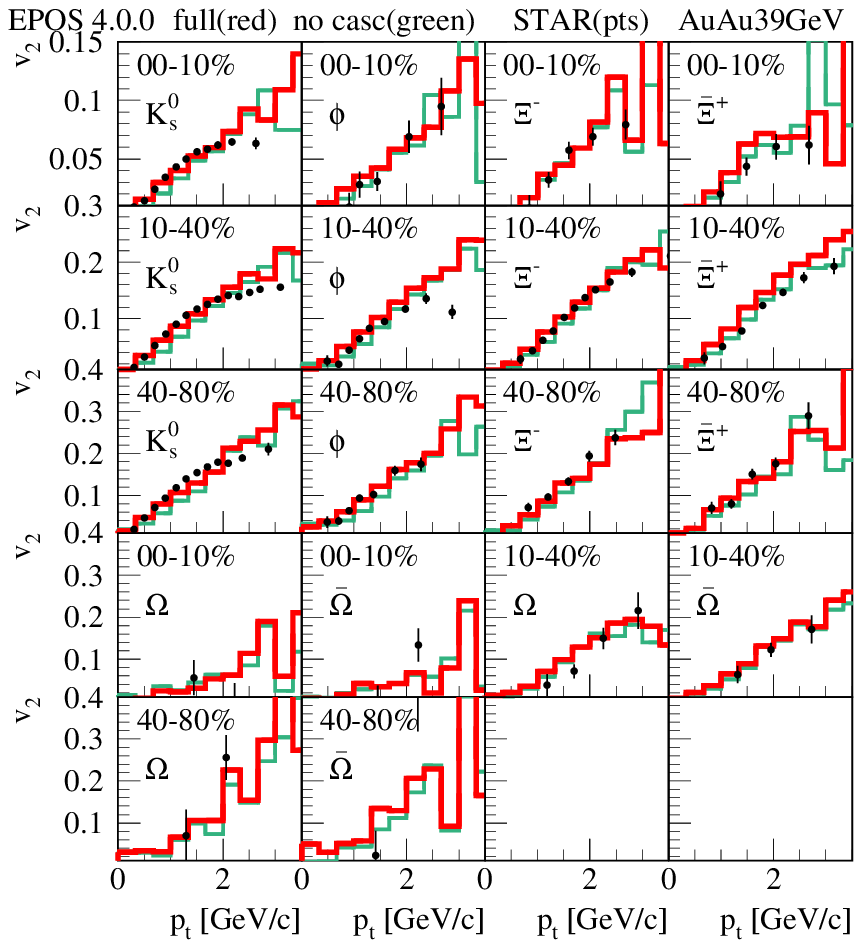}
\par\end{centering}
\centering{}\caption{Transverse momentum dependence of $v_{2}$ of identified particles
in AuAu collisions at 39.4 GeV at central rapidity for different centralities.
EPOS4 simulation, full simulations (thick red lines) and without hadronic
cascade (thin green lines), are compared to data from STAR \cite{STAR:2015-V2-identified-7to62}
(dots). \label{v2-39}}
\end{figure}

\begin{figure}[h]
\begin{centering}
\includegraphics[bb=40bp 35bp 470bp 495bp,clip,scale=0.6]
{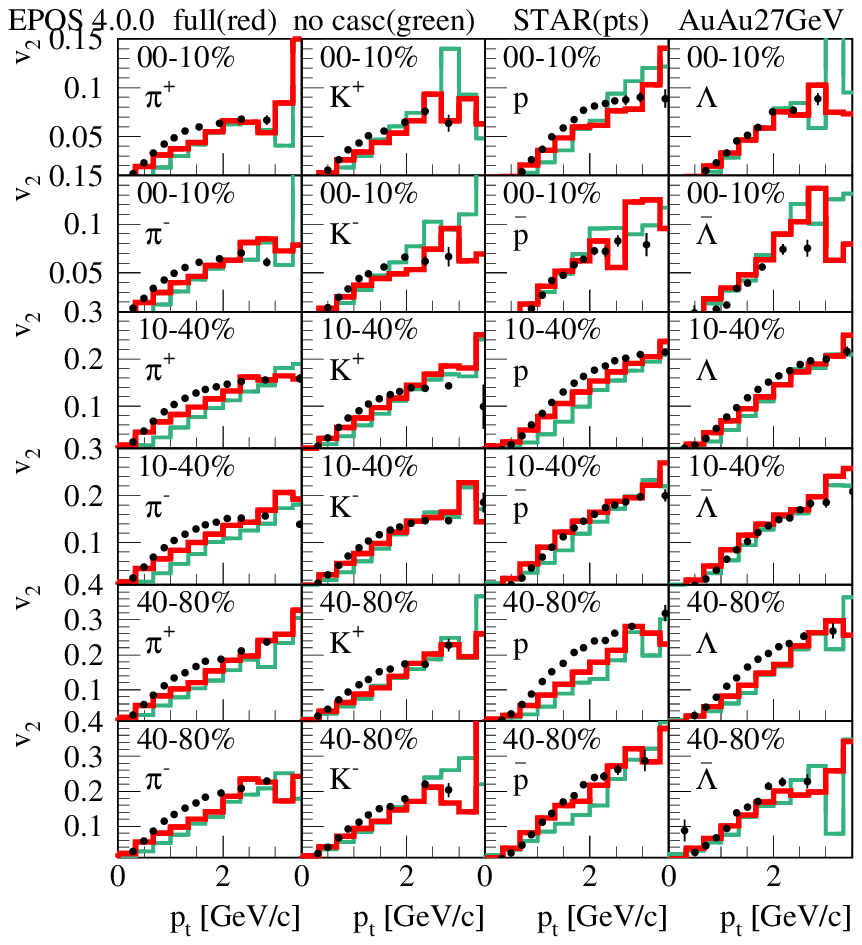}
\par\end{centering}
\begin{centering}
\includegraphics[bb=30bp 35bp 470bp 495bp,clip,scale=0.6]
{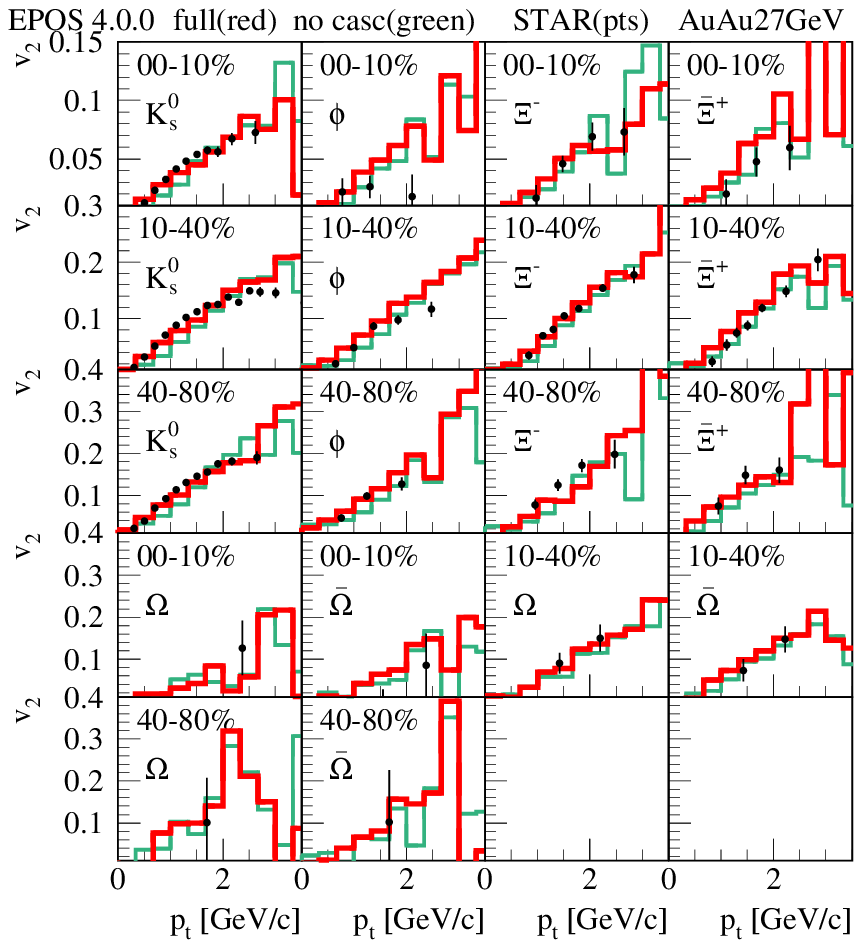}
\par\end{centering}
\centering{}\caption{Transverse momentum dependence of $v_{2}$ of identified particles
in AuAu collisions at 27 GeV at central rapidity for different centralities.
EPOS4 simulation, full simulations (thick red lines) and without hadronic
cascade (thin green lines), are compared to data from STAR \cite{STAR:2015-V2-identified-7to62}
(dots). \label{v2-27}}
\end{figure}

\begin{figure}[h]
\begin{centering}
\includegraphics[bb=40bp 35bp 470bp 495bp,clip,scale=0.6]
{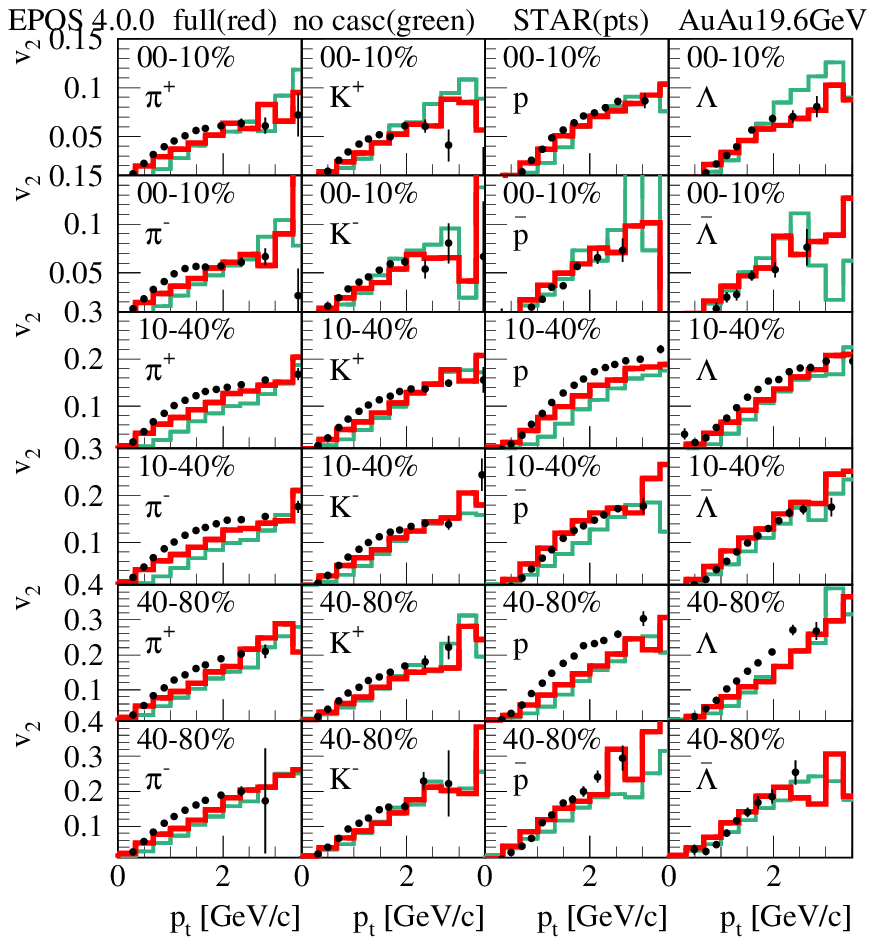}
\par\end{centering}
\begin{centering}
\includegraphics[bb=30bp 35bp 470bp 495bp,clip,scale=0.6]
{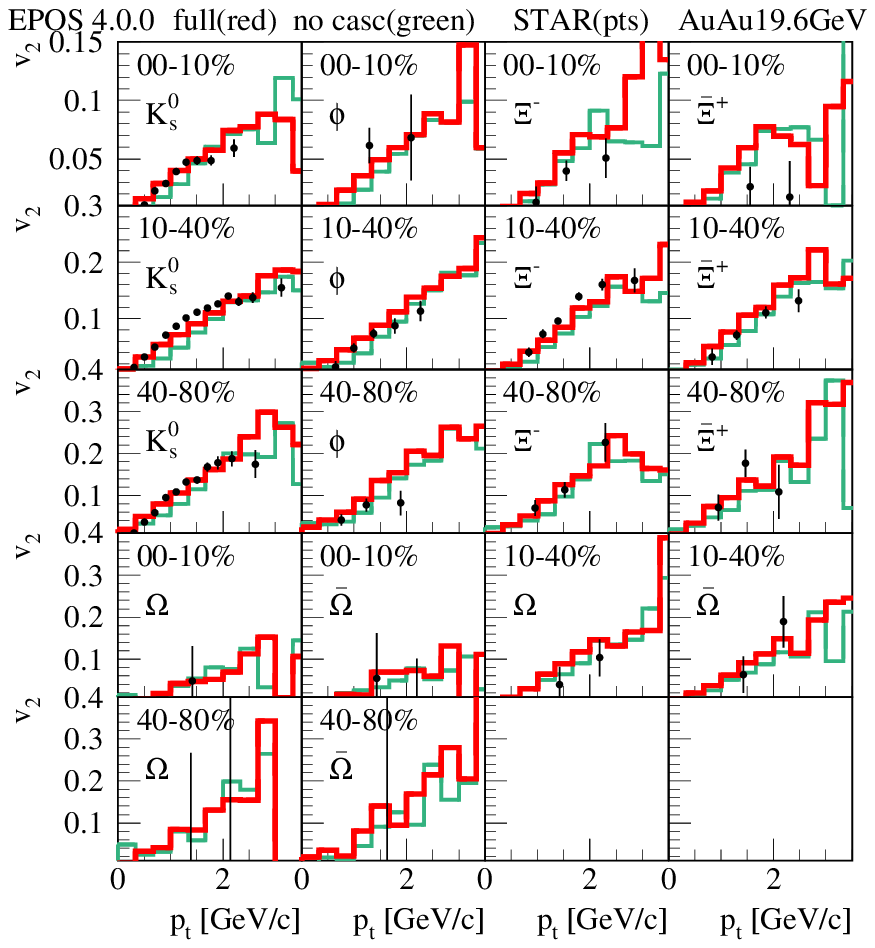}
\par\end{centering}
\centering{}\caption{Transverse momentum dependence of $v_{2}$ of identified particles
in AuAu collisions at 19.6 GeV at central rapidity for different centralities.
EPOS4 simulation, full simulations (thick red lines) and without hadronic
cascade (thin green lines), are compared to data from STAR \cite{STAR:2015-V2-identified-7to62}
(dots). \label{v2-19}}
\end{figure}

\begin{figure}[h]
\begin{centering}
\includegraphics[bb=40bp 35bp 470bp 495bp,clip,scale=0.6]
{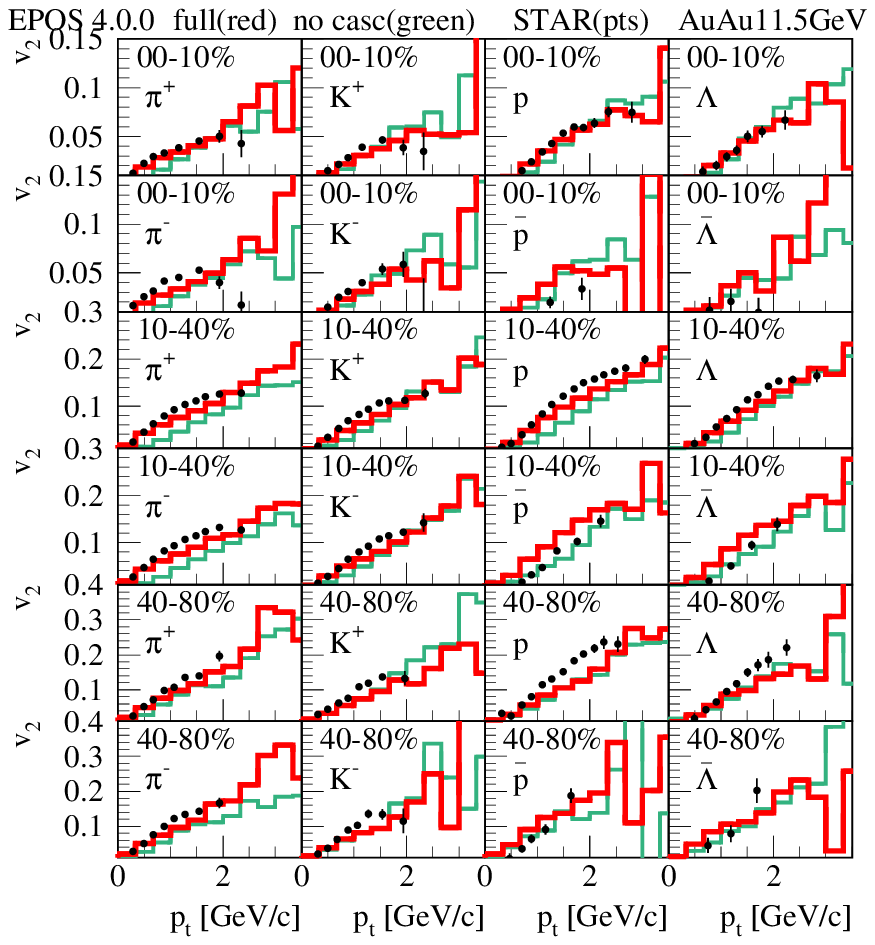}
\par\end{centering}
\begin{centering}
\includegraphics[bb=30bp 35bp 470bp 495bp,clip,scale=0.6]
{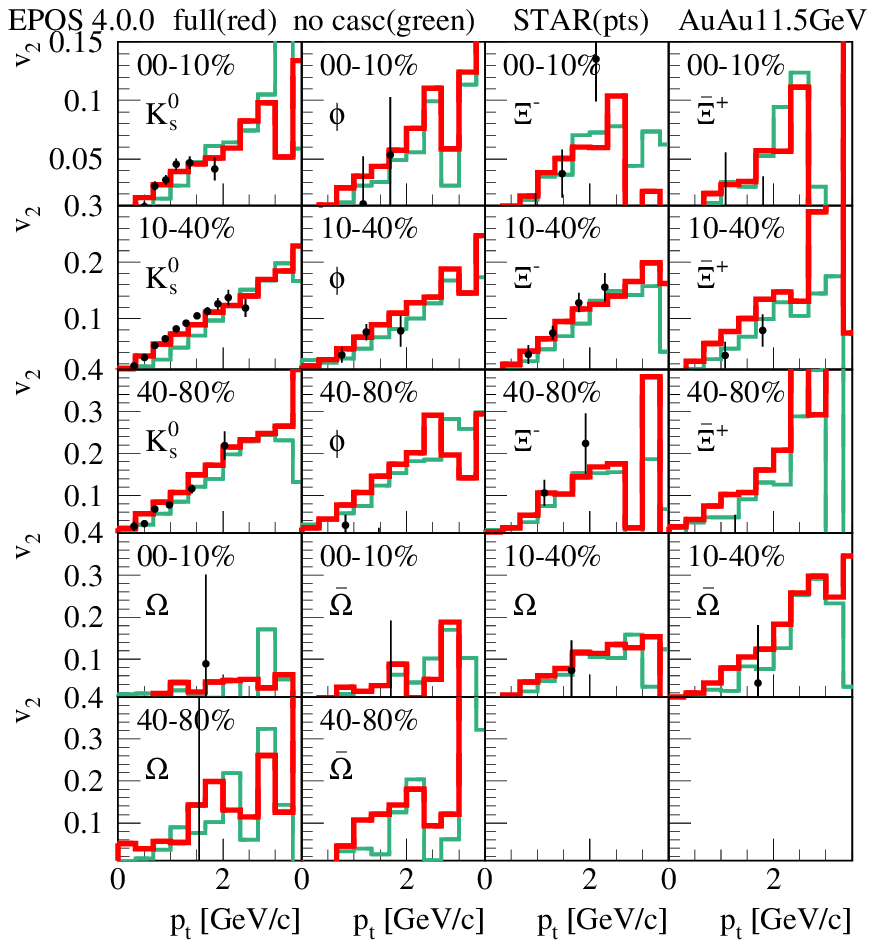}
\par\end{centering}
\centering{}\caption{Transverse momentum dependence of $v_{2}$ of identified particles
in AuAu collisions at 11.5 GeV at central rapidity for different centralities.
EPOS4 simulation, full simulations (thick red lines) and without hadronic
cascade (thin green lines), are compared to data from STAR \cite{STAR:2015-V2-identified-7to62}
(dots). \label{v2-11}}
\end{figure}

\begin{figure}[h]
\begin{centering}
\includegraphics[bb=40bp 35bp 470bp 495bp,clip,scale=0.6]
{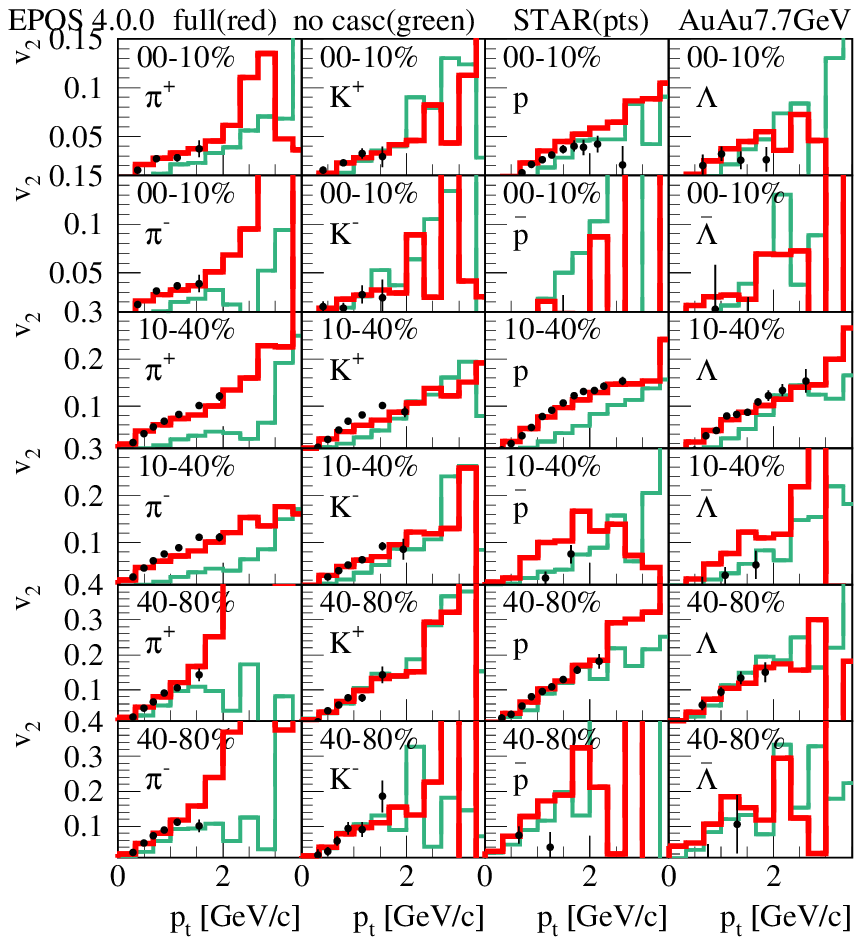}
\par\end{centering}
\begin{centering}
\includegraphics[bb=30bp 35bp 470bp 495bp,clip,scale=0.6]
{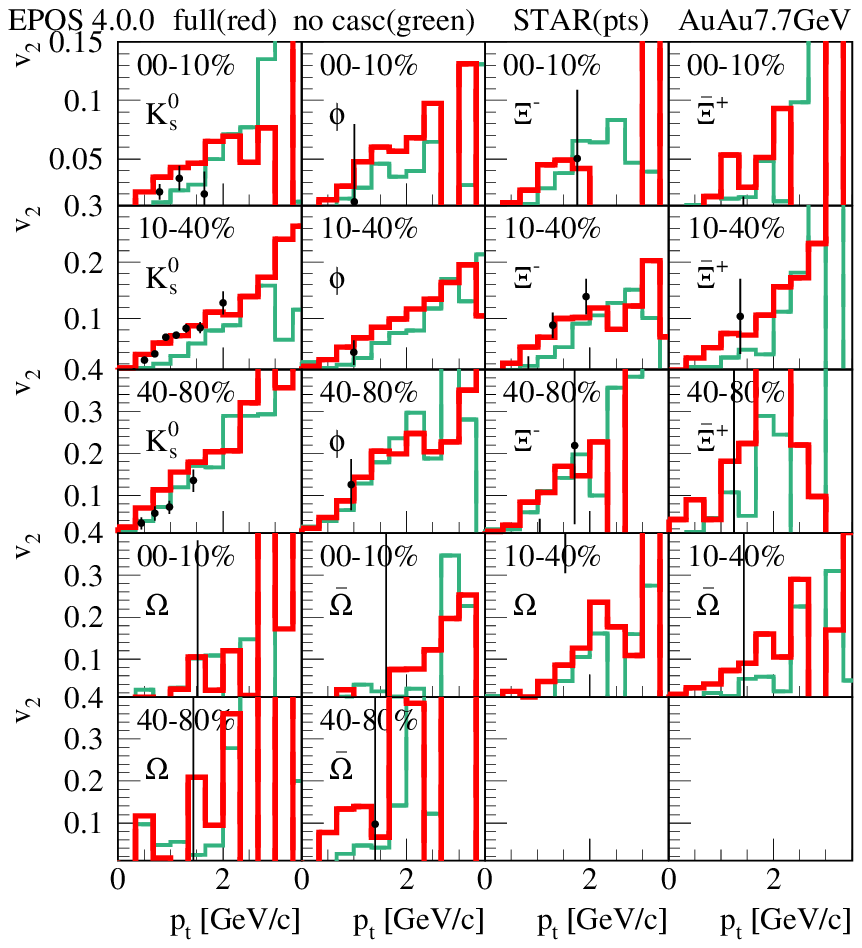}
\par\end{centering}
\centering{}\caption{Transverse momentum dependence of $v_{2}$ of identified particles
in AuAu collisions at 7.7 GeV at central rapidity for different centralities.
EPOS4 simulation, full simulations (thick red lines) and without hadronic
cascade (thin green lines), are compared to data from STAR \cite{STAR:2015-V2-identified-7to62}
(dots). \label{v2-7}}
\end{figure}

In Figs. \ref{v2-62}, \ref{v2-39}, \ref{v2-27}, \ref{v2-19}, \ref{v2-11},
and \ref{v2-7}, the transverse momentum dependence
\noindent 
\noindent 
of $v_{2}$ of identified particles 
within a pseudorapidity range of $\eta<1$
in AuAu collisions at energies
from 62.4 GeV down to 7.7 GeV are shown.  
Different centrality classes are considered: 0-10\%, 10-40\%, and 40-80\%. 
We show full simulations (thick red lines) and those without hadronic
cascade (thin green lines), and they
 are compared to data from STAR \cite{STAR:2015-V2-identified-7to62}
in the Beam Energy Scan at the Relativistic Heavy Ion Collider.
We employ the same event plane method as described in the experimental paper.
For the upper plots, one should note that the ranges on the ordinates change, depending
on the centrality, since $v_{2}$ increases with
decreasing centrality. In general, the simulation results describe
the data reasonably well, also the centrality dependence. The biggest deviation 
is observed for protons for 40-80\%. Also for the pions the agreement is not so great.
But the hyperons ($\Xi^-$ and $\Omega$ and their antiparticles $\bar{\Xi}^+$ and $\bar\Omega$) and as well the $\phi$ and the $K_s$ are relatively close to the data.  
$\quad$

\vspace{1cm}

\section{Summary and Conclusions}

We reviewed briefly the EPOS4 approach, a detailed discussion can
be found in \cite{werner:2023-epos4-overview,werner:2023-epos4-heavy,werner:2023-epos4-smatrix,werner:2023-epos4-micro}.
Most important is the concept of parallel scattering of primary
interactions, which is needed at high energies, to be more precise
above 24 GeV \cite{werner:2023-epos4-smatrix}. We also reviewed
briefly \textquotedblleft secondary interactions\textquotedblright{}
in EPOS4, composed of core-corona separation, hydrodynamic core evolution
with subsequent decay (microcanonically), and final state hadron cascade\\

In the EPOS4 formalism, there is a smooth transition from high to
low energy, certain features change (or disappear) gradually. The
parton ladders become less frequent, they are replaced by soft Pomerons,
and most importantly, the relative importance of particle production
from remnant excitation and decay increases. We discussed how this
affects the core-corona procedure. Below 30 GeV, even central rapidities
are dominated by particle production from remnants, and essentially
all prehadrons go into the core (for central collisions). We
computed the energy densities of the core (the fluid initial condition)
for the different systems, to observe that they drop from about $40\mathrm{\,GeV/fm^{3}}$
at 5.02 ATeV to roughly $5\mathrm{\,GeV/fm^{3}}$ at 11.5 GeV, and
then drop dramatically. At 4 GeV, there is no fluid anymore. \\

With all the model details already being published elsewhere 
\cite{werner:2023-epos4-overview,werner:2023-epos4-heavy,werner:2023-epos4-smatrix,werner:2023-epos4-micro},
the main purpose of the paper is the presentation of a very detailed test, considering a very large set
of experimental data in the RHIC energy domain, covering $p_t$ spectra and also the $p_t$ dependence of the elliptical flow, for identified particles. Spectra and $v_2$ provide complementary information about the fluid expansion. A particular aim 
was the investigation of a possible breakdown of the model at low energies, expected at around 24 GeV 
from theoretical considerations.  \\

We first showed comparisons of simulations with data concerning $p_{t}$
spectra of identified particles, from 39 GeV down to 7.7 GeV. For
the higher energies, down to 19.6 GeV, the simulation results are
relatively close to the data. At lower energies, we observe ``problems''.
At 11.5 GeV, we see for the first time (compared to higher energies)
significant deviations between simulation and data. The most striking
is a large proton excess at low $p_{t}$. At 7.7 GeV, essentially
all spectra from the simulation are too soft, the yields at low $p_{t}$
too high, with the biggest excess observed for protons and $K^{+}$
mesons. So here the model does not work.
These results concerning low $p_{t}$ yields are also summarized, more globally, 
through the integrated yields of different hadronic species 
displayed as a function of energy in the most central collisions.\\

The situation is quite different concerning $v_{2}.$ Here the simulation
results describe the data reasonably well, also the centrality dependence.
There is no ``significant deterioration'' at very low energies, as
in the case of transverse momentum spectra. Although the model is
obviously wrong at low energies, it works for $v_{2}$. \\

The success (at energies $\ge$ 19.6 GeV) and the failure (at energies
below 19.6 GeV) of the model correspond (amazingly well) to our earlier
estimate of where the model should work and where not: it should work
above 24 GeV (roughly estimated). The main problem of the (full)
parallel scenario at low energies is the fact that a given nucleon
hits all target nucleons on its way, although in reality it does not
exist anymore for the final scatterings. This explains the large
proton excess. One of the future projects will be to take that into
account.\vspace{1cm}

\subsection*{Acknowledgements}

IK acknowledges support by the grant GA22-25026S of the Czech Science Foundation. 
MS acknowledges support by the U.S. Department of Energy grant DE-SC0020651. 
J.J. acknowledges support by the MUSES collaboration under National Science 
Foundation (NSF) grant number OAC-2103680.

\vspace{2cm}

\noindent \textbf{\huge{}Appendix}{\huge\par}

\appendix

\section{Core and corona contributions for 27 GeV, 11.5 GeV, and
7.7 GeV }

\begin{figure}[H]
\includegraphics[bb=30bp 30bp 570bp 620bp,clip,scale=0.47]
{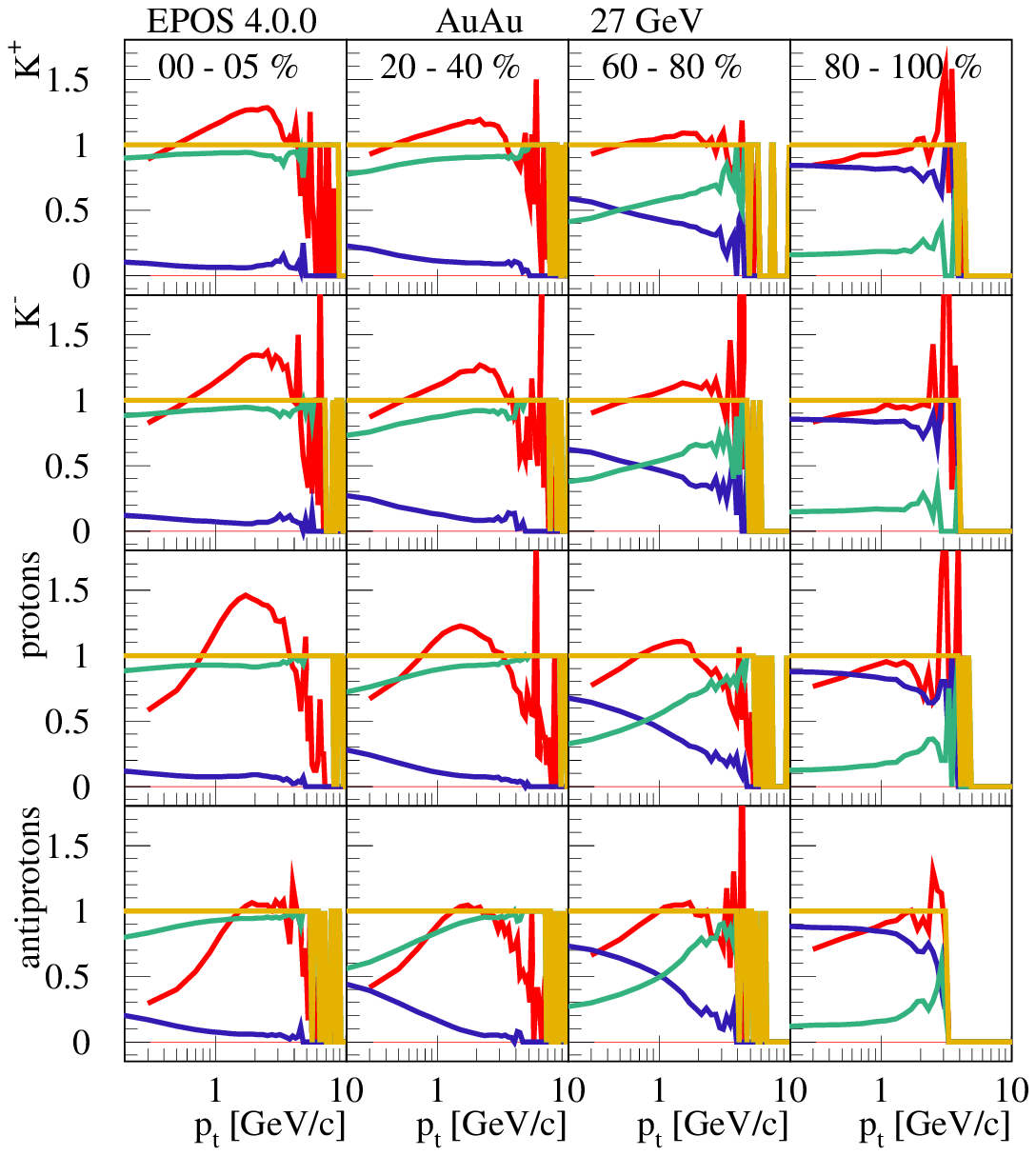}
\caption{The X / core+coron ratio, with X being the corona contribution
(blue), the core (green), and the full contribution (red),
for 4 centrality classes and four different particle species, for
27 GeV.\label{core-corona-4b}}
\end{figure}

\begin{figure}[H]
\includegraphics[bb=30bp 30bp 570bp 620bp,clip,scale=0.47]
{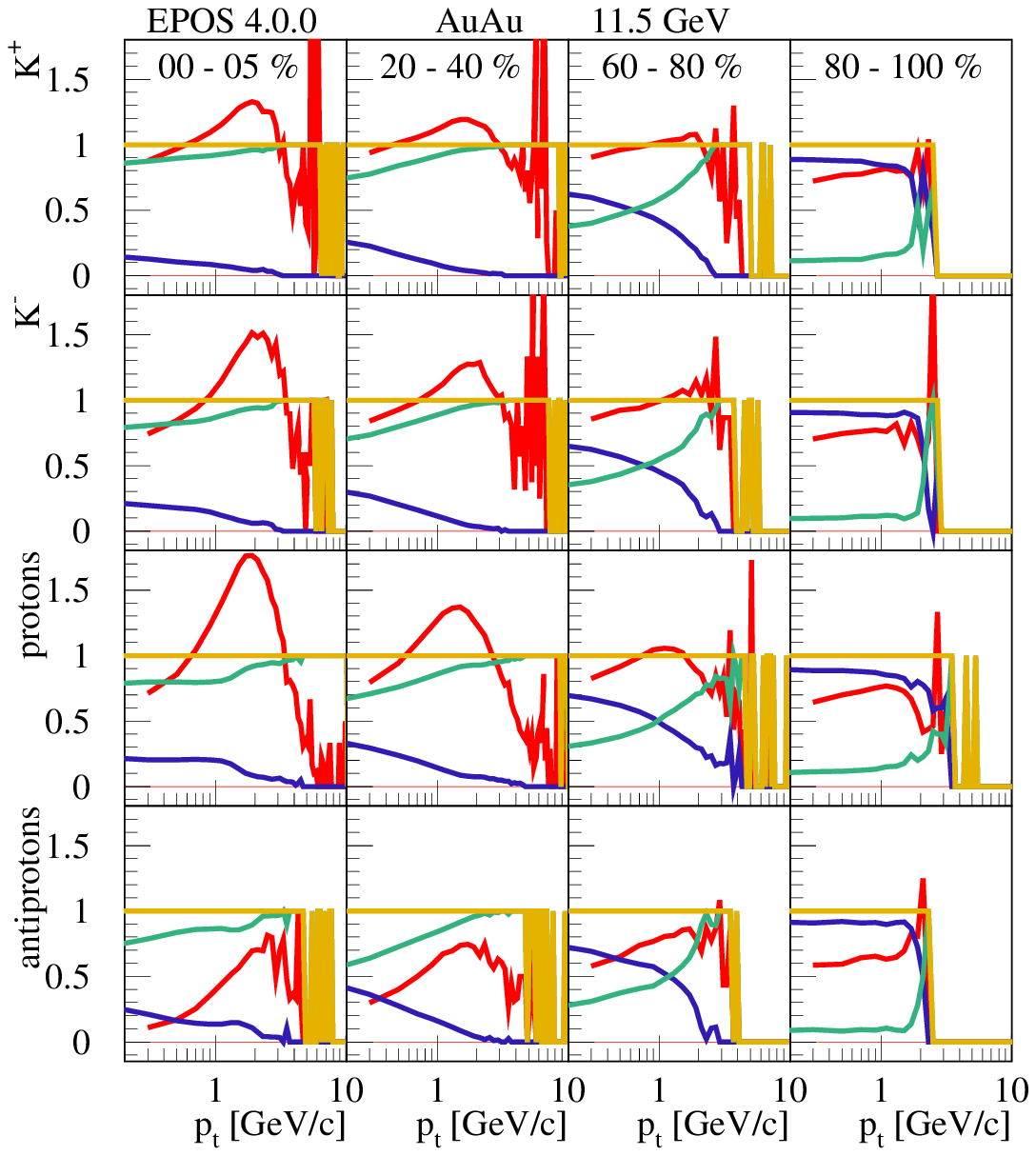}
\caption{The X / core+coron ratio, with X being the corona contribution
(blue), the core (green), and the full contribution (red),
for 4 centrality classes and four different particle species, for
11.5 GeV. \label{core-corona-6a}}
\end{figure}

\begin{figure}[H]
\includegraphics[bb=30bp 30bp 570bp 620bp,clip,scale=0.47]
{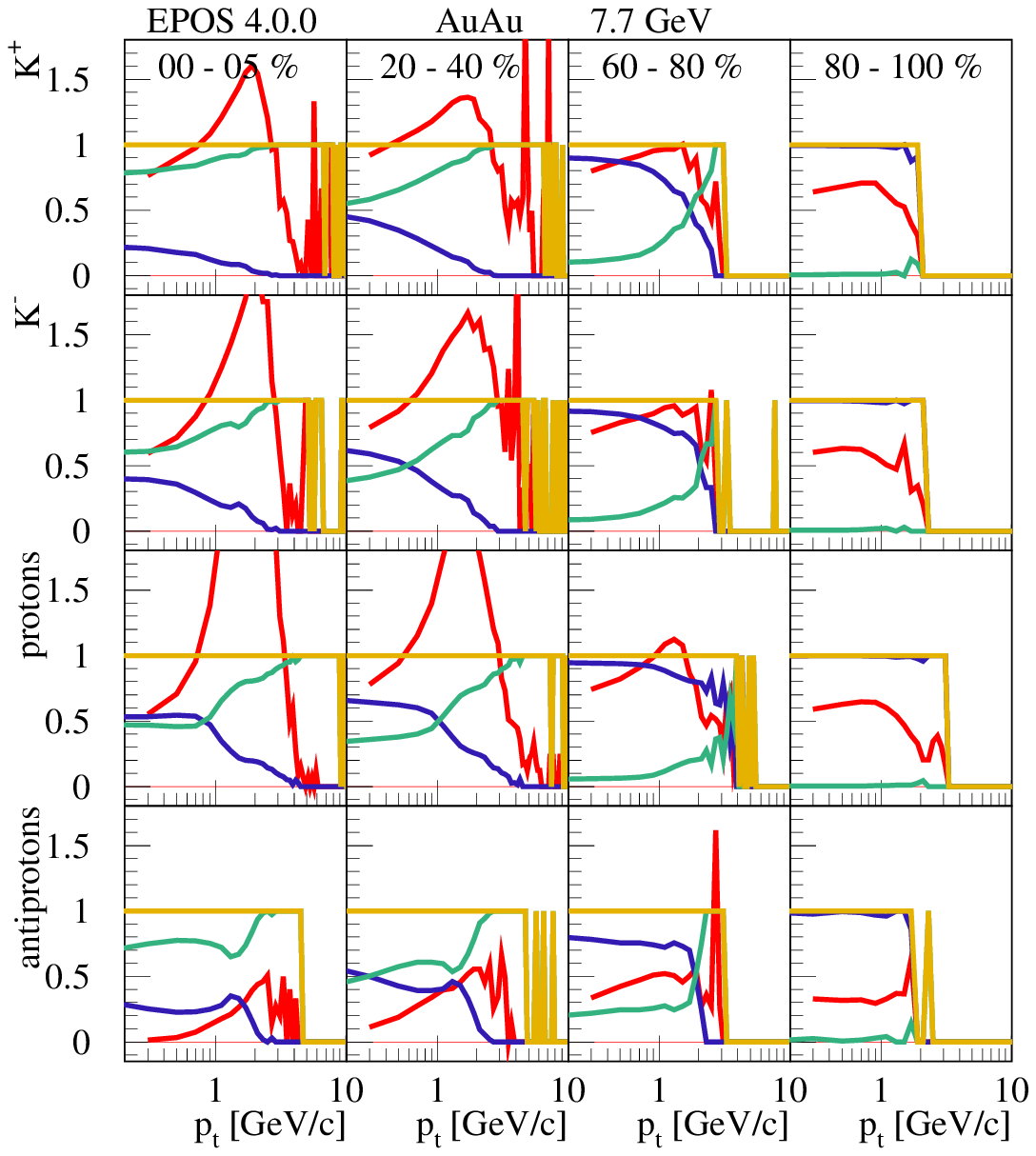}
\caption{The X / core+coron ratio, with X being the corona contribution
(blue), the core (green), and the full contribution (red),
for 4 centrality classes and four different particle species, for
 7.7 GeV. \label{core-corona-6b}}
\end{figure}

In Figs. \ref{core-corona-4b}, \ref{core-corona-6a}, and \ref{core-corona-6b}, we show ratios
X / core+corona versus $p_{t}$, with X being the corona contribution
(blue), the core (green), and the full contribution (red), 
for different hadrons. The four
columns represent four different centrality classes, namely 0-5\%,
20-40\%, 60-80\%, 80-100\%. We show results for AuAu collisions at
27 GeV, 11.5 GeV, and 7.7 GeV.

\end{document}